\documentclass[fleqn,usenatbib]{mnras}

\usepackage{amssymb}	
\usepackage{newtxtext,newtxmath}

\usepackage[T1]{fontenc}

\DeclareRobustCommand{\VAN}[3]{#2}
\let\VANthebibliography\thebibliography
\def\thebibliography{\DeclareRobustCommand{\VAN}[3]{##3}\VANthebibliography}

\usepackage{bibentry}
\usepackage{graphicx}	
\usepackage{amsmath}	
\usepackage{enumitem}
\usepackage{booktabs,tabularx,graphicx,amsmath,makecell,hhline,enumitem,gensymb,multirow,ulem,adjustbox}
\usepackage{bbding,pifont} 
\usepackage{hyperref} 
\usepackage[dvipsnames]{xcolor}
\newcommand{\mycomment}[1]{}

\definecolor{greybox}{RGB}{255, 99, 71}
\definecolor{redcircle}{RGB}{190, 10, 30}
\definecolor{apoor}{RGB}{255, 99, 71}
\definecolor{arich}{RGB}{65, 105, 255}

\definecolor{bimodal}{RGB}{255, 20, 147}
\definecolor{unimodal}{RGB}{138, 43, 226}
\definecolor{multimodal}{RGB}{30, 144, 255}
\definecolor{smeared}{RGB}{50, 205, 50}
\definecolor{GSE}{RGB}{255, 99, 71}

\definecolor{GSE1}{RGB}{100, 149, 237}
\definecolor{GSE2}{RGB}{60, 179, 113}
\definecolor{GSE3}{RGB}{255, 99, 71}

\title[The formation of discs and chemical sequences]{The Milky Way in context: The formation of galactic discs and chemical sequences from a cosmological perspective}

\author[M. D. A. Orkney et al.]{Matthew D. A. Orkney,$^{1,2}$\thanks{E-mail: morkney@icc.ub.edu}
Chervin F. P. Laporte,$^{1,2,3,4}$ Robert J. J. Grand,$^{5}$ and Volker Springel$^{6}$\vspace{0.1cm}\\
$^{1}$Institut de Ci\`{e}ncies del Cosmos (ICCUB), Universitat de Barcelona, Mart\'{i} i Franqu\`{e}s 1, E-08028 Barcelona, Spain\\
$^{2}$Institut d'Estudis Espacials de Catalunya (IEEC), E-08034 Barcelona, Spain\\
$^{3}$LIRA, Observatoire de Paris, Universit\'e PSL, Sorbonne Universit\'e, Universit\'e Paris Cit\'e, CY Cergy Paris Universit\'e, CNRS, 92190 Meudon, France\\
$^{4}$Kavli IPMU (WPI), UTIAS, The University of Tokyo, Kashiwa, Chiba 277-8583, Japan\\
$^{5}$Astrophysics Research Institute, Liverpool John Moores University, 146 Brownlow Hill, Liverpool, L3 5RF, UK\\
$^{6}$Max Planck Institut f{\"u}r Astrophysik, Karl-Schwarzschild-Straße 1, 85748 Garching bei München, Germany \\
}

\date{Accepted 2025 September 10. Received 2025 September 6; in original form 2025 June 6}

\pubyear{2025}

\begin{document}
\label{firstpage}
\pagerange{\pageref{firstpage}--\pageref{lastpage}}
\maketitle

\begin{abstract}
%
We study the formation of chemical sequences in stellar discs of Milky Way (MW)-mass galaxies in a full cosmological context with the {\sc auriga} simulations, with implications for both the MW and external galaxies like M31. The analysis focuses on the conditions giving rise to bi-modal $\alpha$-chemistry and the potential influence of mergers (e.g. Gaia-Enceladus, GSE).
We find a wide diversity of chemical sequences, without correlation between the emergence of dichotomous $\alpha$-chemistry and GSE-like mergers.
The transition between multiple $\alpha$-sequences is sequential, and is mediated by modulations in the star formation rate (SFR). In some cases, this can be caused by the starburst and subsequent quiescence induced by mergers. In others, by exhaustion or violent disruption of the gas disc. Realisations with singular sequences either lack significant modulations in their SFR, or form too late to have a significant high-$\alpha$ sequence.
The metallicity overlap between the high-$\alpha$ and low-$\alpha$ sequences (as seen in the Solar neighbourhood of the MW) arises from accretion of metal-poor gas from the circum-galactic medium. This depends on gas disc thickness, with thinner discs losing their metal-poor extremities. Gas donation from singular gas-rich merger events are incapable of driving long-lived metal dilution ($\Delta\rm{[Fe/H]} \gtrsim 0.3$), and we rule-out this scenario for the low-$\alpha$ sequence in the MW.
Finally, the shape of $\alpha$-sequences in the [Fe/H] versus [Mg/Fe] plane is related to long-term SFR trends, with a downward slanted locus (as observed in the low-$\alpha$ sequence of the MW) owing to a sustained or declining SFR.
\end{abstract}




\section{Introduction} \label{sec:introduction}

Our vantage point within the Milky Way (MW) has allowed for comprehensive spectroscopic surveys into the chemo-dynamic properties of individual stars across large statistical samples and wide volumes (see RAVE \citealt{rave}, SEGUE \& SEGUE-2 \citealt{segue, segue-2}, Gaia-ESO \citealt{gaia-eso}, LAMOST \citealt{lamost}, GALAH \citealt{galah}, APOGEE \citealt{apogee}). This, combined with precise proper motions from \textit{Gaia} \citep{gaia, gaiadr2, gaiadr3}, has led to a broader understanding of how the MW formed and established our Galaxy as a benchmark for studying the underlying physics of galaxy formation as a whole.

High resolution spectroscopic studies of stars in the Solar neighbourhood have revealed that disc stars form a clear ``bi-modal'' distribution in the [Fe/H] versus [$\alpha$/Fe] chemical abundance planes, with two distinct sequences that are separated in their $\alpha$-chemistry but overlap in their metallicity distributions \citep[e.g.][]{fuhrmann1998, haywood2013, anders2014, nidever2014, hayden2015}. This bi-modality varies along with the location within the disc, both in terms of radius and height above or below the plane \citep[e.g. see][]{queiroz2020}. For inner radii (within $\approx6\,$kpc), the sequences form a single track mediated by an under-populated gap \citep[e.g.][]{hayden2015, queiroz2020}. For outer radii (beyond $\approx12\,$kpc), the proportion of stars in the high-$\alpha$ sequence is strongly diminished \citep[e.g.][]{bensby2011, anders2014}.

The high-$\alpha$ and low-$\alpha$ sequences are well correlated with stars in the thick and thin discs respectively, and share similar qualitative properties. Whereas the high-$\alpha$ disc is uniformally old (having formed the majority of its stars earlier than 10\,Gyr ago), centrally concentrated and vertically extended --- the low-$\alpha$ disc is thin, radially extended and formed over a wide time period up to the present day \citep{haywood2013, mackereth2017, bonaca2020, miglio2021, queiroz2023}. This is consistent with extra-Galactic observational evidence that galaxy formation proceeds in an inside-out and top-down manner \citep{franx2008, dokum2013, patel2013, vanderwel2014, lian2024}, a result that is backed up by idealised and fully cosmological simulations \citep[e.g.][]{bournaud2009, brook2012, bird2013}. Yet, it is increasingly believed that the high/low-$\alpha$ and thick/thin discs are distinct populations of their own \citep[e.g. see arguments in ][]{robin1996, juric2008, bovy2012, bovy2016}.  Nonetheless, the populations are often treated as interchangeable and chemical cuts are commonly employed to select the thick and thin disc populations.

Whilst observational data shows that thick and thin discs are a frequent attribute of galaxies as a whole \citep[e.g.][]{mould2005, yoachim2006, tsukui2024}, the same determination has not yet been made for chemical bi-modalities. Extra-galactic observations rely on integrated light, which blends stellar populations across disc regions \citep[e.g.][]{scott2021}. Furthermore, observations at higher redshift increasingly rely on gas-phase chemistry, which may not confer the chemical distribution of the stars. The one exception is our nearest neighbour M31. Early Keck/DEIMOS (Deep Imaging Multi-Object Spectrograph) spectroscopy of 70 M31 stars found no sign of dichotomous chemistry \citep{escala2020}, and near-infrared spectroscopy of around 100 stars from the James Webb Space Telescope (JWST) shows only a single sequence at high-$\alpha$ \citep{nidever2024}. This is an early sign that dichotomous chemistry may not be an ubiquitous quality of massive disc galaxies, which suggests a diversity of formation channels.

If the formation mechanism behind the chemical bi-modality was well known, one could debate whether it was a unique development to the MW. Despite many competing hypotheses, there has been no universal agreement on a dominant mechanism. These hypotheses fall into two categories: secular or episodic formation. An example of a secular mechanism would be the chemical evolution models of \citet{schonrich2009a, schonrich2009}. These find that low-metal and high-$\alpha$ stars form in the centrally-compact proto-Galaxy \footnote{In which a proto-galaxy is defined as the most ancient progenitor to modern-day galaxies, that has undergone minimal star formation or chemical enrichment \citep[see for e.g.][]{belokurov2022}}, and are then driven to the Solar neighbourhood by radial migration \citep[as in][]{sellwood2002, rovskar2008}. There, they overlap with younger, low-$\alpha$ stars in the radially extended thin disc. This scenario is backed up by chemical enrichment models in \citet{sharma2021, chen2023}, $N$-body simulations in \citep{loebman2011}, and could also explain other properties of the disc such as the radial dependence of the metallicity distribution function \citep{loebman2016}. However, other models with radial migration fail to reproduce two chemical sequences \citep{minchev2013, minchev2014, johnson2021, dubay2024}.

Otherwise, the two $\alpha$-sequences may form at the same time but in different Galactic environments. The authors of \citet{clarke2019, beraldo2021} argue that high-$\alpha$ stars could form in the dense intermediate-mass gas clumps that arise from disc fragmentation in the early Universe \citep[see][]{elmegreen2005, elmegreen2007, dekel2009, livermore2012}, within which the star formation rate (SFR) is raised and the $\alpha$-enrichment from core-collapse supernovae is enhanced. Then, low-$\alpha$ stars can form concurrently in a more evenly distributed gas disc. This scenario could simultaneously explain the elevated velocity dispersion of high-$\alpha$ stars via strong scattering processes \citep{bournaud2009, amarante2020b}. However, observations from the CANDELS survey (The Cosmic Assembly Near-infrared Deep Extragalactic Legacy Survey) find that only up to 20 per cent of the integrated SFR comes from within clumps (see \citealt{wuyts2012}, for galaxies over a similar redshift range and mass scale as the proto-MW).

An episodic formation mechanism might suggest that a bi-modality requires a hiatus in the star formation between the formation of the high-$\alpha$ and low-$\alpha$ sequences \citep[as in][]{haywood2016}. This could be due either to a shrinking and then growth of the star-forming gas disc \citep[i.e.][]{grand2018}, or a suppression of the SFR following merger interactions or feedback from AGN \citep{beane2024a, beane2024b}. The pause in the star formation allows the delayed iron enrichment from Type-Ia supernovae to `catch up', thereby reducing the ratio of [$\alpha$/Fe] prior to the onset of the low-$\alpha$ sequence. Indeed, there is tentative evidence for a quenching period around 8\,Gyr ago from closed-box chemical evolution models in \citet{haywood2018} and APOGEE DR17 data in \citet{spitoni2024}.

It could also depend on distinct epochs of gas inflow, as in the two-infall model of \citep{chiappini1997, chiappini2001}. This proposes that the high-$\alpha$ sequence formed with a high SFR from a primordial rapid gas inflow, followed by a lower SFR from subsequent metal-poor gas accretion. The second infall `resets' the ambient metallicity of the gas disc to lower values, thereby also explaining the signature [Fe/H] overlap between the two sequences in the Solar neighbourhood \citep[see][]{grisoni2017, spitoni2019, spitoni2021, spitoni2023}. Prior works have discussed gas-rich mergers as the sources of a second gas infall \citep[e.g.][]{richard2010, snaith2016, mackereth2018, buck2020}, or as a trigger that stimulates disc growth leading to low-$\alpha$ formation \citep[see examples in][]{grand2018}.

The low-$\alpha$ disc may have formed from a prolonged fuelling from the circum-galactic medium (CGM). Studies have identified cold inflowing gas filaments \citep[e.g.][]{angles2017, agertz2021b}, re-accretion of early gas outflows \citep[e.g.][]{khoperskov2021}, and smooth accretion from the CGM \citep[e.g.][]{parul2025} as the dominant sources of gas during the formation of low-$\alpha$ sequences in simulations.

Evidence from the stellar halo has now revealed the signature of an ancient massive merger, hereafter the Gaia-Sausage Enceladus or GSE \citep{helmi2018, belokurov2018}. This merger is predicted to have arrived in the MW around 8-11\,Gyr ago \citep{vincezo2019, belokurov2020, naidu2021, xiang2022}, roughly during the transition between the two $\alpha$-sequences \citep[e.g.][]{linden2017, sahlholdt2022}, with a mass scale similar to that of the present-day Small Magellanic Cloud. Numerous works argue that the GSE played an important role in the history of the MW, with implications for its disc chemistry \citep[e.g.][]{buck2020, buck2023, ciuca2024}, the hot kinematics of the thick disc population \citep{helmi2018, bonaca2020, grand2020, chandra2024} and the warp observed in the outer disc \citep{dillamore2022, dodge2023, deng2024}. While evidence for the GSE merger is strong, its exact mass, infall angle, and impact on the proto-Galaxy remain uncertain. In particular, its exact relation to the chemical bi-modality is still circumstantial. This is a topic we intend to probe further.

In this work, we will investigate the properties and formation of chemical sequences in the {\sc auriga} simulation suite of MW-mass galaxies. Our ultimate goal is to apply insights from a statistical sample of cosmological realisations to the formation of chemical sequences, and to improve our understanding of the MW formation history in a cosmological context. We seek to address the following scientific questions:
\begin{enumerate}
\setlength{\itemindent}{1.35em}
    \item What are the physical channels through which chemical sequences form in disc galaxies?
    \item How do simulated chemical sequences qualitatively compare to that of the MW, and what does this tell us about the physical processes at play?
    \item Are mergers, GSE-like or otherwise, a necessary component in forming a chemical bi-modality at all?
    \item How do GSE-like mergers affect the formation and evolution of the disc, and how is this translated into the present-day structure and kinematic properties of the disc?
\end{enumerate}

We describe our methods and post-processing in section \ref{sec:methods}. Our results are given in Section \ref{sec:results}, where we consider the properties of chemical distributions at $z=0$ in Section \ref{sec:diversity} and the chemical evolution of sequences in Section \ref{sec:SFH}. In Section \ref{sec:bimod_origin} we investigate how distinct $\alpha$-sequences can arise. We discuss how our findings can be interpreted with respect to observations and other simulations in Section \ref{sec:discussion}. Finally, we present our conclusions in Section \ref{sec:conclusions}.

\section{Methods} \label{sec:methods}

\subsection{{\sc auriga} simulation suite}

The simulations analysed in this paper are drawn from the {\sc auriga} project \citep{Auriga, grand2024}, a suite of thirty magneto-hydrodynamic simulations of isolated MW-mass galaxies evolved within a full cosmological context. The initial conditions are generated at $z=127$ using the Gaussian white-noise realisation {\sc panphasia} \citep{Panphasia}, and have a periodic cosmological box with a side length of $100\,$cMpc. The suite uses cosmological parameters from \citet{Planck2014}, which are $\Omega_{\rm m}=0.307$, $\Omega_{\rm b}=0.04825$, $\Omega_{\rm \Lambda}=0.693$ and a Hubble constant of $H_0=100h\,\text{km}^{-1}\,\text{Mpc}^{-1}$, where $h=0.6777$. \par

The {\sc auriga} simulations incorporate a comprehensive suite of sub-grid physics to model galaxy formation in a cosmological setting. These include treatments for a spatially uniform photoionizing UV background, primordial and metal-line cooling, star formation, stellar evolution, feedback from supernovae (SNe), supermassive black hole growth and feedback, as well as magnetohydrodynamics. A detailed overview of these physical prescriptions is provided in \citet{Auriga, grand2024}. The simulations successfully reproduce a variety of key galactic properties expected in a cosmological context \citep{vogelsberger2013, marinacci2014, genel2014}. The {\sc auriga} galaxies themselves exhibit characteristics consistent with observations of MW-like galaxies, including their halo mass–metallicity relations \citep{monachesi2019}, vertical disc oscillations \citep{gomez2017}, rotation curves and SFRs \citep{Auriga}, and detailed chemodynamical structures in their central regions \citep{fragkoudi2020}. Crucially, they also produce the presence of thick and thin disc components \citep{grand2018}, both in terms of geometric, kinematic and chemical properties.

The {\sc auriga} galaxy formation models employ a two-phase interstellar medium, consisting of cold clouds immersed in a hot gaseous background. Star formation occurs stochastically in gas above a threshold density of $0.13\,m_{\rm p}\,\mathrm{cm}^{-3}$, with the target stellar particle mass matching the initial gas cell resolution. Each star particle represents a simple stellar population following a Chabrier initial mass function \citep{chabrier2003}. Mass and metals are returned to the gas via sub-grid models for AGB winds, Type-II, and Type-Ia supernovae. Here, the returns from Type II supernovae are assumed to occur instantaneously, whilst those from Type Ia supernovae follow a time-delay distribution resulting in a characteristic enrichment timescale of $\approx1\,$Gyr. Further details on the physics prescriptions, including the adopted yield tables, are provided in the {\sc auriga} public release \citep{grand2024} and introductory paper \citep{Auriga}. \citet{grand2024} also discusses how these choices compare to the physics implemented in the TNG simulations \citep{nelson2019}.

Throughout this work, we use the ``level-4'' resolution versions of each {\sc auriga} simulation. Here, the high-resolution target regions of the cosmological box are resolved with DM particle masses of  $\sim3\times10^5\,\text{M}_{\odot}$ and initial baryonic gas cell masses of $\sim5\times10^4\,\text{M}_{\odot}$. For convenience, we hereafter refer to the {\sc auriga} simulations as Au-$i$, where $i$ indicates the particular halo.

\subsection{Post-processing} \label{sec:post}

The group and subhalo properties are calculated using the {\sc subfind} halo finder \citep{subfind}, and the virial masses and radii are derived using a sphere of mean density $200\times$ the critical density of the universe. The evolution of halo structure is tracked across timesteps using the {\sc lhalotree} merger tree algorithm \citep{springel2005}. When necessary, inter-snapshot subalo properties are estimated using fitted splines. These are 3-D cubic splines for subhalo orbits, and a PCHIP (Piecewise Cubic Hermite Interpolating Polynomial) 1-D monotonic cubic interpolation for subhalo properties. \par

The {\sc auriga} simulations track nine chemical species. We use Magnesium (Mg) for our estimation of $\alpha$-abundance because it has been shown to be a reliable tracer of the $\alpha$-process, with a high separation between the thick and thin discs \citep{jonsson2018}. All chemical abundances in this work are normalised to Solar values using \citet{asplund2009}, with further corrections of order 0.4\,dex used to lower the metallicity ratio ([Fe/H]). This work is often focused on the distribution of stars in the [Fe/H] versus [Mg/Fe] elemental abundance ratios, and for brevity we will hereafter refer to this simply as the `chemical abundance plane'.

We focus exclusively on stars in the disc, following selection criteria based on both the stellar orbital circularity (using the definition for orbital circularity given in \citealt{abadi2003}) and metallicity as described in the methods section of \citet{orkney2023}. To summarise, we calculate the circularity distribution of all star particles and assume all retrograde stars belong to the stellar halo. Then, we mirror this distribution around $\eta=0$ to find the probability that prograde stars belong to the halo. Prograde star particles are probabilistically assigned to the halo using a random number generator, and the remaining stars are classified as disc. Additionally, we enforce that the retrograde and prograde halo components share an equivalent metallicity distribution function, though we note this has little implication on the final selections.

Unless stated otherwise, we include stars from both \textit{in-situ} and \textit{ex-situ} sources, where the \textit{in/ex-situ} designation is calculated based on whether the star particle is bound to the host halo at the earliest available snapshot. This is necessary to account for the presence of any `\textit{ex-situ} disc'  \citep[see][]{gomez2017}. We note that the \textit{ex-situ} contributions neither materially alter our reported results, nor do they create features in chemical abundance space that might be misinterpreted as independent chemical sequences (with Au-1 being the one exception). Furthermore, it is unlikely that the MW disc is dominated by \textit{ex-situ} stars \citep[e.g.][]{ruchti2011}. The galaxies are aligned at every snapshot on the angular momentum of the youngest (formed in the last 3\,Gyr) and innermost ($R_{\rm G}<0.1\times R_{200}$, where $R_{\rm G}$ is the galacto-centric radius) \textit{in-situ} star particles. \par

We calculate merger mass ratios based on the ratio between the total {\sc subfind} masses (dark matter, gas and stars) as opposed to the commonly used ratio of $M_{200}$ virial masses. This is because the virial mass will often include massive substructures, and as a consequence the time-evolution of the mass ratio is more stochastic. The quoted merger mass ratio is then found as the maximum ratio after the satellite has joined the FoF (Friends of Friends) group of the host structure. This usually occurs just after the satellite joins the FoF group, because gravitational tides and ram pressure stripping will begin to degrade it thereafter.

\begin{figure*}
\centering
  \setlength\tabcolsep{2pt}%
    \includegraphics[keepaspectratio, trim={0.0cm 0.0cm 1.5cm 0.0cm}, width=1.0\linewidth]{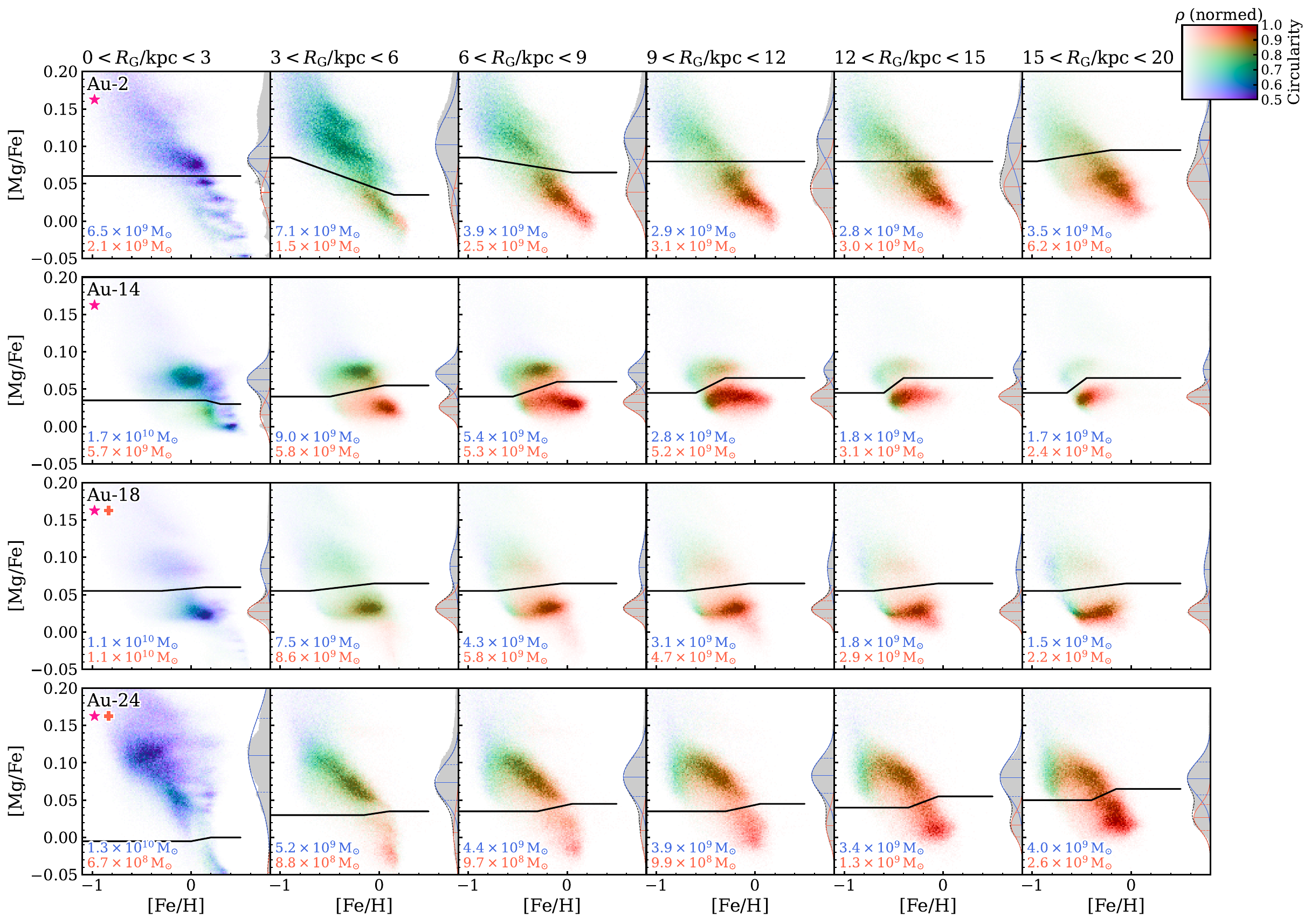}\\
\caption{Examples of chemical sequences in four {\sc auriga} galaxies (where the simulation name is indicated in the top-left corner of the left-most panels), over a range of radial bins. Stars with halo-like kinematics have been excluded, following the methods described in Section \ref{sec:methods}. The coloured 2-D histograms indicates the normalised mass-weighted density of star particles in units of M$_{\odot}\,\rm{dex}^{-1}\,\rm{dex}^{-1}$. Whereas stars in the MW are typically depicted with a logarithmic weighting, we opt for a linear weighting to better emphasise the main overdensities. The histogram colour represents the mean orbital circularity within each pixel, as indicated by the colourbar in the top-right, where red colouration corresponds to thin-disc kinematics. Each panel includes grey 1-D histograms that show the [Mg/Fe] abundance ratio distributions, where we overplot Gaussian profile fits to the high- (\textcolor{arich}{\textbf{blue}}) and low-$\alpha$ (\textcolor{apoor}{\textbf{red}}) selections. These selections are shown with a black line.}
\label{fig:metal_planes}
\end{figure*}

\section{Results} \label{sec:results}

\subsection{A diversity of chemical sequences in {\sc auriga}} \label{sec:diversity}

\begin{figure*}
\centering
  \setlength\tabcolsep{2pt}%
    \includegraphics[keepaspectratio, trim={0.0cm 0.0cm 0.0cm 0.0cm}, width=0.8\linewidth]{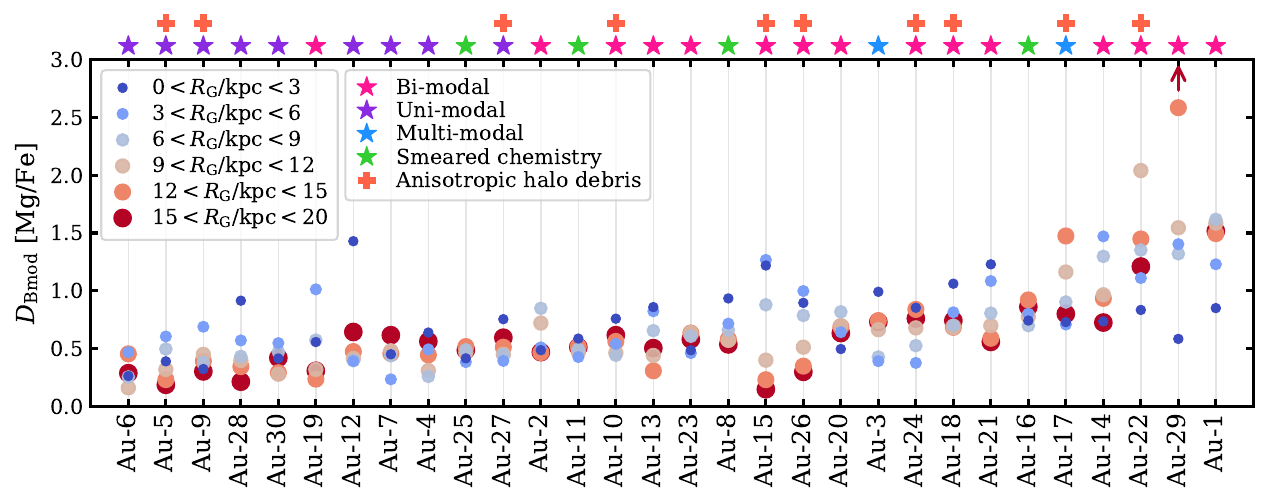}\\
\caption{A measure of the bi-modality strength across the {\sc auriga} suite, as defined by a modified Bhattacharyya distance in Equation \ref{equ:B_D}. Stars are split into a range of radial bins, as given in the left legend. The ordering of simulations along the $x$-axis is determined by their average value of $D_{\rm Bmod}$. The coloured markers along the top $x$-axis denote the type of chemical distribution as judged by eye (see the right legend), with an additional \textcolor{GSE}{$\mathord{\text{\ding{58}}}$} symbol to highlight galaxies that have a stellar halo debris feature (as found in the grey-dashed box of figure 3 in \citealt{fattahi2019}, galaxies that have a stellar halo with a component of velocity anisotropy $\beta>0.8$ and contribution fraction $>0.5$). These features were linked to specific massive mergers with radial infalls \citep[and see also][]{orkney2023}. There is no strong correlation between $D_{\rm Bmod}$ and the presence of anisotropic halo debris.}
\label{fig:bimods}
\end{figure*}

We show stellar mass-weighted histograms in the chemical abundance plane for a sample of {\sc auriga} galaxies in Figure \ref{fig:metal_planes}, where the full suite is included in an \href{https://drive.google.com/drive/folders/1vVSkF2Q65KrIaz1Gn2xBbBpKVPeSBIK2?usp=sharing}{online repository} \footnote{We echo the acknowledgment in \citet{grand2018} that the $\alpha$-abundances in {\sc auriga} do not necessarily align with the chemistry expected for the MW, and refer the reader to their discussion section where these concerns are explored.}. The four chosen examples were selected to exhibit different behaviours; \textbf{Au-2}: with a slanted and narrow track through the chemical abundance plane, reminiscent of the distribution observed in the MW at sub-Solar radii \citep[e.g. figure 6 of ][]{queiroz2020}; \textbf{Au-14}: with a clear bi-modality exhibiting a centrally-dominant high-$\alpha$ sequence, formed in a galaxy that experiences multiple gas-rich mergers on shallow infall trajectories; \textbf{Au-18}: with a clear chemical dichotomy that forms in a realisation with a GSE-like merger occurring at a lookback time of $\tau=9.3\,$Gyr, and a significant metallicity overlap between each chemical sequence; \textbf{Au-24}: with another chemical dichotomy forming in a realisation with a GSE-like merger, this time at $\tau=8.5\,$Gyr. This simulation exhibits a slanted high-$\alpha$ sequence similar to the chemical distribution in the MW, but a less massive low-$\alpha$ sequence that has a narrow metal distribution ($\Delta\rm{[Fe/H]}\approx0.3$) and hardly any overlap with the metallicity of the high-$\alpha$ sequence, much unlike the MW.

Each panel features a black line, determined by eye, which separates the stars into high-$\alpha$ and low-$\alpha$ selections. Manual selections like these are commonly employed in both observational data and simulated analogues, and our reported results are not sensitive to the exact selection choices. We adjust the selections with each radial bin in order to account for the radial dependence of the chemical sequences in some realisations, which ensures a minimal contamination between the high-$\alpha$ and low-$\alpha$ sequences. The total stellar mass in each selection is indicated in the lower-left corner of each panel. For the purposes of this investigation, we \textit{always} separate the stars into a high-$\alpha$ and low-$\alpha$ selection even when just one dominant sequence is present. All chemical sequences display some level of granularity in their distribution, and so it is always possible to define an appropriate separation even in the case of single sequences.

We further divide the chemical abundance plane into six radial bins spanning $0<R_{\rm G}/\rm{kpc}<20$, which brackets the majority of disc stars at $z=0$. We note that there are also trends with height above and below the disc plane, as seen in the MW \citep[e.g.][]{nidever2014, minchev2014, hayden2015, queiroz2020} and also in {\sc auriga} at a higher resolution level \citep{grand2018}. Whilst we do not show these trends explicitly, we find that the proportion of stars in each $\alpha$-sequence is dependent on the $|z|$-height, with stars at higher [Mg/Fe] typically gaining mass dominance at greater $|z|$-heights (especially around and within Solar radii). This is in agreement with observational data of the MW from APOGEE \citep[e.g.][]{hayden2015}, and is therefore likely to be a general outcome of galaxy formation at MW-mass scales rather than a property that is unique to the MW.

The histograms are coloured by the mean orbital circularity of stars within each pixel (see the definition in \citealt{orkney2023}), revealing some interesting features. Firstly, stars in the high-$\alpha$ selection tend to have lower orbital circularities than those in the low-$\alpha$ selection, in-keeping with expectations from our own Galaxy (see for example figure 5 in \citealt{chandra2024} and figure 11 in \citealt{khoperskov2024}). This pattern is consistent with a top-down, inside-out formation scenario: the older high-$\alpha$ stars formed when the galaxy was more radially compact but vertically extended, resulting in, on average, less circular orbits. In line with this picture, high-$\alpha$ stars tend to contribute more mass at lower orbital radii than their low-$\alpha$ counterparts. We will show that this is indeed the case for one example simulation in Section \ref{sec:disc_metallicity_variation}.

Secondly, stars at lower [Fe/H] tend to have lower orbital circularities even at a fixed [Mg/Fe]. This behaviour can also be seen in the MW: see the low-$\alpha$ sequence below a metallicity of approximately $\rm{[Fe/H]}=-0.3$ in figure 6 of \citealt{chandra2024}. One interpretation would be that these low-metal stars formed at older times, meaning they have undergone more kinematic perturbations that might heat their orbits, or that they formed at a time when the disc was of a lower orbital circularity. However, both here and in the MW, this wide metallicity spread in the low-$\alpha$ sequence is more accurately a proxy of the underlying radial metallicity gradient than any temporal evolution (i.e. \citealt{frankel2018, frankel2020, patil2024, cerqui2025}). This is because the stellar disc has kinematically cooled and grown to greater radii during the formation of the low-$\alpha$ sequence, such that the radial metallicity gradient dominates over the temporal metallicity evolution. We will argue in Section \ref{sec:disc_metallicity_variation} that the lower circularity of these low-metal and low-$\alpha$ stars owes to their formation from recently accreted gases that have not yet fully settled into the rotating gaseous disc.

\begin{figure*}
\centering
  \setlength\tabcolsep{2pt}%
    \includegraphics[keepaspectratio, trim={0.0cm 0.0cm 0.0cm 0.0cm}, width=1\linewidth]{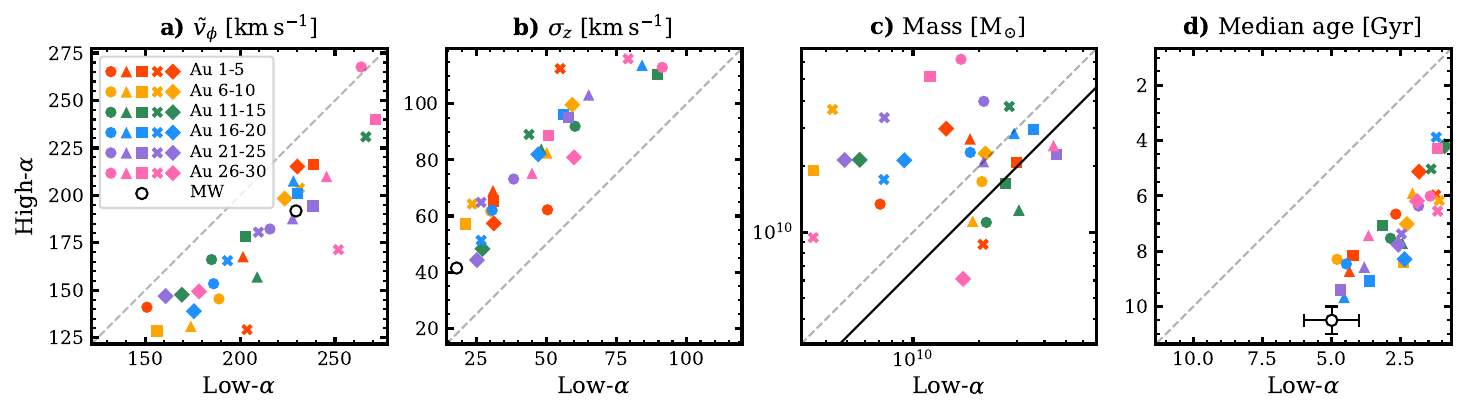}\\
\caption{A comparison of four salient properties for the high- and low-$\alpha$ selections described in Section \ref{sec:bimods} (for stars in the radial range $0<R_{\rm G}/\rm{kpc}<20$). In order; \textbf{a)} The median azimuthal velocity, \textbf{b)} The standard deviation of the velocity in the $z$-direction, \textbf{c)} the summed stellar mass, \textbf{d)} the median of the stellar age. The values for each {\sc auriga} simulation are provided by unique markers as indicated in the legend. In black, we show expected values for the MW, where kinematic properties are taken from \citet{anguiano2020}, and the high/low-$\alpha$ mass ratio value is taken from \citet{khoperskov2024}.}
\label{fig:bimod_props}
\end{figure*}

\subsubsection{Quantifying chemical bi-modalities} \label{sec:bimods}

From a cursory examination of the full suite, there is clearly a wide variation in the shape and number of $\alpha$-sequences. To aid comparison, we manually classify the chemical distributions into four groups:
\begin{itemize}
\setlength{\itemindent}{0.5em}
    \item \textcolor{bimodal}{$\text{\ding{72}}$} A bi-modality.
    \item \textcolor{unimodal}{$\text{\ding{72}}$} A uni-modality.
    \item \textcolor{multimodal}{$\text{\ding{72}}$} A multi-modality, for cases where there are more than two substantial loci in the chemical abundance plane.
    \item \textcolor{smeared}{$\text{\ding{72}}$} Smeared chemistry, for cases with a diffuse smear of stars across the chemical abundance plane, that do not adhere to obviously independent sequences.
\end{itemize}

In some instances, the classification would change depending on the radial bin considered, and so we prioritise the classification that seems to fit the best overall. Coloured stars corresponding to each of our four classifications are appended to the first column of panels in Figure \ref{fig:metal_planes}, and will be used elsewhere in this work where appropriate. We also include a \textcolor{GSE}{$\mathord{\text{\ding{58}}}$} symbol to indicate simulations that possess a radially anisotropic stellar halo feature loosely comparable to the GSE in the MW. These realisations are taken from the grey-dashed box selection in Figure 3 of \citet{fattahi2019}, which includes stellar haloes that have features corresponding to a velocity anisotropy of $\beta>0.8$ and a stellar halo contribution fraction of $>0.5$. The majority of the stellar mass in these features owes to ancient massive mergers with radial infalls \citep[and see also][]{orkney2023}, though not all of these mergers are GSE-like in the totality of their properties (i.e. their infall time or merger mass ratio). The mergers in Au-18 and Au-24 are perhaps the most comparable to expectations of the GSE.

For a more quantitative analysis, we fit a Gaussian profile to the manual selections used in Figure \ref{fig:metal_planes}, and then estimate the overlap between the two profiles by using the Bhattacharyya distance \citep{Bhattacharyya1946} ($D_{\rm B}$):
\begin{equation}
D_{\rm B}(p,q) = \frac{1}{4} \ln{\left [\frac{1}{4} \left (\frac{\sigma_{p}^2}{\sigma_{q}^2} + \frac{\sigma_{q}^2}{\sigma_{p}^2} + 2 \right ) \right ]} + \frac{1}{4} \frac{(\mu_{p} - \mu_{q})^2}{\sigma_{p}^2 + \sigma_{q}^2},
\label{equ:B_D1}
\end{equation}
where $p,q$ correspond to two normal distributions, $\mu$ is the mean and $\sigma^2$ is the variance.

We use the assumption that the chemical sequences are well-described by Gaussian profiles, despite the lack of direct physical justification for this choice. Nevertheless, the Gaussian models provide good fits in nearly all cases (e.g., see the histograms along the right $y$-axis in Figure \ref{fig:bimods}). We also remind the reader that our manual selections divide the chemical plane into the most representative high-$\alpha$ and low-$\alpha$ populations, even in cases where the distribution appears predominantly uni-modal or smeared. In the case of smeared distributions, the Gaussian fits are far poorer.

This simple estimate will yield higher values in cases where the two Gaussians have less overlap. However, there may be cases where one of the two sequences has only a negligible mass fraction, and this should never be considered as a `strong' bi-modality. Therefore, we modify $D_{\rm B}$ as such:
\begin{equation}
D_{\rm Bmod}(p,q) = D_{\rm B}(p,q) \times \sqrt{\frac{\min(A_{p},A_{q})}{\max(A_{p},A_{q})}},
\label{equ:B_D}
\end{equation}
where $A$ is the Gaussian amplitude, meaning the final value is reduced in the case of uneven Gaussian amplitudes.

We calculate $D_{\rm Bmod}$ over each radial bin and display the results in Figure \ref{fig:bimods}. This shows that there is a wide variation in the `goodness' of bi-modalites within {\sc auriga}, both between realisations and also at different radii. Generally, the values of $D_{\rm Bmod}$ are well correlated with our visual classifications. Bi-modal realisations have higher values and uni-modal realisations have lower values, but realisations with `smeared' chemistry are not well-modeled by two Gaussians and should be discounted.

The figure also demonstrates that radial variation, even between neighbouring radial bins, can have a significant impact on the goodness of the chemical bi-modality. This can be due to both a variation in the mass budget in high-$\alpha$ and low-$\alpha$ sequences, but also due to the changing chemical abundances of stars. An example of the latter effect can be seen in Au-15, where two distinct sequences present at $3<R_{\rm G}/\rm{kpc}<6$ begin to converge towards a single mean [Mg/Fe] value at progressively higher radii.

\subsubsection{Correlation with GSE-like halo debris}

For the reasons discussed in our introduction, it may be expected that realisations with GSE-like mergers would be more likely to have two distinct chemical sequences. However, from Figure \ref{fig:bimods} we find that simulations hosting GSE-like debris features (marked with a \textcolor{GSE}{$\mathord{\text{\ding{58}}}$} symbol) have chemical abundance distributions at both low and high values of $D_{\rm Bmod}$. Furthermore, there are examples of galaxies with strong chemical dichotomies that do not experience GSE-like mergers (i.e. Au-14, Au-29).

Whilst GSE-like mergers do not seem to be a prerequisite in forming multiple chemical sequences, there may still be a requirement for some kind of merger activity. As an example, Au-25 has a particularly quiescent accretion history with the last $>1:30$ mass ratio merger occurring before a lookback time of $\tau=11\,$Gyr. The evolution through the chemical abundance plane is smooth and almost featureless. This is in opposition to results from \citet{prantzos2023}, which finds that a chemical bi-modality can develop from secular evolution even in the absence of mergers. We will return to this idea in Section \ref{sec:SFH}.

\subsubsection{Properties of each chemical sequence} \label{sec:bimod_props}

\begin{figure*}
\centering
  \setlength\tabcolsep{2pt}%
    \includegraphics[keepaspectratio, trim={0.0cm 0.0cm 0.0cm 0.0cm}, width=1\linewidth]{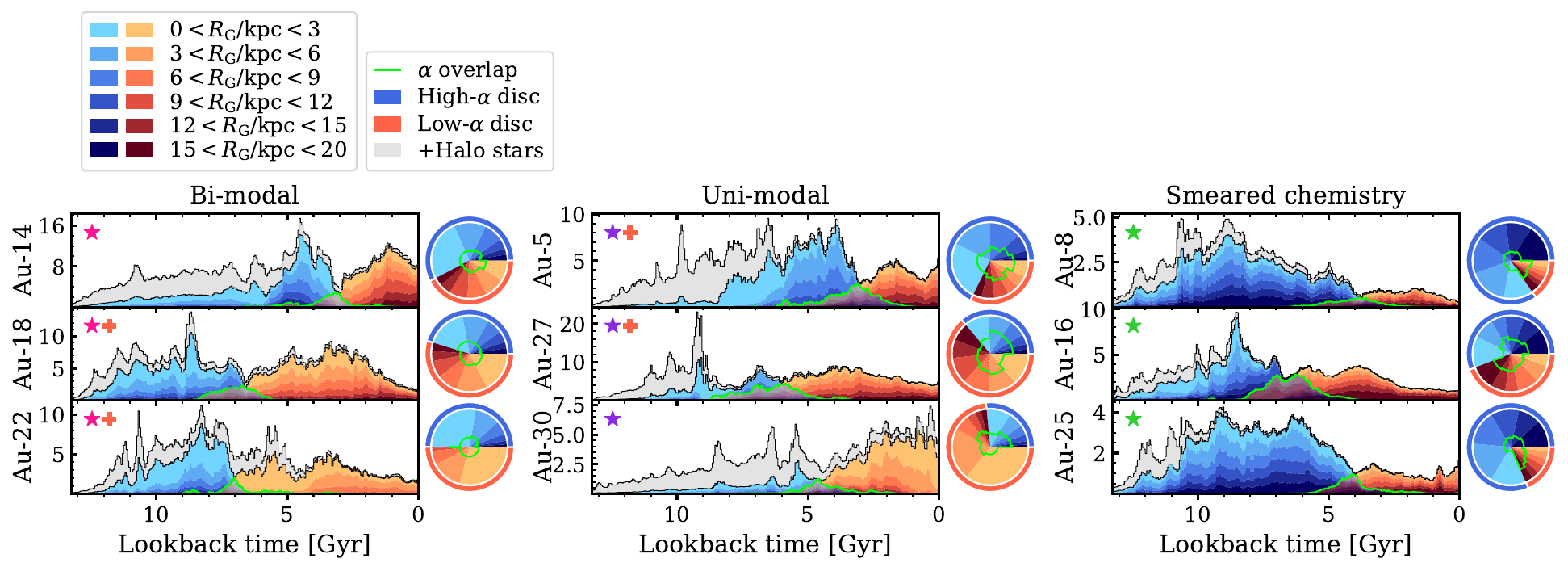}\\
\caption{Included are a selection of {\sc auriga} realisations with bi-modal (column 1), uni-modal (column 2) and smeared chemistry (column 3). The entire suite can be seen in Appendix \ref{AppendixSFH}. Each panel shows the star formation history in units of $\rm{M}_{\odot}\,\rm{yr}^{-1}$. The \textcolor{arich}{\textbf{blue}} and \textcolor{apoor}{\textbf{red}} histograms are based on the high-$\alpha$ and low-$\alpha$ selections defined in \ref{sec:bimods}. These selections are subdivided into stacked histograms corresponding to distinct radial bands (based on the stellar positions at $z=0$), as described in the left legend. The total overlapping region is highlighted with a green outline, which shows the contribution of stars that formed coevally. A light-grey histogram shows the additional contribution of stars on halo-like orbits. The pie charts on the right-hand-sides of each panel show the total mass ratios within each radial band. The coeval fraction is once again represented with a green line, but here it is calculated per radial band.}
\label{fig:SFH}
\end{figure*}

We present a comparison of four different properties for the high-$\alpha$ and low-$\alpha$ selections in Figure \ref{fig:bimod_props}: a) the median azimuthal velocity, b) the standard deviation of the velocity in the $z$-direction, c) the total stellar mass and d) the median stellar age. These properties are sampled for stars at $z=0$, over the radial range $0<R_{\rm G}/\rm{kpc}<20$, and once again exclude stars with halo-like kinematics following the methods described in \ref{sec:methods}. The figure shows that, whilst many of the galaxies contain no dichotomous chemical distribution, there are still very predictable overall trends which are shared with the MW. The low-$\alpha$ selections rotate at higher velocities (with Au-26 being the one exception), are kinematically cooler in the vertical direction, and are younger by 2\,Gyr or more. These results are an inevitable outcome of the progressive enrichment in MW-mass galaxies, whereby high-$\alpha$ stars are born at higher lookback times. The stellar disc at those times is often rotating at a lower velocity and was kinematically hotter, and/or has been subjected to a greater amount of kinematic heating (e.g. mergers and secular evolution).

We include reference values for the MW in black. The kinematic properties are taken from APOGEE and \textit{Gaia} DR2 data in \citet{anguiano2020}, who decomposed the Galaxy based on chemical selections of the thick and thin discs. In comparison, the {\sc auriga} simulations possess stellar discs that are kinematically hotter. This may owe to the numerical limitations inherent to simulations that resolve stellar distributions with discrete mass elements (e.g. particle noise), and cosmological simulations often struggle in reproducing the cooler kinematics observed in the MW \citep[e.g.][]{house2011, sanderson2020}. The mass ratio between our high-$\alpha$ and low-$\alpha$ selections has a near random scatter both above and below the $1:1$ dashed-grey line. We compare against the high-$\alpha$ to low-$\alpha$ mass ratio estimate of $4:6$ from the work of \citet{khoperskov2024}, which is based upon an orbit superposition approach. We compare the median age distributions of each sequence with loose predictions that are cited throughout the academic literature. The comparative late formation of the {\sc auriga} sequences may either be an indication that the MW itself is an outlier with an especially front-loaded assembly history, or that the {\sc auriga} chemical evolution favours later disc assembly.

A key result in panel d) is the wide variation of formation times for both high-$\alpha$ and low-$\alpha$ sequences. The formation of a bi-modality can then be considered to be the time when a low-$\alpha$ sequence has grown a significant mass fraction, which is qualitatively similar to the median low-$\alpha$ formation times shown here. This highlights that chemical bi-modalities depend on the unique assembly and star formation history of each galaxy rather than intrinsic universal formation epochs, a topic we explore in the next section.

\subsection{The time formation of chemical sequences} \label{sec:SFH}

The star formation histories for a selection of {\sc auriga} galaxies are presented in Figure \ref{fig:SFH}, where we consider examples that have bi-modal, uni-modal and smeared chemistry. The full suite can be found in Appendix \ref{AppendixSFH}. Each star formation history is represented with two overlaid histograms for the high-$\alpha$ (\textcolor{arich}{\textbf{blue}}) and low-$\alpha$ (\textcolor{apoor}{\textbf{red}}) selections, including all stars in the range $0<R_{\rm G}/\rm{kpc}<20$ at $z=0$.

The lookback times where the two distributions overlap is outlined in green. Whilst many realisations appear to have a significant fraction of coeval stars, these are not necessarily indicative of a coeval formation of the high-$\alpha$ and low-$\alpha$ sequences. In fact, the formation of two chemical sequences in {\sc auriga} progresses mostly sequentially in all cases. Instead, the fraction of coeval stars are primarily made up of the intermediate-$\alpha$ abundances between the higher and lower sequences. Regardless, it is immediately evident that there are a negligible number of stars in the low-$\alpha$ selections forming at $\tau>9\,$Gyr, similar to what is observed in the MW (though see \citealt{borbolato2025}).

These two histograms are subdivided into six stacked histograms, representing the radial regions considered throughout this work. The total mass budgets in each region are shown in the pie charts appended to the right hand side of each panel, which once again includes a green line to show the overlap between high-$\alpha$ and low-$\alpha$ stars (though here, the overlapping fraction is restricted to stars within each radial bin). In most cases, the stars found in the innermost radial range of $0<R_{\rm G}/\rm{kpc}<3$ form either the majority or the plurality of the total stars for both high-$\alpha$ and low-$\alpha$ selections (though with the low-$\alpha$ stars in `smeared chemistry' examples proving an exception). Yet, it is also common that the overlapping fraction is minimised in this innermost region. This is due to a combination of a few factors. Firstly, the SFR is the most bursty in the inner few kpc, mainly due to its sensitive response to subhalo interactions and inflows. Secondly, gas mixing in inner regions occurs on shorter timescales, such that there is a narrow chemical abundance range at any one time. Thirdly, there is a lower fraction of stars that migrated from other birth radii where the disc chemistry might be different (which we will show in Section \ref{sec:migration}). These effects lead to a higher-definition or `clumpy' distribution of stars in the chemical abundance plane.

We stack additional histograms in light-grey, where the thickness represents all remaining stars that do not have disc-like orbits at $z=0$. These can be stars that formed before the development of the stellar disc, stars that naturally formed in regions outside of the disc, or disc stars that have transitioned to halo-like kinematics due to kinematic heating. In some cases, large numbers of disc stars are scattered into the halo and onto higher radial orbits during a major merger event. Signs of these major mergers can be seen with sudden and short-lived spikes in the SFR, such as Au-27 at $\tau=9.5\,$Gyrs. In the most extreme cases, the scattering depletes the mass budget of stars in the high-$\alpha$ sequence, which in turn reduces the strength of any potential bi-modality.

The transitionary epoch between high-$\alpha$ and low-$\alpha$ selections are often preceded by either a period of rising and falling SFR, or a short-lived merger-induced starburst. This is especially evident in realisations with strong dichotomous chemistry, i.e. the `bi-modal' examples. We now explain how the interplay between SFR and chemical enrichment can contribute to the formation of a bi-modality: A period of rapid star formation results in a pileup of stars at the current $\alpha$-enrichment of the proto-galaxy. The ensuing $\alpha$-enrichment via rapidly exploding Type-II SNe further encourages the formation of stars with raised [$\alpha$/Fe] abundances. Both of these mechanisms combined result in the formation of a high-$\alpha$ sequence. Then, as the SFR falls, the iron enrichment from Type-Ia SNe begins to take effect (with a lag of approximately 1\,Gyr). The $\alpha$-to-iron ratio drops to a new level, and the low-$\alpha$ sequence begins to form. This is, in essence, a description of the $\alpha$-knee as in \citet{tinsley1979, tinsley1980, mcwilliam1997}.

Whilst some `uni-modal' realisations share this SFR feature, there are circumstances that have inhibited the formation of a bi-modality. For example, in Au-27 the merger-induced starburst at $\tau=9.5\,$Gyr occurred when the host galaxy had not grown much beyond 3\,kpc. As a result, some chemical dichotomy is present in the innermost regions but does not reach as far as the Solar radius. The same scenario can be seen occurring for Au-12 in Appendix \ref{AppendixSFH}. There is no significant rise and fall in the SFR outside of the galactic centre, and this is shown to greater depth in Section \ref{sec:SFH2}.

Across the entire suite, there is a great variation in the star formation history for realisations with bi-modal and uni-modal chemistry. Certainly, there are multiple routes to forming different chemical morphologies. In contrast, the `smeared chemistry' examples display very similar characteristic histories: an early radial growth of the disc, and then a gradual reduction in the SFR until the present day. This early radial growth is a consequence of high angular momenta accretions from the cosmic web during the early galactic assembly. The relatively steady SFR over all radial bins, and the scarcity of disruptive radial merger events when compared with the full suite, results in a smooth chemical enrichment through time. As such, these realisations lack a chemical bi-modality.

\begin{figure}
  \setlength\tabcolsep{2pt}%
    \includegraphics[keepaspectratio, trim={0.0cm 0.5cm 0.0cm 0.0cm}, width=1\columnwidth]{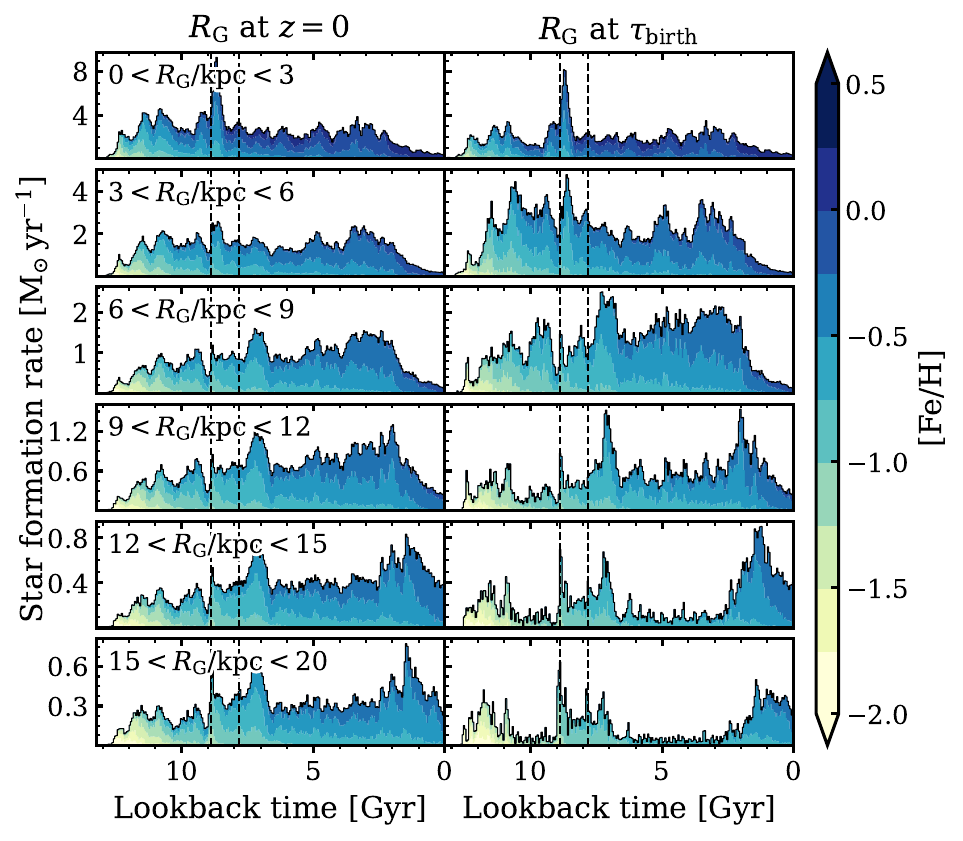}\\
\caption{Each row shows the \textit{in-situ} SFR of Au-18 for stars in different radial regions (as indicated in the top-left corners). The left column is for stars at their $z=0$ positions, whereas the right column is for positions closest to the snapshot that the star was born in (approximately the birth radius). In all panels, the histograms are subdivided into stars of different metallicities, as described in the colourbar. Vertical black dashed lines mark the pericentres of important mergers. Additional plots for the full suite are included \href{https://drive.google.com/drive/folders/1QFg0YvuOLpgyXGf-nUKu0tqos9BnZFSl?usp=sharing}{online}.}
\label{fig:SFHr}
\end{figure}

We focus on the star formation history of Au-18 in Figure \ref{fig:SFHr}, showing only the \textit{in-situ} stars. As in Figure \ref{fig:SFH}, there is a peak in the star formation history at $\tau=9\,$Gyr that is most prevalent in the innermost radial region, which corresponds to a starburst that is induced by the accretion of a GSE-like merger. This is one example of starburst-induced formation of the chemical thick disc that is described in \citet{grand2018}. However, these stars contribute a much lower mass fraction to the high-$\alpha$ sequence in the Solar neighbourhood ($6<R_{\rm G}/\rm{kpc}<9$).

The high-$\alpha$ stars in the Solar neighbourhood arise mostly from a second SFR peak at around $\tau=7\,$Gyr. Curiously, this peak is due to the passage of an extremely minor satellite galaxy with a merger mass ratio of less than $1:100$ at the time of infall, and a peak \textit{total} mass of just $8\times10^9\,\rm{M}_{\odot}$. It passes beyond the edge of the gas disc at $\tau = 7.8\,$Gyr (at which time the disc is mostly confined to within $10\,$kpc) and briefly disrupts its symmetry, causing star-forming gas to flow beyond the Solar region on one side of the galaxy. This yields a delayed star formation peak that is most prevalent around the Solar neighbourhood, but specifically \textit{not} at inner radii as with other merger-induced starbursts.

From the decomposition with [Fe/H], there is no obvious sign of a long-lived dilution in the metallicity, as might be expected following the two-infall model of \citet{chiappini1997, chiappini2001}. Instead, there is a sudden and short-lived dilution corresponding to the starburst at $\tau=9\,$Gyr, though this occurs during the formation of the high-$\alpha$ sequence and is not relevant for the later formation of the low-$\alpha$ sequence. Otherwise, there is a gradual progression from metal-poor to metal-rich, with almost a dex of variation at all times and across most radial regions.

The impact of radial migration can be seen when comparing the first and second columns of Figure \ref{fig:SFHr}. There is an extinction in the SFR in the two highest radial bins at birth time ($12<R_{\rm G}/\rm{kpc}<15$, $15<R_{\rm G}/\rm{kpc}<20$) between $7>\tau/\rm{Gyr}>2$. Before this time period, the compact gas disc is regularly disturbed by merger passages, which act to temporarily propagate star-forming gases to higher radii. For the duration of the extinction time period, the accretion history is extremely quiescent and the gas disc remains undisturbed. The few stars that do form have a low and narrow range of metallicities. By $z=0$, this star formation valley has been repopulated by migrating stars from the inner radial regions. In turn, the range of metallicities broadens and extends to higher values. We will discuss the affects of radial migration further in Section \ref{sec:migration}. We note that this star formation valley at the highest radii does not contribute to the formation of the two $\alpha$-sequences in Au-18, as it does in some other {\sc auriga} simulations \citep{grand2018}.

\subsubsection{The bi-modality as a consequence of star formation rate} \label{sec:SFH2}

\begin{figure*}
\centering
  \setlength\tabcolsep{2pt}%
    \includegraphics[keepaspectratio, trim={0.0cm 0.0cm 0cm 0.0cm}, width=1\linewidth]{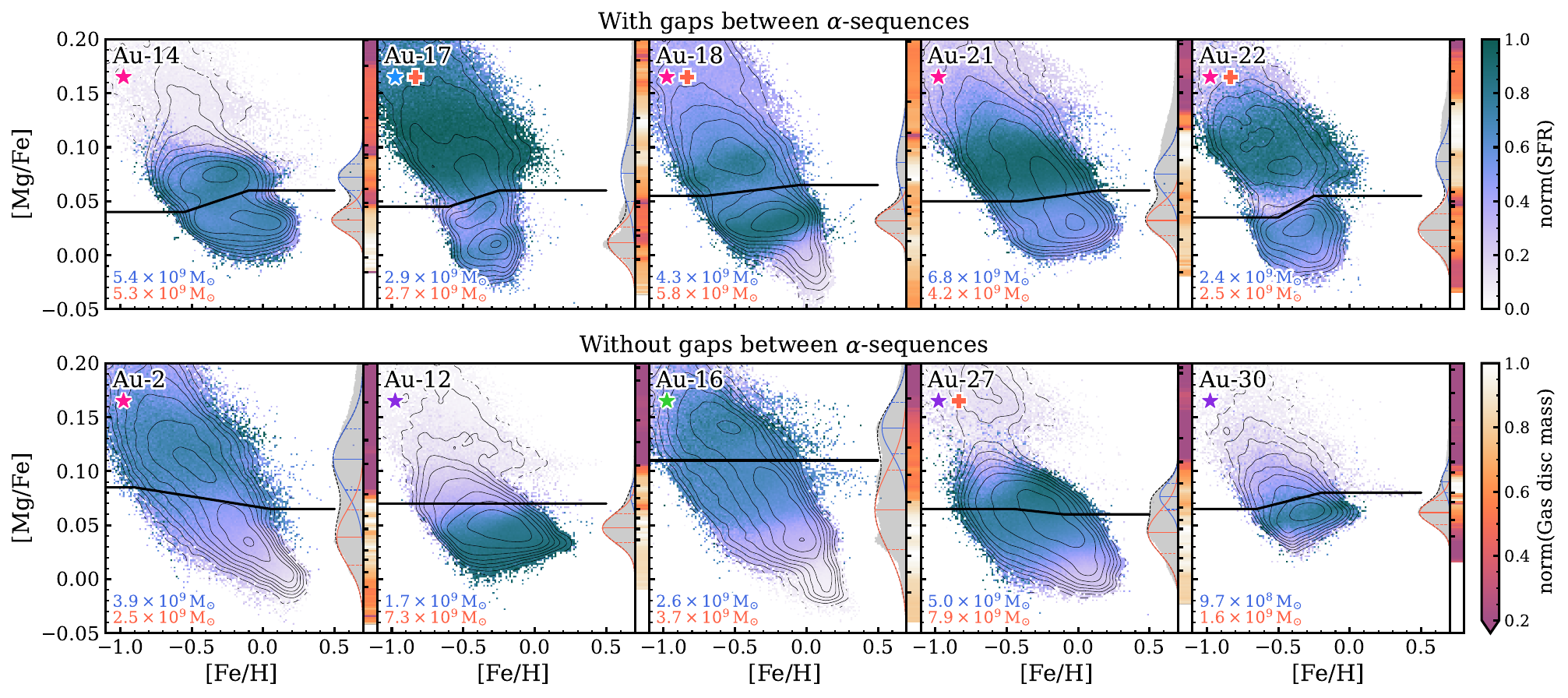}\\
\caption{The chemical abundance planes for a selection of {\sc auriga} galaxies, showing only disc stars in the Solar region ($6<R_{\rm G}/\rm{kpc}<9$). The histograms are coloured by the mean instantaneous SFR during the birthtimes of each star particle, where the SFR is evaluated over 50\,Myr intervals. The stellar mass distribution is overlaid with black contours. Included on the right-hand-side of each panel is a coloured bar that represents the mass in the gas disc at a time corresponding to the median stellar age in slices of [Mg/Fe]. This is intended to show how the gas disc varies along with the formation of $\alpha$-sequences. Median time intervals of 1\,Gyr are marked with black ticks. The formatting is otherwise as in Figure \ref{fig:metal_planes}. \textit{Upper panels:} Examples of galaxies which experience an under-populated `gap' between their high-$\alpha$ and low-$\alpha$ sequences. \textit{Lower panels:} Examples where there is no obvious gap. Additional plots for the full suite are included \href{https://drive.google.com/file/d/1SL7nUtWxv77hnkhj5YFarc_pEe8ZT_VI/view?usp=sharing}{online}.}
\label{fig:metal_planes_SFR}
\end{figure*}

A key feature of the chemical bi-modality in the MW is that the high-$\alpha$ and low-$\alpha$ sequences are mediated by an under-populated `gap', most prominent at metallicities of $\rm{[Fe/H]}<-0.25$. In the {\sc auriga} simulations, several realisations display comparable gaps between the two $\alpha$-sequences, including Au-1, Au-14, Au-17, Au-18, Au-21, Au-22, and Au-29. Both Au-1 and Au-29 are strongly affected by recent major mergers, so for clarity we opt to focus on the other examples here.

The chemical abundance planes for these simulations are shown in the upper panels of Figure \ref{fig:metal_planes_SFR}, where the pixels are color-coded by the mean instantaneous SFR at the birth time of each star particle (this SFR is calculated with a histogram of the stars found within $6<R_{\rm G}/\rm{kpc}<9$ at $z=0$). This reveals that the high-$\alpha$ sequences are associated with a high SFR, which sharply declines in the chemical region corresponding to the gap. The SFR then increases again for stars within the low-$\alpha$ sequence. The exception is Au-14, where the two sequences are close in their [Mg/Fe] and so the reduced intermediary SFR cannot be seen.

There are multiple mechanisms by which the SFR can be reduced. These include a disruption of the gaseous disc by satellite fly-bys or mergers, feedback from stellar formation processes and/or AGN, or simply an exhaustion of the local gas reservoir. We select the mass in the gas disc with a crude cut of $\eta>0.7$ on the orbital circularity and $\rho>10^{5}\,\rm{M}_{\odot}\,\rm{kpc}^{-3}$ on the density. We show this mass in the purple-orange bar on the right side of each panel in Figure \ref{fig:metal_planes_SFR}. This bar represents the normalised gas disc mass at the median birth time of the corresponding stars in each [Mg/Fe] bin. Since the evolution of the stellar [Mg/Fe] is decreasing almost monotonically with time, this effectively provides an analogue of the evolution of mass in the gas disc. Intervals of 1 Gyr are indicated using ticks along the axis.

\begin{figure*}
\centering
  \setlength\tabcolsep{2pt}%
    \includegraphics[keepaspectratio, trim={0.0cm 0.0cm 0.0cm 0.0cm}, width=0.825\linewidth]{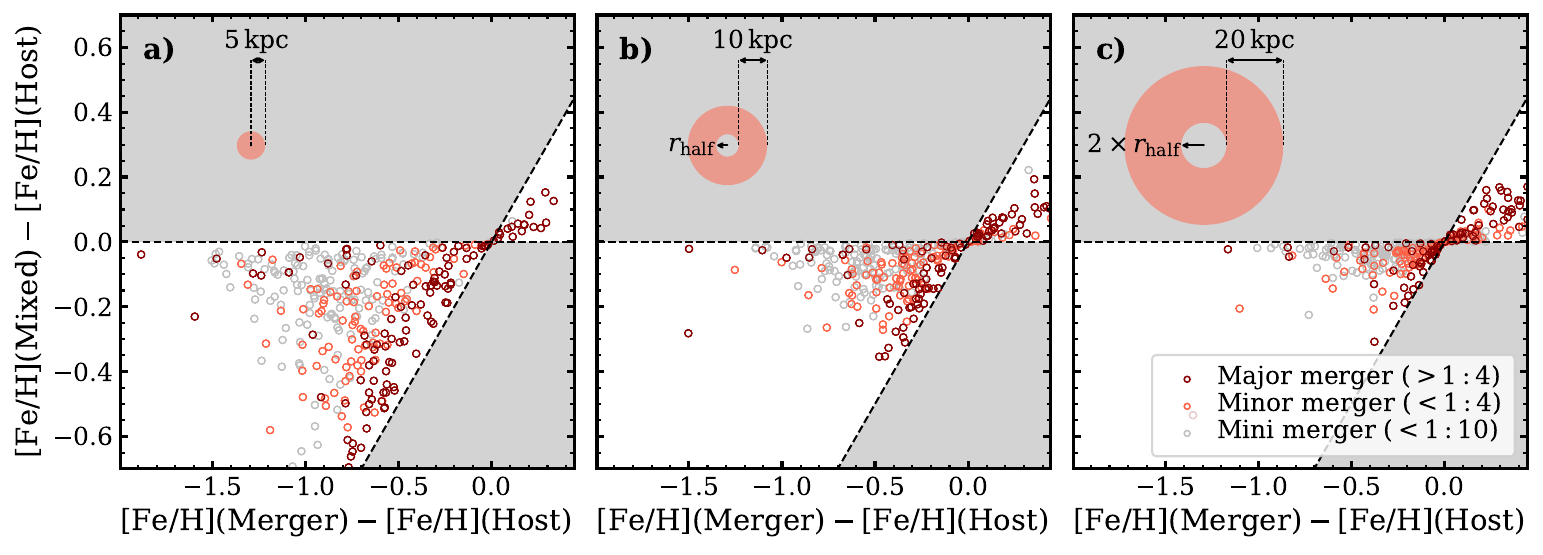}\\
\caption{On the $x$-axis, the difference between the gas metallicity of merging subhaloes (determined at the snapshot before crossing over the virial threshold of the host halo) and the gas metallicity of the host galaxy (the main progenitor galaxy, at the same time as the merging subhalo). On the $y$-axis, the difference between the `mixed' gas metallicity and the gas metallicity of the host, where the mixed metallicity is calculated differently for the first, second and third panels: \textbf{a)} the merger gas is assumed to mix with the host gas in the inner 5\,kpc, \textbf{b)} with the gas contained between the stellar $r_{\rm half}$ and $r_{\rm half}+10\,$kpc, and \textbf{c)} with the gas contained between $2\times r_{\rm half}$ and $2\times r_{\rm half}+20\,$kpc. These mixing volumes are illustrated with schematics in the top-left corners. Mergers of different mass ratios are indicated with separate colours, down to a minimum merger gas mass of $10^8\,\rm{M}_{\odot}$. Forbidden regions are shaded grey.}
\label{fig:mixing}
\end{figure*}

In all cases, the reduction in the SFR during the gap is coincident with a drop in the gas disc mass. This is a result of gas exhaustion following the epoch of elevated SFR. For Au-14, the gas disc also faces violent disruption both before and during the gap due to the accretion of multiple gas-rich mergers. For Au-21, the raised SFR and gas disc mass are both caused by two simultaneous major mergers with mass ratios of approximately $1:2$. For Au-18, there is an instantaneous drop in the gas disc mass just \textit{before} the high-$\alpha$ sequence begins to form (i.e., at higher [Mg/Fe]). This is due to the previously discussed GSE-like merger briefly disrupting the gas disc. There is only a very minor rise and fall of the gas disc mass surrounding the gap itself, indicating that the two chemical sequences in Au-18 are formed by variations in the star formation efficiency rather than changes in the available gas reservoir. As we discussed in Section \ref{sec:SFH}, star formation bursts in this realisation are stimulated by the action of minor mergers at approximately $\tau=9$ and $\tau=7\,$Gyr, the earliest of which is a GSE-like merger.

We draw special attention to Au-17 and Au-22, for which the gas disc both loses mass and rapidly shrinks into a compact structure less than 3\,kpc wide. This results in a total suppression of star formation around the Solar neighbourhood, as can also be seen in Figure \ref{fig:SFH}. After a quiescent period of several Gyrs, the disc is partially re-established with gases from the CGM and from low-mass mergers. This mechanism was first described in \citet{grand2018}, for which the higher-resolution variants of Au-6 and Au-23 are found to form multiple chemical sequences through this same process. Interestingly, we do not find clearly distinct chemical sequences in Au-6 at the lower resolution that is examined in this work, indicating that these mechanisms are sensitive to subtle variations in resolution.

In the lower panels, we show examples of {\sc auriga} simulations that do not exhibit strong gaps between the high-$\alpha$ and low-$\alpha$ sequences. In the case of Au-2 and Au-16, transitions in the SFR correlate with bumps in the [Mg/Fe] histograms (shown on the right side of each panel). However, these transitions are more gradual, and the SFR continues to decline during the formation of the low-$\alpha$ sequence. In the case of Au-27, Au-12, and Au-30, the SFR was very low in ancient times, and so the conditions leading to a SFR gap never arose.

In general, different overdensities in the chemical abundance plane are well correlated with step-like transitions in the SFR. As such, the formation of chemical bi-modalities are dependent on the star formation history of the host galaxy, which in turn is sensitive to the stochastic assembly history. This, then, explains why the formation times of chemical sequences in Figure \ref{fig:bimod_props} display such great diversity.

\subsection{How do the high-$\alpha$ and low-$\alpha$ sequences arise?} \label{sec:bimod_origin}

\subsubsection{Merger gas dilution}

One key feature of the MW is that the high-$\alpha$ and low-$\alpha$ sequences share a substantial overlap in their [Fe/H] chemistry around the Solar neighbourhood. This would seem contrary to the expectation that the Galactic gas metallicity is continuously enriched by SNe. As discussed in Section \ref{sec:introduction}, one explanation is that the Galaxy underwent two periods of gas infall, or the ``two-infall model'' \citep{chiappini1997}. In this model, the Galactic metallicity is reset to a less-enriched state by a second infall of primordial gas. In recent years, this infall has been frequently attributed to the GSE merger event, which is predicted to have been a massive gas-rich merger that occurred around the same time as the formation of the high-$\alpha$ sequence ($\approx 10\,$Gyr ago). Here, we investigate whether comparable merger events in {\sc auriga} can yield a similar effect.

GSE-like mergers could be important for the creation of a low-$\alpha$ sequence by providing gaseous fuel and by diluting the gas metallicity. At this point, we stress that we are focussed on the kinds of gas dilution that can explain the overlapping metallicities between two $\alpha$-sequences as is observed in the MW. Therefore, the gas dilution must be sufficient long-lived that a significant number of stars are formed, must be of at least a magnitude $\approx0.5\,$dex, and must apply to the Solar neighbourhood.

To further explore these possibilities, we create a toy model for the metallicity dilution of all subhaloes across the {\sc auriga} suite with a gas content of $10^8\,\rm{M}_{\odot}$ or more. The most straightforward estimate would be to calculate the combined [Fe/H] ratio for both the host and the merging subhalo. However, the gas from a subhalo may not mix evenly with all of the gas in the host galaxy. The dilution effect will vary depending on whether the merger is able to effectively penetrate the hot gaseous halo and deliver its gas directly into the galactic centre, or if its gas is stripped at higher radii and accretes onto the edge of the star-forming gas disc (which is what happens in the higher-resolution version of Au-23 in \citealt{grand2018}). These different outcomes will depend on the infall trajectory of the subhalo, on the ram-pressure stripping, on the respective matter density distributions, and more. Then, the magnitude of the chemical dilution will depend on four different properties; i) the gas mass of the merger, ii) the mass of host gas that is diluted, iii) the metallicity of the merger gas, and iv) the metallicity, and likewise the metallicity gradient, in the host gas.

In an attempt to account for such possibilities, we consider three different mixing scenarios for the donated gas:
\begin{itemize}
\setlength{\itemindent}{0.5em}
    \item \textbf{a)} The donated gas mixes with the central host gas (within 5\,kpc).
    \item \textbf{b)} The donated gas mixes with the host gas around the Solar neighbourhood (between the stellar half mass radius at the time of the merger ($r_{\rm half}$), and $r_{\rm half}+10\,$kpc).
    \item \textbf{b)} The donated gas mixes with the host gas towards the edge of the gas disc (between $2\times r_{\rm half}$, and $2\times r_{\rm half}+10\,$kpc).
\end{itemize}
These toy scenarios, especially a), assume an improbably small mixing volume. This is by design, and is intended to maximise the magnitude of the gas dilution.

We show the results in Figure \ref{fig:mixing}, where on the $x$-axis we plot the difference between the mass-weighted average metallicity for the merger and host gas, and on the $y$-axis we plot the metallicity difference between the new ``mixed'' gas and the original host gas. The mergers are divided across three groups corresponding to their pre-infall merger mass ratio; major ($>1:4$), minor ($<1:4$) and mini ($<1:10$).

Whilst major mergers have the highest proportional gas masses, they also tend to be more chemically enriched. The effect of this can be seen in Figure \ref{fig:mixing}; the major merger scatter points tend to group-up at lower values of $\rm{[Fe/H](Merger)}-\rm{[Fe/H](Host)}$. There are some mergers with gas that is \textit{more} enriched than that of the host. In the cases of panels b) and c), this is primarily because the mixing volume excludes the central enriched gas of the host. However, there are still some examples of super-enriched mergers in panel a), and these are extremely ancient mergers ($\tau>12\,$Gyr) that come from a time when merger mass ratios are typically much higher, and when galaxy formation was more stochastic.

The dilution is sensitive to the gas mixing volume, and this is because of the gas metallicity gradient in the host galaxy. The galactic centre is the most enriched, meaning it has a greater potential for dilution, and this is the only scenario that sees dilutions of 0.5\,dex or more. In comparison, the gas at higher radii is far less enriched, and there is often no opportunity for dilution at all ($\rm{[Fe/H](Merger)}-\rm{[Fe/H](Host)}\simeq0$).

Some realisations may experience multiple gas-rich and metal-poor mergers, and this would lead to greater combined dilution. One such example is Au-14, which undergoes three minor mergers at $8>\tau/\rm{Gyr}>6$ followed by three mini mergers at $6>\tau/\rm{Gyr}>5$. The combined gas supplied by these merging groups is $1.67\times10^{10}\,\rm{M}_{\odot}$ and $1.07\times10^{10}\,\rm{M}_{\odot}$ respectively, and has important repercussions on the formation of the stellar disc. However, Au-14 is a late-forming galaxy and all of these mergers occur before or during the emergence of the high-$\alpha$ sequence, and are therefore not responsible for a metallicity dilution between the formations of the high-$\alpha$ and low-$\alpha$ sequences.

Metallicity dilution can also result from the inward transport of metal-poor gas located at the edge of the gas disc, driven inward by gravitational torques from interacting subhaloes. This mechanism is investigated by \citet{bustamante2018}, who find that major mergers are the most effective at inducing such gas inflows. However, the resulting metallicity dilution is around $0.1$\,dex for major mergers, and even lower for minor mergers. Nonetheless, the dilution from this mechanism should be considered alongside our toy mixing model.

\subsubsection{GSE-like merger gas donation}

\begin{figure}
  \setlength\tabcolsep{2pt}%
    \includegraphics[keepaspectratio, trim={0.2cm 0.5cm 0.2cm 0.0cm}, width=\columnwidth]{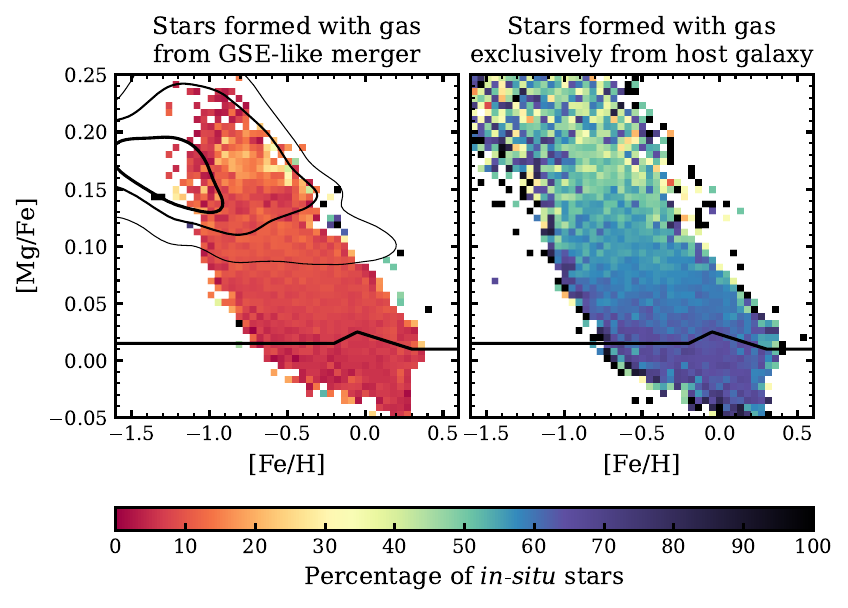}\\
\caption{\textit{Left panel:} A 2-D histogram showing the percentage of \textit{in-situ} stars which formed from the gas that was donated by the GSE-like merger in Au-24. Overlayed are black contours showing the 1,2,3$\sigma$ limits of the merger gas chemical abundances, taken at a time before infall.
\textit{Right panel:} The same 2-D histogram, but this time showing the percentage of stars that formed from the host gas (i.e., gas that was not donated from any merger above a merger mass ratio of $1:30$, but rather smooth accretion from the CGM).
All stars are from the radial range $3<R_{\rm G}/\rm{kpc}<20$ at $z=0$, as this is the range where the chemical sequences are most prevalent. The black line marks the division between the high-$\alpha$ and low-$\alpha$ selections.}
\label{fig:tracers}
\end{figure}

To determine how many stars form from merger-donated gas, we analyse a novel reproduction of Au-24 in which we have activated Monte-Carlo tracer particles as in \citet{genel2013}. These track the gas flow and enables us to determine where star-forming gas originated. This realisation was chosen because there is a merger with an infall time of $\tau=9.1\,$Gyr and a mass ratio of $1:8$, broadly comparable to expectations of the GSE. This merger brings in a gas mass of $9.3\times10^{9}\,\rm{M}_{\odot}$ with an average $\rm{[Fe/H](gas)}=-1.1$.

In the left panel of Figure \ref{fig:tracers}, we show the fraction of stars across the chemical abundance plane that formed from the gas associated with this GSE-like merger. The chemistry of the pre-merger gas is overlaid with black contours. The fraction of stars that formed from the merger gas is $\sim10$ per cent for stars in the locus of the high-$\alpha$ sequence, and then drops to $<10$ per cent for stars in the locus of the low-$\alpha$ sequence. We also note that there are no other mergers in Au-24 which contribute a greater overall fraction than this. We have excluded the inner 3\,kpc from this comparison, and this is because there is a higher-metallicity group of stars with a fraction of $\approx20$ per cent in this region. This tendency is typical of GSE-like mergers, as the low impact parameters naturally funnels a higher proportion of their gas toward the galactic centre, where metal enrichment is naturally high.

The fractional variation across the chemical abundance plane is relatively low, indicating that the merger gas rapidly becomes well-mixed with the host gas prior to forming stars. There is a region at $\rm{[Mg/Fe]}\approx0.17$ in which the fraction is around 15 per cent, which represents the proportionally low number of stars that form before the merger gas is well mixed. These stars share a similar chemistry to that of the merger gas.

In the right panel of Figure \ref{fig:tracers}, we show the mass fraction of stars that formed from the host gas. This is defined as gas that was not bound to mergers of mass ratio $>1:30$, and should therefore represent stars that formed from local gas, gas that accreted from the smooth CGM, or gas from a spectrum of mini mergers. The result is that the disc chemistry is comprised from host gas at around 60 per cent or higher, where the remaining fraction is comprised of gas from the GSE-like merger (left panel) and several other massive mergers (not shown).

\begin{figure*}
\centering
  \setlength\tabcolsep{2pt}%
    \includegraphics[keepaspectratio, trim={0.0cm 0.0cm 0.0cm 0.0cm}, width=0.95\textwidth]{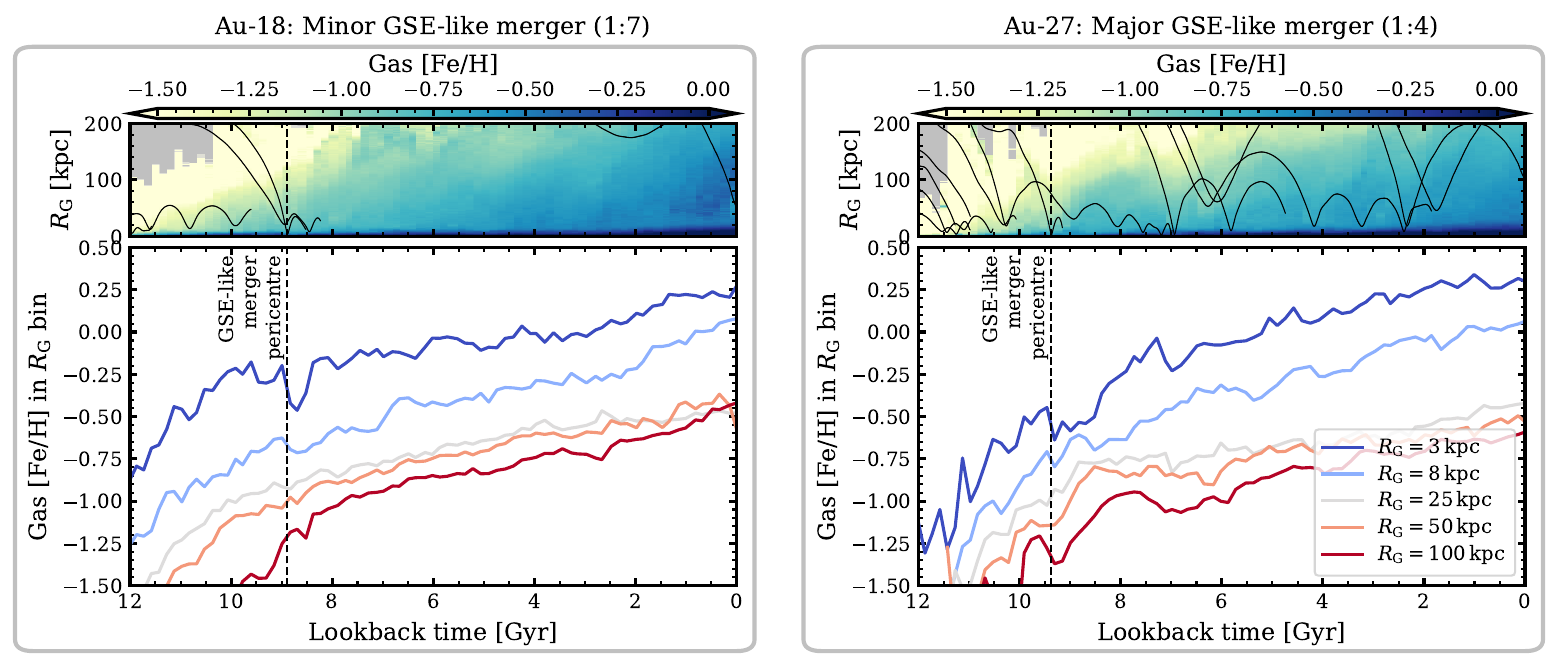}\\
\caption{\textit{Upper panels:} The mass-weighted metallicity of the gas surrounding the main progenitor galaxy, shown with galactocentric radius versus lookback time. The infall paths of merging galaxies are plotted with solid black lines (merger mass ratios $>1:20$). The bin size is $0.4\,$kpc. \textit{Lower panel:} The mass-weighted gas metallicity at a selection of constant radii, as indicated in the legend. The verticle black dashed line marks the pericentre time of a GSE-like merger. Additional plots for the full suite are included \href{https://drive.google.com/drive/folders/1zfx4371eM_Ta7WzV8N6N-7bs3CTFFrAB?usp=sharing}{online}.}
\label{fig:CGM}
\end{figure*}

In Figure \ref{fig:CGM}, we show the evolution of the mass-weighted gas metallicity in the CGM in radial bins out to 200\,kpc (upper panels) and for lines of constant radius (lower panels). We select Au-18 and Au-27 for this comparison, where both realisations experience a GSE-like merger but only Au-18 exhibits a chemical dichotomy in its disc stars. The trends shown here are representative of the broader behaviour observed across most of the {\sc auriga} suite.

Au-18 experiences a GSE-like merger with an infall time of $\tau=9.3\,$Gyr and a mass ratio of $1:7$. This merger brings $7.7\times10^9\,\rm{M}_{\odot}$ of gas with $\rm{[Fe/H](gas)}=-1.17$, a full 0.66\,dex lower than the gas metallicity in the inner 10\,kpc of the host galaxy that it merges with. This same realisation was previously investigated in \citet{ciuca2024}, where it is shown that the merger causes a sudden but short-lived drop in the stellar metallicity in Solar neighbourhood stars of $\approx0.25\,$dex (see their figure 4).

We see a much lower gas dilution in our analysis, where a dilution of up to $\approx0.25\,$dex is only achieved in the galactic centre (3\,kpc). There, the dilution is short-lived and is entirely recovered within a few 100\,Myrs. As we will consider later in Section \ref{sec:migration}, some of the stars that form from this gas will ultimately migrate into the Solar neighbourhood and will therefore contribute to the feature described in \citet{ciuca2024}. Nonetheless, this metal dilution is not relevant for the overlapping [Fe/H] values of the two chemical sequences in Au-18. This is because the dilution occurs during the formation of the high-$\alpha$ sequence, and the gas metallicity has entirely returned to pre-dilution levels before the time period that the low-$\alpha$ sequence forms.

The GSE-like merger in Au-27 has an infall time of $\tau=9.7\,$Gyr and a higher mass ratio in excess of $1:4$. This merger brings $1.6\times10^{10}\,\rm{M}_{\odot}$ of gas with $\rm{[Fe/H](gas)}=-0.98$, which is 0.41\,dex lower than the gas metallicity in the inner 10\,kpc of the corresponding host galaxy. Despite the greater quantity of donated gas, the only perceptible metallicity dilution occurs at 3\,kpc and is of a magnitude 0.2\,dex. In both realisations considered here, the metallicity of galactic gas around the Solar neighbourhood (the 8\,kpc line) is minimally impacted by the GSE-like merger accretion.

These results demonstrate the same point made in Figure \ref{fig:tracers}: the gas donated by a GSE-like merger \textit{is not the reason why low-$\alpha$ sequences overlap with the metallicity of high-$\alpha$ sequences}. Whilst there is a measurable metallicity dilution, it is not of a great enough magnitude to explain the overlapping chemical sequences. This is not to say that the merger is unimportant in other aspects, such as its effect on the SFR within the host galaxy, its kinematic influence on the gas \citep[e.g.][]{merrow2024}, or on pre-existing disc stars \citep[e.g.][]{orkney2025}. The figure also reveals another significant detail: there is a reservoir of consistently metal-poor gas in the wider CGM (beyond 25\,kpc), which remains at $\rm{[Fe/H]}<-0.5$ even up to $z=0$. We will discuss how this can impact the development of the low-$\alpha$ sequence, next.

\subsubsection{Chemical variation across the disc.} \label{sec:disc_metallicity_variation}

\begin{figure*}
\centering
  \setlength\tabcolsep{2pt}%
    \includegraphics[keepaspectratio, trim={0.0cm 0.0cm 0.0cm 0.0cm}, width=\textwidth]{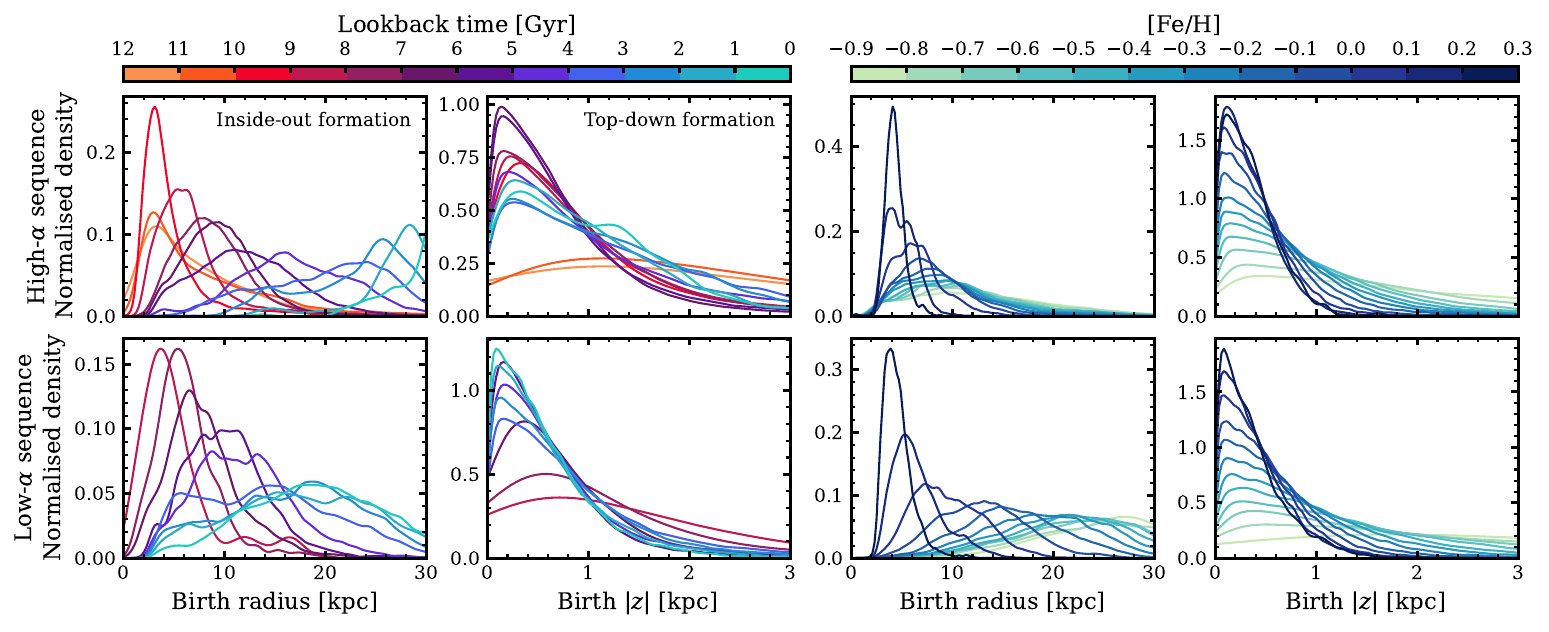}\\
\caption{This plot shows the normalised density functions of stars as a function of birth galacto-centric radius and birth height above/below the disc plane in one example realisation. The left four panels segregate the stars by their birth time, and the right four panels do so by their [Fe/H]. The upper panels show stars within the high-$\alpha$ selection, whereas the lower panels show stars within the low-$\alpha$ selection. The trends here are seen across all realisations, no matter what their chemical sequences look like.}
\label{fig:birthprops}
\end{figure*}

If gas donation from mergers is relatively unimportant for the creation of $\alpha$-sequences in {\sc auriga}, then why do some {\sc auriga} realisations manifest dichotomous chemical distributions with substantial overlaps in [Fe/H] (e.g. Au-14 and Au-18 in Figure \ref{fig:metal_planes})?

To investigate this further, we extract the precise birth locations of stellar particles in one example {\sc auriga} realisation, whilst noting that the resulting trends are general across the suite. Then, we derive the galactocentric radius and height above or below the disc plane for all disc stars at the time of their birth. In Figure \ref{fig:birthprops}, we show normalised Gaussian Kernel Density Estimates for the stars that formed in the high-$\alpha$ (top panels) and low-$\alpha$ (bottom panels) sequences.

The left set of panels shows the stars as partitioned by their birth time, demonstrating the typical inside-out (first column) and top-down (second column) formation that has been widely predicted for galaxy formation. The right set of panels shows the stars as partitioned by their metallicities. This reveals consistent trends for both high-$\alpha$ and low-$\alpha$ sequences; i) metal-poor stars form at greater radii, ii) metal-poor stars form at greater heights above and below the disc. This is an intuitive result, galaxies naturally form a negative metallicity gradient in their gas with both radius and height. This is imprinted on the kinematic properties of the stars at $z=0$: stars from the metal-poor ends of each sequence have slightly lower mean orbital circularities in Figure \ref{fig:metal_planes}.

\begin{figure*}
\centering
  \setlength\tabcolsep{2pt}%
    \includegraphics[keepaspectratio, trim={0.25cm 0.5cm 0.25cm 0.0cm}, width=\linewidth]{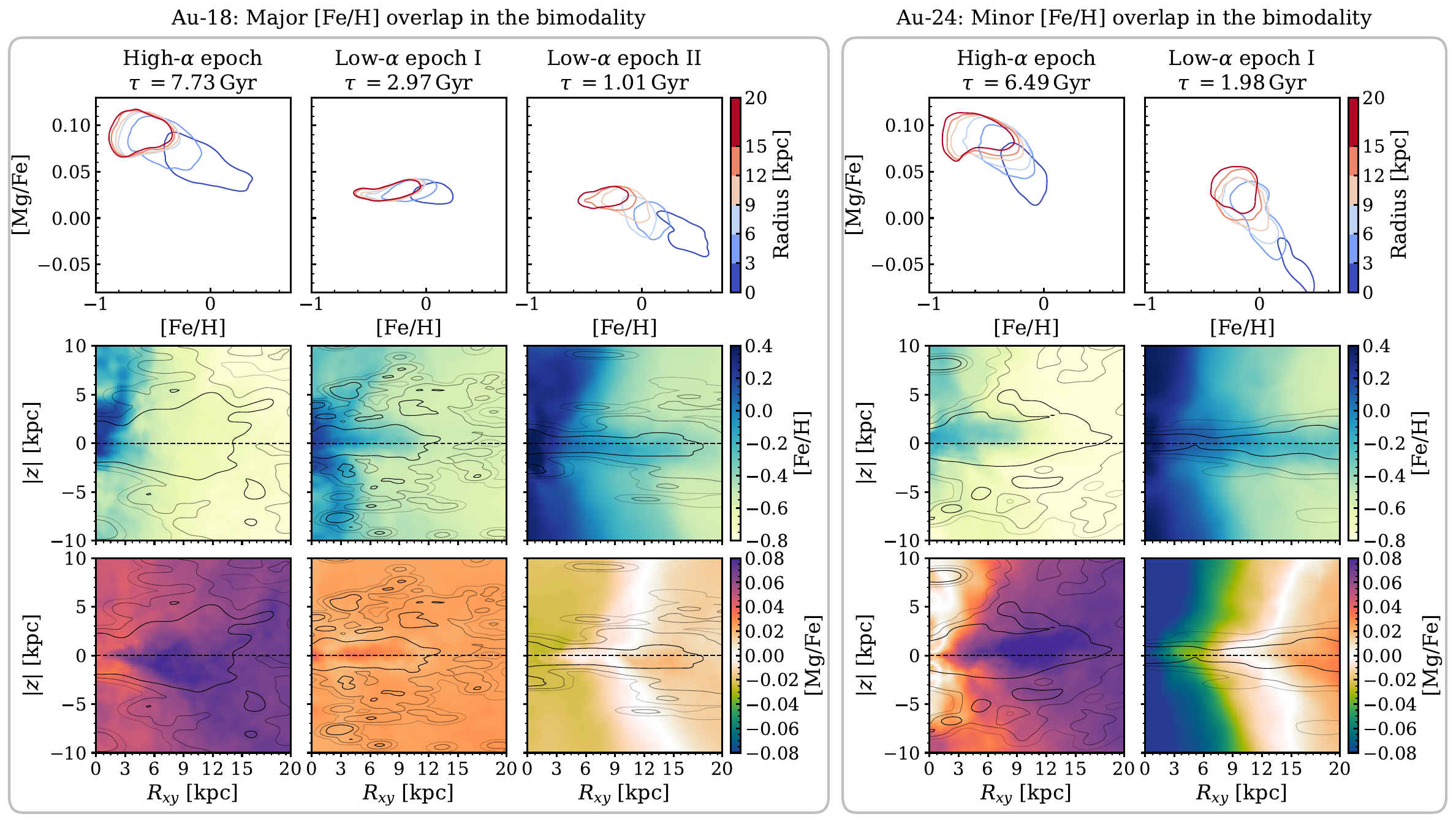}
\caption{This figure shows the chemistry of the galactic gas and stars at various critical epochs for two example realisations, Au-18 and Au-24. \textit{Upper panels}: The 1$\sigma$ contour encircling the stars that formed in each time epoch (where the time epoch is indicated in the title, and stars are selected from a birth time interval of $\tau \pm 0.5\,$Gyr). The stars are grouped by their $z=0$ orbital radius, as indicated in the colourbar. \textit{Middle panels}: Mass-weighted [Fe/H] abundance ratio for galactic gas in the $R{xy}$ versus $|z|$ plane. The overlaid black contours indicate the 1,2,3$\sigma$ of the instantaneous SFR in the gas cells. This roughly corresponds with the gas density, due to the star formation criteria used in {\sc auriga} \citep{grand2024}. \textit{Bottom panels}: The same, but now for the [Mg/Fe] abundance ratio.} \label{fig:gasdisc}
\end{figure*}

\begin{figure*}
\centering
  \setlength\tabcolsep{2pt}%
    \includegraphics[keepaspectratio, trim={0.0cm 0.5cm 0.0cm 0.0cm}, width=\linewidth]{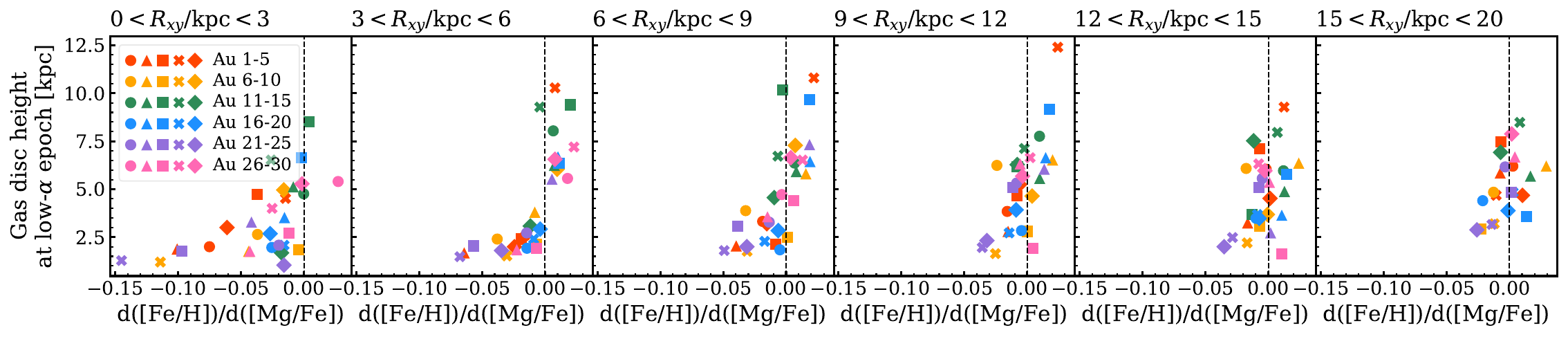}
\caption{An estimate of the gas disc height versus the gradient of low-$\alpha$ sequences in the chemical abundance plane, for a series of cylindrical bins. The gas is selected from a single representative lookback time, whereas stellar properties are gathered at $z=0$ from a birth time interval of $\tau \pm 1\,$Gyr}\label{fig:locii}
\end{figure*}

We expand upon this result in Figure \ref{fig:gasdisc}, which shows histograms of the chemistry in different regions of the gas disc for Au-18 and Au-24 at times corresponding to the formation of their high-$\alpha$ and low-$\alpha$ sequences. These histograms are shown in terms of distance in the $x\text{-}y$ plane and height above and below the disc. Black contour lines show the $1,2,3\sigma$ contours of the SFR in the gas, and we emphasise that the SFR in {\sc auriga} is a function of the gas density \citep[see][]{Auriga}. Here, Au-18 is chosen as an example galaxy for which there is a large [Fe/H] overlap between the two $\alpha$-sequences, and Au-24 is chosen as an example galaxy for which there is almost no overlap (see Figure \ref{fig:metal_planes}). In the upper panels, we include plots of the chemical abundance plane for the contemporaneous stars as a helpful reference.

During the formation of the high-$\alpha$ sequences, the gas discs in both examples are thick and flared, and so there is star formation occurring up to and beyond $10\,$kpc above and below the disc plane. The central gas disc is enriched with metals, but the surrounding gases are approaching metal-poor values of $\left < \rm{[Fe/H]} \right > \simeq-0.8$.

During the formation of the low-$\alpha$ sequence, a difference emerges between the two examples. The gas in Au-18 is now less flared, but there are still dense metal-poor gases at heights up to $\sim10\,$kpc above and below the disc plane, leading to continued star-forming activity in those regions. On the other hand, the star formation in Au-24 is almost exclusively concentrated in the disc plane. Here, the gas is metal-rich and forms a tight radial metallicity gradient. There is still metal-poor gas above and below the disc plane, but it is no longer sufficiently dense to form stars. The explanation is that the gas disc has become stable at late times, ultimately collapsing into a thin plane in which the metal-poor and metal-rich gases are mixed together.

From closer examination of Au-18 in Figure \ref{fig:metal_planes}, there is a faint tail of stars below the main overdensity of the low-$\alpha$ sequence, and these have an extremely narrow metallicity range of $\Delta\rm{[Fe/H]}\approx 0.2$. This tail is formed entirely from stars in the final 2\,Gyr, by which time the gas disc in Au-18 has settled into a configuration that is almost as thin and stable as in Au-24. We show this epoch in the column titled ``Low-$\alpha$ epoch II''. The behaviour is seen across the whole simulation suite: a narrow metal distribution is correlated with a thinner gas disc, whereas a wide metal spread is correlated with a thicker gas disc. Then, it is apparent that a wide spread of metal values in these simulations is dependent on the presence of dense metal-poor gases at elevated heights above and below the disc plane. Therefore, we argue that [Fe/H] overlaps in {\sc auriga} come not from metallicity dilution following a merger event, but from a natural variation in gas chemistry and its geometric distribution throughout time.

In the lower panels of Figure \ref{fig:gasdisc}, we include the same plots but now weighted by the gaseous [Mg/Fe] abundance ratio. When the [Fe/H] and [Mg/Fe] gas abundance ratios are considered together, we can explain the radial dependence of the chemical sequences shown in the upper set of panels. In both examples, there are modest [Fe/H] gradients and [Mg/Fe] gradients during the formation of the high-$\alpha$ sequence. This leads to $\alpha$-sequences with chemistry that is strongly radially dependent. Both examples have a metallicity gradient during the `low-$\alpha$ epoch I', but only Au-24 has a corresponding gradient in its [Mg/Fe] abundances. The outcome is that the low-$\alpha$ sequence of Au-18 has a radial dependence on [Fe/H] but not on [Mg/Fe]. For comparison, the $\alpha$-sequences in the MW show no radial evolution for the high-$\alpha$ sequence, and evolution in both [Fe/H] and [Mg/Fe] for the low-$\alpha$ sequence (see figure 10 of \citealt{imig2023}).

\subsubsection{The shape of chemical sequences} \label{sec:shape}

\begin{figure*}
\centering
  \setlength\tabcolsep{2pt}%
    \includegraphics[keepaspectratio, trim={0.0cm 0.5cm 0.0cm 0.0cm}, width=\linewidth]{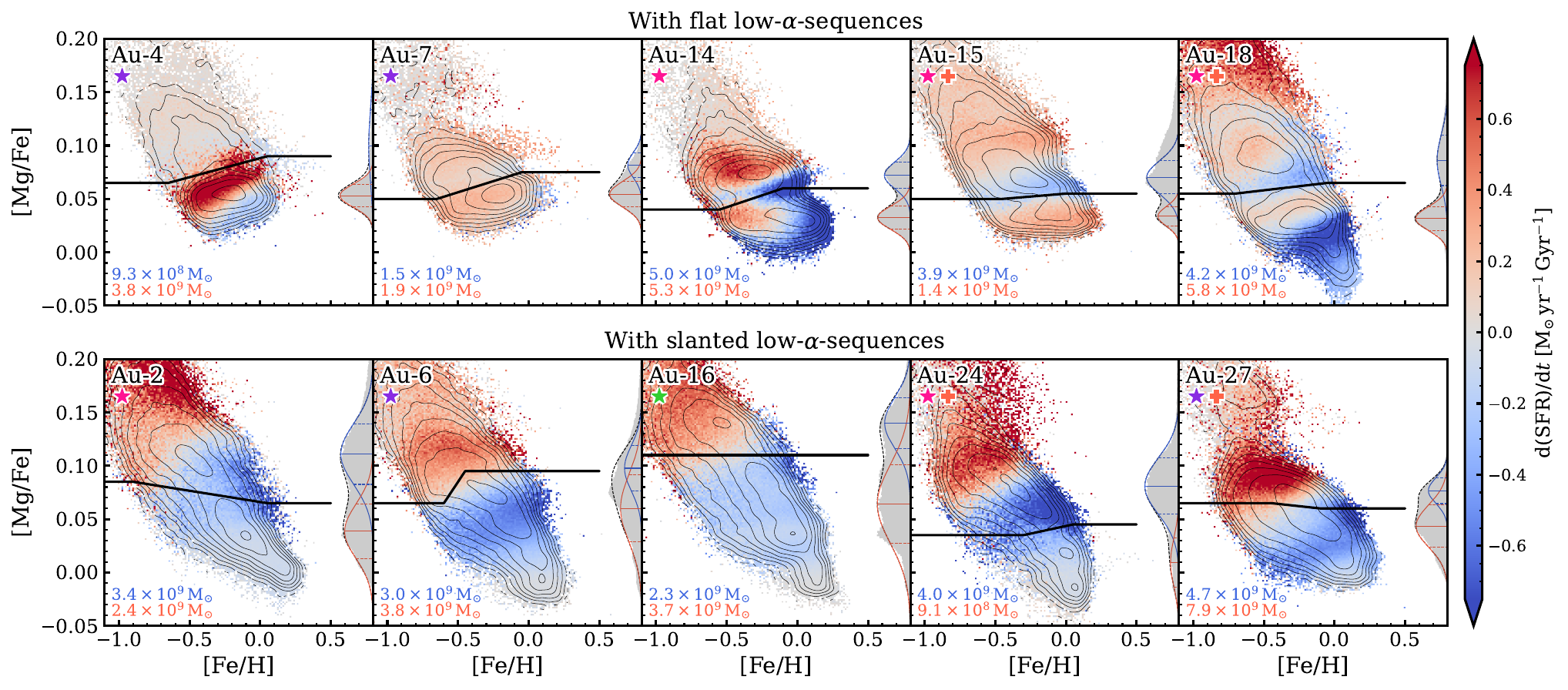}
\caption{The chemical abundance planes for a selection of {\sc auriga} galaxies, showing only disc stars that were born in the Solar region ($6<R_{\rm G}/\rm{kpc}<9$). The histograms are coloured by the gradient of the local SFR associated with each star particle, evaluated over a time period of $\tau_{\rm birth}\pm1$\,Gyr. We calculate the SFR for stars that were born in the Solar region rather than stars found in the Solar region at $z=0$. This is so that the gradient of the SFR better reflects the birth environment of each star particle, which may change due to radial mixing and migration. The stellar mass distribution is overlaid with black contours. The formatting is otherwise as in Figure \ref{fig:metal_planes}. \textit{Upper panels:} Examples of galaxies with a flat distribution in their $\alpha$-sequences. \textit{Lower panels:} Examples where there is a downward slanted distribution in their $\alpha$-sequences. Additional plots for the full suite are included \href{https://drive.google.com/drive/folders/1-v1nL457GmcYIzuH0XC4_geHhqcq-0TX?usp=sharing}{online}.}\label{fig:dSFR}
\end{figure*}

\begin{figure*}
\centering
  \setlength\tabcolsep{2pt}%
    \includegraphics[keepaspectratio, trim={0.0cm 0.5cm 1.25cm 0.0cm}, width=\linewidth]{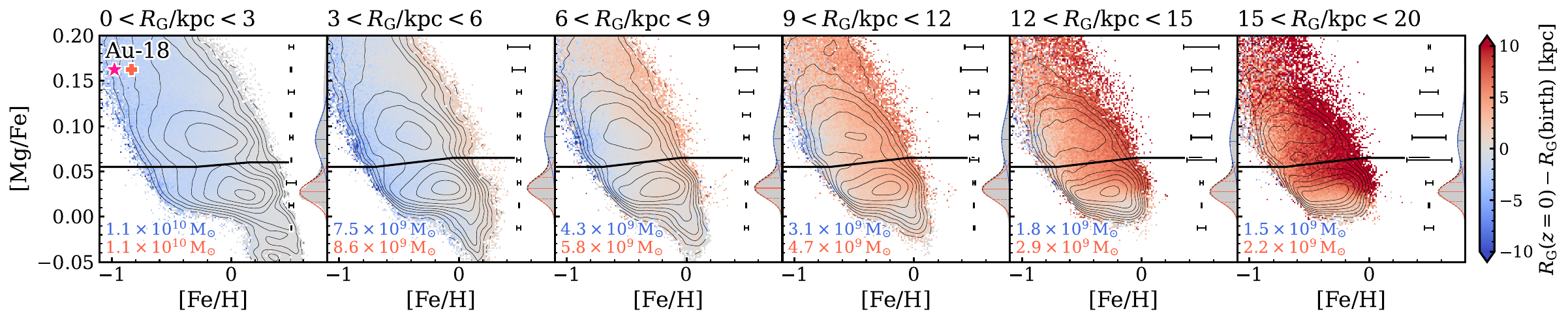}
\caption{In this plot, we show the radial migration of stars in the chemical abundance plane for one example galaxy (Au-18). The pixel colour shows the mean change in orbital radius, from the closest birth snapshot to $z=0$. The stellar mass distribution is overlaid with black contours. Errorbars to the right of each panel illustrate how the spread of the [Fe/H] distribution has increased due to radial migration, calculated as the difference between the 5th and 95th mass-weighted percentiles for stars at their birth radii and at their $z=0$ radii. Additional plots for the full suite are included \href{https://drive.google.com/drive/folders/1_8R_RkelCSCA2LA4kDQYN_IUNiSsnGvT?usp=sharing}{online}.}\label{fig:radialmigration}
\end{figure*}

In Figure \ref{fig:metal_planes}, there are sequences that form a flat pancake (fixed [Mg/Fe] over a range of [Fe/H]; Au-14 and Au-18) and sequences that slant down from left-to-right (an inverse relation between [Fe/H] and [Mg/Fe]; Au-2 and Au-24). We find that the slant of chemical sequences is correlated with several properties of the gas disc, as was already seen for two examples in Figure \ref{fig:gasdisc}. There, a thinner and more radially extended gas disc is associated with a downward slanted chemical sequence.

We show how the slope of low-$\alpha$ sequences relates to the height of the gas disc across the whole {\sc auriga} suite in Figure \ref{fig:locii}. Here, we have chosen representative snapshots of the low-$\alpha$ formation epochs that avoid moments of disruptive merger interactions. The gas disc height is defined as the average vertical distance within which more than 10 per cent of the gas is denser than the {\sc auriga} star formation threshold of $\rho_{\rm SF}=0.13\,m_{\rm p}\,\rm{cm}^{-3}$. We calculate this value in a series of cylindrical bins corresponding to the radial cuts used throughout this work. This definition is designed to accommodate for the very clumpy gas distribution often found at greater vertical heights, which would be ignored in a volume-averaged gas density. Then, the slant of the chemical sequence is determined as the gradient of a first order polynomial fit to the stars that formed at a birth time interval of $\tau_{\rm epoch} \pm 1\,$Gyr. Our stellar selection is based on the $z=0$ properties.

The figure shows a clear relation: thinner gas discs are associated with low-$\alpha$ sequences that have a negative relation between [Mg/Fe] and [Fe/H]. These relations become increasingly negative as we sample stars further toward the galactic centres, because the gaseous abundance gradients are steepest there.

We attribute the origin of this relation to long-lived trends in the SFR. The $\alpha$-elements are produced by Type-II SNe of short-lived massive stars, which rapidly enrich the ISM. Iron is mainly produced by the Type-Ia SNe of longer-lived binary stars, and can take $\sim 1\,$Gyr to enrich the ISM. Therefore, an increasing SFR will raise or maintain the [Mg/Fe] ratio because the iron enrichment is lagging behind the $\alpha$-enrichment. A SFR that is decreasing or unchanging over timescales of a Gyr or more will drive the same ratio down. Therefore, chemical sequences with steep downward `slants' should correlate with an unchanging or decreasing SFR.

We provide evidence that this occurs in {\sc auriga} in Figure \ref{fig:dSFR}, which shows the mean gradient of the radially-binned SFR in the chemical abundance plane for a selection of realisations. Here, we are calculating the SFR for stars that were born in the range $6<R_{\rm G}/\rm{kpc}<9$, though the same results are recovered when considering stars found in the same radial range at $z=0$. The upper five panels show examples where the $\alpha$-sequences are flat. There, we see that the SFR is rising for both high-$\alpha$ and low-$\alpha$ sequences. For realisations with bi-modal chemistry (see the pink coloured markers in the top-left), we can also see that the SFR declines for the under-populated region between the two sequences. This is similar to the result shown in Figure \ref{fig:metal_planes_SFR}.

The lower five panels show examples where the $\alpha$-sequences are `slanted' from top-left to bottom-right (an inverse relation between [Fe/H] and [Mg/Fe]). Each example has an initially growing SFR at high-$\alpha$, corresponding to the growth of the proto-galaxy at early times. Then, the SFR is either unchanging or decreasing throughout the formation of the low-$\alpha$ sequence. These trends are consistent across the rest of the {\sc auriga} suite and for other radial ranges not shown here. This behaviour is the same as is reported in the {\sc eagle} simulation suite by \citet{mason2024}, in which the formation of ``knees'' in the [$\alpha$/Fe] versus [Fe/H] plane are attributed to a sustained decline in the SFR.

Considering both Figures \ref{fig:locii} and \ref{fig:dSFR} together helps to build a more complete picture of the processes behind the morphology of $\alpha$-sequences. A declining SFR means a reduction in turbulent stellar feedback processes, which enables the development of a thin and stable gas disc. 

\subsubsection{Radial migration} \label{sec:migration}

A potential complication in the formation of galactic chemical sequences is the tendency for stars to migrate away from their initial birth sites (radial migration). This is caused by a change in their orbital angular momenta, originating from interactions with non-axisymmetric disc structure (e.g. spiral arms \citealp{sellwood2002, rovskar2008}, central bars \citealp{minchev2010, minchev2011}, or mergers and subhaloes \citealp{minchev2014b, carr2022}).

The impact of radial migration on the {\sc auriga} level-3 simulations was previously discussed in \citet{grand2018}, where it was found to have little importance for the development of chemical sequences. This is consistent with idealised simulations that show radial migration is not the dominant factor with respect to the bi-modality \citep{minchev2014}. For completeness, we have performed our own analysis on all of the level-4 simulations, for which we show just one example in Figure \ref{fig:radialmigration}.

This figure is generally representative of the rest of the suite, and exhibits a few key features. Firstly, there is very little bulk migration occurring for the high-$\alpha$ and low-$\alpha$ sequences around the Solar neighbourhood ($6<R_{\rm G}/\rm{kpc}<9$). This shows that the bi-modality at these radii does not arise through the radial migration of stars, but specifically from the gas chemistry and SFR local to the Solar neighbourhood.

Secondly, the outward migration of stars becomes increasingly prevalent at regions beyond the Solar neighbourhood, with the high-$\alpha$ sequence having migrated the most. This would be because stars that formed in the high-$\alpha$ sequence have had a longer time over which to migrate, and also because this population is more likely to have exclusively formed in the compact proto-galaxy.

Thirdly, there is a strong trend whereby stars at lower [Fe/H] have migrated from higher birth radii, and stars at higher [Fe/H] have migrated from lower birth radii. This is an inevitable outcome of the negative metallicity gradient present in the gaseous disc throughout time. Whilst this radial migration or mixing is not responsible for forming the bi-modality itself, it does cause a broadening of the [Fe/H] distribution over radii near the Solar region. We include horizontal black errorbars to show the magnitude of this broadening. In the Solar region, the broadening is only significant for the most ancient high-$\alpha$ stars.

This result is qualitatively in agreement with the broadening of [Fe/H] distributions reported in idealised simulations of a MW-mass galaxy in \citet{grand2015}, which was also caused by radial migration processes. Similarly, \citet{haywood2013} use observational data to argue that radial migration would not have significantly contaminated the Solar neighbourhood in the MW.

\section{Discussion} \label{sec:discussion}

Since its discovery, the origins of the chemical bi-modality have remained elusive. Nonetheless, the wealth of observational data in recent times have provided us with important clues. In this work, our intention has been to use the {\sc auriga} simulation suite to explore some of these clues in detail using a representative statistical sample of MW-like realisations. From this, we were able to gain a better understanding of the kinds of processes responsible for forming chemical sequences, and more specifically the most likely scenario for our own Galaxy. This provides insight into not only the chemical evolution of the MW, but of its overall formation history too. Furthermore, it is relevant for the many other disc galaxies (which may or may not resemble the chemical structure of the MW) that will be uncovered by future spectroscopic surveys of the Local Volume in the era of Extremely Large Telescopes.

Here, we discuss the implications of our results on the formation time of the two $\alpha$-sequences in the MW, the roles that the GSE merger may or may not have played, and provide our own interpretation for the timeline of events that built the MW disc and other notable stellar populations.

\subsection{A consecutive formation of the chemical sequences}

We have found that, in all cases except for one unusual outlier (Au-1), the formation of $\alpha$-sequences is sequential. Any apparent temporal overlap between two sequences is due to the spread in chemical abundances for mono-age populations rather than a faithful coeval formation. This, then, is accurate to first-order to what is observed in the MW, with many studies finding that the high-$\alpha$ sequence formed rapidly before $\sim8\,$Gyr ago and the low-$\alpha$ sequence formed over a much longer period of time thereafter \citep[e.g.][]{fuhrmann1998, bensby2005, fuhrmann2011, haywood2013, mackereth2017, hayden2017}.

However, there is also observational evidence for a fraction of low-$\alpha$ stars that formed  at lookback times older than 10\,Gyr ago, in consort with the high-$\alpha$ sequence \citep[e.g.][]{beraldo2021, gent2024, borbolato2025}. The authors in \citet{gent2024} show that up to 20 per cent of stars in the low-$\alpha$ sequence formed prior to 9\,Gyr ago using Gaia-ESO data for main sequence turnoff stars (MSTOs) and subgiants. Similarly, \citet{borbolato2025} use the {\sc StarHorse} code \citep{santiago2016} to determine stellar ages for Solar neighbourhood stars in both APOGEE and LAMOST spectroscopic survey data, finding the fraction of low-$\alpha$ stars with ages older than 11\,Gyrs to be over a third in the case of APOGEE and nearly a half in the case of LAMOST. These suggest the early Galaxy was very capable of achieving low-$\alpha$ chemistry even before the accretion of the GSE at around 10\,Gyr ago.

One possible explanation is that these stars were born within intermediate-mass gas clumps that are predisposed to form at high redshifts when gas fractions are higher \citep[see][]{bournaud2009, clarke2019}. The sub-grid physics model in {\sc auriga} does not resolve the formation of such gas clumps, and further studies would be needed to explore the impact of these physics in fully cosmological simulations.

Observational studies find that the high-$\alpha$ disc transitioned to almost exclusively low-$\alpha$ formation very rapidly at approximately 8-9\,Gyr ago \citep{linden2017, sahlholdt2022, borbolato2025}. This, then, is just 1-2\,Gyr after the likely accretion time of the GSE merger \citep[e.g.][]{belokurov2018, helmi2018, mackereth2019, Kruijssen2019}. However, these investigations are often restricted to the Solar neighbourhood and/or limited stellar populations \citep[e.g. see the review of][]{soderblom2010}, and by high uncertainty in stellar age determinations. Likewise, the exact accretion time of the GSE is still uncertain to within a precision of 1-2\,Gyrs. In this work, we find that gas from GSE-like mergers was not instrumental for the formation low-$\alpha$ sequences via their gas donation or metallicity dilution, but could still have been important in other respects. For example, a massive injection of gas from a singular direction establishes a high angular momentum in the gas, and may also help to align the disc with the angular momentum of wider-scale CGM gas, which ultimately aids in the build-up of an extended stellar disc at higher radii \citep[e.g. see the investigation into Au-18 in][]{merrow2024}.

To better understand the timeline of disc formation, it will be crucial to obtain stellar age estimates with higher precisions and for wider samples. This will be addressed by upcoming asteroseismology on thousands of stars in ESA's half-sky PLATO survey \citep{PLATO}, or for a million stars in the further future by the all-sky Chronos mission \citep{chronos}.

\subsection{The possible roles of the GSE merger} \label{sec:discuss_GSE}

With the ambiguity surrounding the early formation history of the MW, the GSE merger event stands out as an isolated pillar of certainty. Therefore, it is natural to develop explanations that revolve around this event. The estimated infall time of the GSE is remarkably close to the estimated transition time between the two $\alpha$-sequences, suggesting that it may be responsible in some way.

\subsubsection{Gas fuelling and dilution}

The GSE would be an opportune vehicle for the donation of fresh metal-poor gas into the ancient Galaxy, as called for by the ``two-infall'' model of \citet{chiappini1997, chiappini2001}. This could reset the gas metallicity to low values whilst also providing the fuel to support the formation of the low-$\alpha$ sequence. The chemical evolution models in \citet{spitoni2019, spitoni2021, spitoni2023} show that a metallicity dilution of approximately a full dex is able to reproduce the low-$\alpha$ sequence in the MW. Improving stellar age estimates and extensions to existing catalogues have allowed for more detailed chemical evolution models, highlighting some features of the low-$\alpha$ disc that are difficult to explain with a two-infall model \citep[see][]{dubay2025}.

In our investigation, we have found that singular merger events \textit{do not} induce a long-lived and significant ($\Delta[\rm{Fe/H}]>0.5$) dilution in the gas discs of their hosts. The metal-poor gases donated by minor mergers are rapidly mixed into the surrounding CGM with low impact on the overall metallicity, whereas major mergers increasingly approach parity with the metallicity of their host and so rarely yield a dilution greater than $\sim0.3\,$dex. In fact, we find some realisations with GSE-like mergers that exhibit no metallicity overlap between their two $\alpha$-sequences, or do not have multiple sequences at all. Furthermore, the gas from most mergers forms only a minor component of subsequent \textit{in-situ} disc stars (with massive major mergers of mass ratio $>1:2$ being an exception).

This is not to suggest that mergers are entirely incapable of inducing dilution. For instance, \citet{grand2018} demonstrated that dilution occurs at the edge of the gas disc around the time of a late gas-rich merger in the high-resolution version of Au-23. In the version of Au-23 used in this study, we observe that dilution begins prior to the arrival of the merger gas, suggesting that the merger's gravitational influence may draw metal-poor gas from the CGM onto the disc's outskirts (which is then supplemented by the merger gas itself), thereby indirectly causing dilution. Nonetheless, this dilution is insufficient to account for the metallicity overlap observed in the Milky Way's chemical sequences, and the properties of this merger event are inconsistent with those expected for the GSE.

We show that the greatest metal dilution occurs when metal-poor merger gases mix with the highly-enriched central gas of the host, in agreement with \citet{buck2020} who find that merger-induced metal dilution in the {\sc nihao} simulations was maximised in the inner 3\,kpc. Simulations have also shown that gravitational torques from satellites can stimulate the inflow of metal-poor CGM gases into galactic centres \citep{dimatteo2007, bustamante2018, moreno2021, faria2025}, contributing to a central starburst and further metallicity dilution. Yet, observations of the MW indicate that the metallicity overlap between the two $\alpha$-sequences is \textit{least} pronounced from within $\sim6\,$kpc \citep[e.g.][]{queiroz2020}. Therefore, whilst these effects may have occurred in the MW, they were clearly not the primary drivers in the development of the chemical sequences. Then, it is unlikely that the metallicity overlap in the chemical sequences of the MW was caused by a highly radial infall such as the GSE, if it was caused by a merger event at all.

Contrary to our interpretations, \citet{buck2020} argue that metallicity dilution from gas-rich minor mergers (of less than $1:10$ mass ratio) explains the overlapping metallicities of bi-modal chemical sequences. However, we show that the metal dilution following those mergers is insufficient. The metallicity overlap is as great as 0.5\,dex (see their figure 1). Yet, from inspection of their figure 7, the metal dilution following gas-rich mergers is typically far less than half of this, and is short-lived in most cases. Then, this would be consistent with the findings presented in this work. In fact, our result is general and should hold for any reasonable galaxy formation model in a $\Lambda$CDM Universe. In Appendix \ref{AppendixB} we demonstrate this further using simple arguments involving empirical relations (e.g. stellar halo mass relation \citealp{moster2013}) and observations (e.g. the stellar mass-metallicity relation \citealp{kirby2013}).

\subsubsection{Influence on the star formation rate}

In Section \ref{sec:SFH}, we showed that {\sc auriga} galaxies with particularly pronounced bi-modal chemistry exhibit a lull in their star formation following merger-driven, centrally concentrated starbursts. Indeed, this is a behaviour also seen among other cosmological simulations like {\sc simba} \citep{rodriguez2019}, {\sc IllustrisTNG} \citep{hani2020}, {\sc vintergatan} \citep{segovia2022}, and {\sc fire-2} \citep{li2025}. In {\sc auriga}, the enhanced SFR has either fully or partially depleted the star-forming gas, ultimately leading to a temporary suppression of the star formation. This is reflected in observational data from SDSS (the Sloan Digital Sky Survey) finding evidence that starbursts correlate with features of merger interactions \citep{luo2014}. This is a possible scenario for the MW, where the two $\alpha$-sequences are divided by an under-populated gap, and there is a clear star formation quiescence at $10\text{-}8\,$Gyr ago (e.g. \citealt{haywood2016, haywood2018, spitoni2024}, and see also figure 7 in \citealt{snaith2015}).

Our finding is consistent with \citet{beane2024a}, where idealised simulations show that a brief star formation hiatus of a few hundred Myr could arise due to gas starvation following a GSE-like merger (though with great sensitivity to the infall inclination of said merger). The hiatus does not necessarily need to owe to the influence of a merger, and in a follow-up work they propose AGN-induced quenching coincident with the formation of a stellar bar as another possible mechanism (\citealp{beane2024b}, and see also \citealp{haywood2018}). Additionally, \citet{grand2018} demonstrate than the time-evolution in the size of the gas disc can lead to the formation of a chemical bi-modality over the pertinent radii. Disc-galaxy evolution models in \citet{noguchi2018} also predict that a star formation hiatus of approximately 2\,Gyr can create a bi-modality. Finally, simulations in \citet{lian2020} predict a starburst followed by rapid quenching could be responsible for forming the thick disc. These are compatible with observations of the MW from APOGEE, which finds evidence of a $\sim2\,$Gyr quenching period roughly 9\,Gyr ago \citep[see][]{haywood2016}.

\subsection{Fuelling from the CGM}

We find that, in most cases, the plurality of gas that forms the disc comes from a smooth accretion from the CGM and/or mini mergers. This is a universal feature for galaxies that grow from the cooling of CGM gas. The exceptions are realisations that experience late-time massive major mergers with mass ratios exceeding $1:2$, but these realisations normally have uni-modal chemistry, and are not considered a plausible scenario for the MW.

This is a result that is backed up by idealised simulations in \citet{khoperskov2021}, in which the low-$\alpha$ sequence forms from a mixture of metal-poor CGM and enriched gases that were ejected by stellar feedback during the formation of the high-$\alpha$ sequence. Furthermore, they find that the formation of two chemical sequences is not dependent on any one particular merger history or on radial migration.

Other fully cosmological simulations also highlight the importance of the CGM. Analysis of the FIRE-2 cosmological simulations in \citet{parul2025} showed that bi-modal sequences arise when metal-poor gas accretes onto the outskirts of the disc, owing to both merger donation and the CGM. However, their realisation with a high fraction of merger-donated gases (36 per cent) is only weakly bi-modal, and they conclude that merger-donated gases do not necessarily lead to the development of a distinct chemical sequence.

Investigations into the higher resolution versions of {\sc auriga} in \citet{grand2018} show that late-time minor mergers can penetrate the hot gas halo and deliver cold metal-poor gas directly onto the galaxy. While our own analysis shows that the merger gas neither dominates the low-$\alpha$ sequence nor reduces the metallicity by more than $\sim 0.3\,$dex, it nonetheless plays an important role in triggering the growth of the gas disc. The merger may also incite the condensation of surrounding metal-poor CGM gas onto the disc, leading to an indirect dilution effect.

In the {\sc vintergatan} cosmological simulations, a disc with bi-modal chemical sequences forms in the presence of a GSE-like merger \citep{agertz2021, agertz2021b}. In particular, their low-$\alpha$ sequence exhibits a `reset' in its metallicity. A detailed investigation revealed that the gas donated by the GSE-like merger was unimportant for both the formation of the low-$\alpha$ sequence and the metallicity dilution. Instead, they owe to an accreting metal-poor and cold gas filament that is initially misaligned with respect to the gas disc. The GSE-like merger is nonetheless essential because its interaction with this accreting gas filament acts to excite star formation within it. Similarly, we find that mergers are important for triggering star formation, even if their merger mass ratios are less than $1:100$. This is a result also echoed in \citet{parul2025}.

\subsection{Our interpretation for the MW history}

Based on our research and what we have learned from our analysis of the {\sc auriga} simulation suite, which is broadly consistent across different analytical and empirical models, we can formulate an interpretation for the history of our Galaxy:

The MW began to form early, with an inside-out and top-down formation of the stellar disc. These early stars formed during a period of intense SFR from a turbulent ISM. Whilst the bulk of the high-$\alpha$ sequence formed at this time, there may have been sub-dominant populations of low-$\alpha$ stars forming in dense gas clumps. The proto-galaxy would have undergone a series of high mass-ratio mergers at this time, perhaps including the speculative \textit{kraken} \citep[i.e.][]{Kruijssen2019, massari2019}. These mergers contribute gaseous fuel that would have been important for developing the early gas disc and initiating star formation in the high-$\alpha$ sequence, but would have been relatively unimportant in terms of their direct chemical impact (see \citealt{orkney2022} and arguments in Section \ref{sec:bimod_origin}).

At around 10\,Gyr ago, the GSE merger accretes onto the proto-MW. Following reasoning in the literature \citep[e.g.][]{Kruijssen2019, kruijssen2020, lane2023}, this was a minor merger with a mass ratio $<1:4$. This event initiates a combination of physical processes (inward funnelling of CGM gas, donation of fresh cold gas, gas shocks) which conspire to induce a central star formation burst. This burst bolsters the number of stars in the high-$\alpha$ sequence, but also acts to deplete the star-forming gas, and is followed by a reduction of the SFR that lasts $\mathcal{O} 1\,$Gyr. The GSE has other implications on the fledgling MW disc, kinematically heating its stars and shaping what is now the thick disc (e.g. see \citealt{grand2020}).

Assuming a variety of masses and metallicities for the GSE progenitor galaxy and contemporary proto-MW, we show in Appendix \ref{AppendixB} that the GSE merger would have been incapable of driving a substantial metallicity `reset' between the metal-rich end of the high-$\alpha$ sequence and the metal-poor end of the low-$\alpha$ sequence. Instead, continuous feeding from the CGM is responsible for maintaining a metal-poor zone at the height and radial extremities of the gas disc (see Figure \ref{fig:gasdisc}). The metal-poor ends of the low-$\alpha$ sequence is comprised from stars that form out of these gases.

With a relatively quiescent history thereafter, the MW disc becomes thinner and more stable, and the SFR gradually declines as the gas density in the centre dwindles. This is shown in the decline of the local SFR from around 6\,Gyrs ago in \citet{ruizlara2020}, and around 2.5\,Gyrs ago in \citet{Fernandez2025}, though with significant fluctuations in both cases. The outcome is the development of a radial gradient in the [$\alpha$/Fe] ratio, which grows from the inside-out as the enrichment from SNe Ia starts to dominate over the enrichment from SNe II. As we show in Section \ref{sec:shape}, this gradient can be imprinted on the low-$\alpha$ sequence as a downward `slant' (a negative relation between [Fe/H] and [Mg/Fe], and see also \citealt{mason2024}).

Several key elements are still missing from this interpretation. First, there is a lack of high-quality observational analogues of the MW at redshifts $z>2$. Current surveys continue to face challenges in resolving galaxies at the necessary mass scales at such early epochs. However, gravitational lensing techniques \citep{jones2023} offer a promising avenue for overcoming these limitations by magnifying faint and distant systems, though further progress is needed to fully exploit their potential. It is also crucial to identify observational evidence of dense, star-forming gas clumps and the specific environments in which they emerge. Finally, improved age statistics for stellar populations in the MW is essential, though upcoming missions like PLATO \citep{PLATO} may provide some relief on this front.

Our investigation is also relevant to M31, for which early spectroscopic results suggest a single chemical sequence similar to the MW high-$\alpha$ disc \citep{escala2020, nidever2024}. Inspection of Figure 4 in \citet{nidever2024} reveals a narrow chemical track that spans from high-$\alpha$, low-metallicity to low-$\alpha$, high-metallicity stars. Compared to our analysis, this distribution closely resembles realisations that form an early thin gaseous disc (e.g., Au-2 and Au-24). This appears to conflict with observational evidence from dust geometry indicating that the dominant disc component of M31 has a thickness comparable to the MW thick disc \citep{dalcanton2023}. However, these kinematics may be shaped by a possible major merger 2-4,Gyr ago \citep[see arguments in][]{hammer2018, Bhattacharya2024, tsakonas2025}.

A recurring theme is that decrypting the earliest phases of our Galaxy’s formation will increasingly rely on constraints from high-redshift observations. The coming decade of MW exploration will not be solely a Galactic Archaeology enterprise, but very likely an extra-galactic effort.

\section{Conclusions} \label{sec:conclusions}

We have performed an investigation into disc formation in the {\sc auriga} simulations of MW-mass galaxies. Our primary focus is the emergence of chemical sequences in the [Fe/H] versus [Mg/Fe] chemical abundance plane, including bi-modalities as is seen in the MW for disc stars around the Solar neighbourhood and in the Galactic centre. Whilst none of the simulations are an ideal match for the MW, we have isolated the formation mechanisms behind various properties which can be used to constrain or probe ideas about our Galaxy. Our key results are as follows:

\begin{itemize}
\setlength{\itemindent}{0.5em}
    \item Half of the thirty {\sc auriga} simulations exhibit bi-modal distribution in their [Mg/Fe] distribution, with a further two having multi-modal distributions. Regardless of the presence of a bi-modality, high-$\alpha$ and low-$\alpha$ populations usually adhere to the same typical bulk properties --- those being hotter kinematics (with higher $\sigma_z$ by an average of $33\,\rm{km}\,\rm{s}^{-1}$), slower rotational velocities (by an average of $32\,\rm{km}\,\rm{s}^{-1}$), and older ages for the high-$\alpha$ stars (by an average of 4.6\,Gyrs).
    \item There is no clear correlation between simulations with pronounced bi-modal distributions and those with GSE-like debris in their stellar haloes. This shows that the formation of two chemical sequences is not dependent on a GSE-like merger.
    \item Singular merger events, in particular those that are GSE-like, do not dilute the metallicity of the CGM by more than $\approx0.3\,$dex, and usually by even less. The gas donated by GSE-like mergers forms only a minority of subsequent disc stars. The only exceptions are in the cases of very massive mergers (merger mass ratios of $>1:2$, and total masses of $>2\times10^{10}$), though these are ruled out for the MW. Therefore, the GSE gas was not responsible for the overlapping metallicities of the high and low-$\alpha$ sequences, and we have reinforced this with empirical estimates. Other merger events such as \textit{kraken} will have been just as unimportant, even if they were of higher mass ratio \citep[see also arguments in][]{orkney2022}.
    \item Instead, the plurality of gas fuel for the formation of low-$\alpha$ sequences comes from a smooth accretion of the CGM. Sequences with low-[Fe/H] stars come from metal-poor gas that is freshly accreted onto the extremities of the gas disc. In contrast, gas discs that are thin form fewer low-[Fe/H] stars because the enriched and metal-poor gases mix together to form an enriched radial chemical gradient.
    \item The break between two chemical sequences is driven by a temporary quenching or reduction of the SFR. This can be due to multiple reasons, including gas exhaustion, post-merger quenching, or a violent disruption of the gas disc. As such, massive mergers such as the GSE can potentially contribute to the development of chemical bi-modalities due to their merger-induced starbursts and the following quiescence. Other mechanisms may also be important, such as AGN-induced shutdowns or bar formation (e.g. \citealt{beane2024a, beane2024b}), though these may themselves be triggered by a merger event. Disentangling these processes is beyond the scope of this work.
    \item The distribution of chemical sequences in the [Fe/H] versus [Mg/Fe] property planes can be explained by trends in the SFR. Increasing SFRs lead to positive relations between [Fe/H] and [Mg/Fe], whereas decreasing or steady SFRs lead to the inverse.
    \item As in prior studies, we find that radial migration is in effect but is not the underlying cause for the development of multiple chemical sequences.
\end{itemize}

From these findings, we can explain several features observed in the MW bi-modality:

\begin{itemize}
\setlength{\itemindent}{0.5em}
    \item The gap between the high-$\alpha$ and low-$\alpha$ sequences is produced by a hiatus in the star formation, possibly due to the accretion of the GSE merger.
    \item The slanted-shaped locus in the low-$\alpha$ sequence is due to the decaying SFR over the past $\sim6\,$Gyrs.
    \item The overlap in the metallicities of high-$\alpha$ and low-$\alpha$ stars in the Solar neighbourhood arises due to a fraction of low-$\alpha$ stars forming in recently accreted metal-poor gases from the CGM.
    \item The lack of a metal overlap in the inner Galaxy implies that the low-$\alpha$ sequence formed from a thin gas disc at those radii. This suggests the gas disc was thin but strongly flared during much of the low-$\alpha$ disc formation.
\end{itemize}

Our results highlight the importance of obtaining high-quality observational data for MW analogues at higher redshifts ($z \geq 2$). The James Webb Space Telescope (JWST) has recently begun probing such galaxies \citep[e.g.][]{mowla2024, tan2024, tan2024b}, and surveys like JADES (JWST Advanced Deep Extragalactic Survey) may already be detecting MW analogues as early as $z \approx 8.2$ \citep{rusta2024}.

Early spectroscopic studies of M31 suggest that its chemistry may not be bi-modal \citep{escala2020, nidever2024}. As observations of other Local Volume galaxies continue to improve, understanding the formation of chemical abundance patterns beyond the MW will become increasingly important. Our investigation revealed a broad diversity in the chemical evolution of MW-mass disc galaxies, which are highly sensitive to factors such as overall mass assembly, star formation history, and even the specifics of individual merger events. To better constrain the physical and dynamical processes that set chemical trends in MW-like galaxies, detailed extragalactic observations within the Local Volume, as well as the CGM of high-redshift galaxies, will prove essential in the era of Extremely Large Telescopes \citep[see][]{hammer2021, padovani2023}.

\section{Acknowledgements}

We thank the anonymous referee for their thoughtful comments, which have greatly contributed to this manuscript.
This work has used resources from the MareNostrum 4 supercomputer at the Barcelona Supercomputing Center (BSC). Much of our analysis was performed on the Virgo supercomputer at the Max Planck Computing and Data Facility (MPCDF). Further analysis and simulation work was performed on the NYX supercomputer at the Universit\"{a}t de Barcelona (ICCUB).
CL \& MO acknowledge funding from the European Research Council (ERC) under the European Union’s Horizon 2020 research and innovation programme (grant agreement No. 852839). CL also acknowledges funding from the Agence Nationale de la Recherche (ANR project ANR-24-CPJ1-0160-01).
The author acknowledges financial support from the CEX2024-001451-M grant funded by MICIU/AEI/10.13039/501100011033.
RJJG is supported by an STFC Ernest Rutherford Fellowship (ST/W003643/1).

\section*{Data Availability}

The {\sc auriga} simulation data is publicly available to download via the Globus platform as described in section 4 of \citet{Auriga}.
Any novel post-processed data is available upon request.



\bibliographystyle{mnras}
\bibliography{references} 

@ARTICLE{Auriga,
       author = {{Grand}, Robert J.~J. and {G{\'o}mez}, Facundo A. and {Marinacci}, Federico and {Pakmor}, R{\"u}diger and {Springel}, Volker and {Campbell}, David J.~R. and {Frenk}, Carlos S. and {Jenkins}, Adrian and {White}, Simon D.~M.},
        title = "{The Auriga Project: the properties and formation mechanisms of disc galaxies across cosmic time}",
      journal = {\mnras},
         year = 2017,
        month = may,
       volume = {467},
       number = {1},
        pages = {179-207},
          doi = {10.1093/mnras/stx071},
archivePrefix = {arXiv},
       eprint = {1610.01159},
 primaryClass = {astro-ph.GA},
       adsurl = {https://ui.adsabs.harvard.edu/abs/2017MNRAS.467..179G}
}

@ARTICLE{fattahi2019,
       author = {{Fattahi}, Azadeh and {Belokurov}, Vasily and {Deason}, Alis J. and {Frenk}, Carlos S. and {G{\'o}mez}, Facundo A. and {Grand}, Robert J.~J. and {Marinacci}, Federico and {Pakmor}, R{\"u}diger and {Springel}, Volker},
        title = "{The origin of galactic metal-rich stellar halo components with highly eccentric orbits}",
      journal = {\mnras},
         year = 2019,
        month = apr,
       volume = {484},
       number = {4},
        pages = {4471-4483},
          doi = {10.1093/mnras/stz159},
archivePrefix = {arXiv},
       eprint = {1810.07779},
 primaryClass = {astro-ph.GA},
       adsurl = {https://ui.adsabs.harvard.edu/abs/2019MNRAS.484.4471F}
}

@article{Bhattacharyya1946,
 ISSN = {00364452},
 URL = {http://www.jstor.org/stable/25047882},
 author = {A. Bhattacharyya},
 journal = {Sankhyā: The Indian Journal of Statistics (1933-1960)},
 number = {4},
 pages = {401--406},
 publisher = {Springer},
 title = {On a Measure of Divergence between Two Multinomial Populations},
 urldate = {2024-08-27},
 volume = {7},
 year = {1946}
}

@ARTICLE{abadi2003,
       author = {{Abadi}, Mario G. and {Navarro}, Julio F. and {Steinmetz}, Matthias and {Eke}, Vincent R.},
        title = "{Simulations of Galaxy Formation in a {\ensuremath{\Lambda}} Cold Dark Matter Universe. I. Dynamical and Photometric Properties of a Simulated Disk Galaxy}",
      journal = {\apj},
         year = 2003,
        month = jul,
       volume = {591},
       number = {2},
        pages = {499-514},
          doi = {10.1086/375512},
archivePrefix = {arXiv},
       eprint = {astro-ph/0211331},
 primaryClass = {astro-ph},
       adsurl = {https://ui.adsabs.harvard.edu/abs/2003ApJ...591..499A}
}

@ARTICLE{agertz2021,
       author = {{Agertz}, Oscar and {Renaud}, Florent and {Feltzing}, Sofia and {Read}, Justin I. and {Ryde}, Nils and {Andersson}, Eric P. and {Rey}, Martin P. and {Bensby}, Thomas and {Feuillet}, Diane K.},
        title = "{VINTERGATAN - I. The origins of chemically, kinematically, and structurally distinct discs in a simulated Milky Way-mass galaxy}",
      journal = {\mnras},
         year = 2021,
        month = jun,
       volume = {503},
       number = {4},
        pages = {5826-5845},
          doi = {10.1093/mnras/stab322},
archivePrefix = {arXiv},
       eprint = {2006.06008},
 primaryClass = {astro-ph.GA},
       adsurl = {https://ui.adsabs.harvard.edu/abs/2021MNRAS.503.5826A}
}

@ARTICLE{agertz2021b,
       author = {{Renaud}, Florent and {Agertz}, Oscar and {Andersson}, Eric P. and {Read}, Justin I. and {Ryde}, Nils and {Bensby}, Thomas and {Rey}, Martin P. and {Feuillet}, Diane K.},
        title = "{VINTERGATAN III: how to reset the metallicity of the Milky Way}",
      journal = {\mnras},
         year = 2021,
        month = jun,
       volume = {503},
       number = {4},
        pages = {5868-5876},
          doi = {10.1093/mnras/stab543},
archivePrefix = {arXiv},
       eprint = {2006.06012},
 primaryClass = {astro-ph.GA},
       adsurl = {https://ui.adsabs.harvard.edu/abs/2021MNRAS.503.5868R}
}

@ARTICLE{grand2020,
       author = {{Grand}, Robert J.~J. and {Kawata}, Daisuke and {Belokurov}, Vasily and {Deason}, Alis J. and {Fattahi}, Azadeh and {Fragkoudi}, Francesca and {G{\'o}mez}, Facundo A. and {Marinacci}, Federico and {Pakmor}, R{\"u}diger},
        title = "{The dual origin of the Galactic thick disc and halo from the gas-rich Gaia-Enceladus Sausage merger}",
      journal = {\mnras},
         year = 2020,
        month = sep,
       volume = {497},
       number = {2},
        pages = {1603-1618},
          doi = {10.1093/mnras/staa2057},
archivePrefix = {arXiv},
       eprint = {2001.06009},
 primaryClass = {astro-ph.GA},
       adsurl = {https://ui.adsabs.harvard.edu/abs/2020MNRAS.497.1603G}
}

@ARTICLE{orkney2023,
       author = {{Orkney}, Matthew D.~A. and {Laporte}, Chervin F.~P. and {Grand}, Robert J.~J. and {G{\'o}mez}, Facundo A. and {van de Voort}, Freeke and {Fattahi}, Azadeh and {Marinacci}, Federico and {Pakmor}, R{\"u}diger and {Fragkoudi}, Francesca and {Springel}, Volker},
        title = "{Exploring the diversity and similarity of radially anisotropic Milky Way-like stellar haloes: implications for disrupted dwarf galaxy searches}",
      journal = {\mnras},
         year = 2023,
        month = oct,
       volume = {525},
       number = {1},
        pages = {683-705},
          doi = {10.1093/mnras/stad2361},
archivePrefix = {arXiv},
       eprint = {2303.02147},
 primaryClass = {astro-ph.GA},
       adsurl = {https://ui.adsabs.harvard.edu/abs/2023MNRAS.525..683O}
}

@ARTICLE{asplund2009,
       author = {{Asplund}, Martin and {Grevesse}, Nicolas and {Sauval}, A. Jacques and {Scott}, Pat},
        title = "{The Chemical Composition of the Sun}",
      journal = {\araa},
         year = 2009,
        month = sep,
       volume = {47},
       number = {1},
        pages = {481-522},
          doi = {10.1146/annurev.astro.46.060407.145222},
archivePrefix = {arXiv},
       eprint = {0909.0948},
 primaryClass = {astro-ph.SR},
       adsurl = {https://ui.adsabs.harvard.edu/abs/2009ARA&A..47..481A}
}

@ARTICLE{gomez2017,
       author = {{G{\'o}mez}, Facundo A. and {Grand}, Robert J.~J. and {Monachesi}, Antonela and {White}, Simon D.~M. and {Bustamante}, Sebastian and {Marinacci}, Federico and {Pakmor}, R{\"u}diger and {Simpson}, Christine M. and {Springel}, Volker and {Frenk}, Carlos S.},
        title = "{Lessons from the Auriga discs: the hunt for the Milky Way's ex situ disc is not yet over}",
      journal = {\mnras},
         year = 2017,
        month = dec,
       volume = {472},
       number = {3},
        pages = {3722-3733},
          doi = {10.1093/mnras/stx2149},
archivePrefix = {arXiv},
       eprint = {1704.08261},
 primaryClass = {astro-ph.GA},
       adsurl = {https://ui.adsabs.harvard.edu/abs/2017MNRAS.472.3722G}
}

@ARTICLE{orkney2022,
       author = {{Orkney}, Matthew D.~A. and {Laporte}, Chervin F.~P. and {Grand}, Robert J.~J. and {G{\'o}mez}, Facundo A. and {van de Voort}, Freeke and {Marinacci}, Federico and {Fragkoudi}, Francesca and {Pakmor}, Ruediger and {Springel}, Volker},
        title = "{The impact of two massive early accretion events in a Milky Way-like galaxy: repercussions for the buildup of the stellar disc and halo}",
      journal = {\mnras},
         year = 2022,
        month = nov,
       volume = {517},
       number = {1},
        pages = {L138-L142},
          doi = {10.1093/mnrasl/slac126},
archivePrefix = {arXiv},
       eprint = {2206.09246},
 primaryClass = {astro-ph.GA},
       adsurl = {https://ui.adsabs.harvard.edu/abs/2022MNRAS.517L.138O}
}

@ARTICLE{chiappini1997,
       author = {{Chiappini}, C. and {Matteucci}, F. and {Gratton}, R.},
        title = "{The Chemical Evolution of the Galaxy: The Two-Infall Model}",
      journal = {\apj},
         year = 1997,
        month = mar,
       volume = {477},
       number = {2},
        pages = {765-780},
          doi = {10.1086/303726},
archivePrefix = {arXiv},
       eprint = {astro-ph/9609199},
 primaryClass = {astro-ph},
       adsurl = {https://ui.adsabs.harvard.edu/abs/1997ApJ...477..765C}
}

@ARTICLE{prantzos2023,
       author = {{Prantzos}, Nikos and {Abia}, Carlos and {Chen}, Tianxiang and {de Laverny}, Patrick and {Recio-Blanco}, Alejandra and {Athanassoula}, E. and {Roberti}, Lorenzo and {Vescovi}, Diego and {Limongi}, Marco and {Chieffi}, Alessandro and {Cristallo}, Sergio},
        title = "{On the origin of the Galactic thin and thick discs, their abundance gradients and the diagnostic potential of their abundance ratios}",
      journal = {\mnras},
         year = 2023,
        month = aug,
       volume = {523},
       number = {2},
        pages = {2126-2145},
          doi = {10.1093/mnras/stad1551},
archivePrefix = {arXiv},
       eprint = {2305.13431},
 primaryClass = {astro-ph.GA},
       adsurl = {https://ui.adsabs.harvard.edu/abs/2023MNRAS.523.2126P}
}

@ARTICLE{ruchti2011,
       author = {{Ruchti}, Gregory R. and {Fulbright}, Jon P. and {Wyse}, Rosemary F.~G. and {Gilmore}, Gerard F. and {Bienaym{\'e}}, Olivier and {Bland-Hawthorn}, Joss and {Gibson}, Brad K. and {Grebel}, Eva K. and {Helmi}, Amina and {Munari}, Ulisse and {Navarro}, Julio F. and {Parker}, Quentin A. and {Reid}, Warren and {Seabroke}, George M. and {Siebert}, Arnaud and {Siviero}, Alessandro and {Steinmetz}, Matthias and {Watson}, Fred G. and {Williams}, Mary and {Zwitter}, Tomaz},
        title = "{Observational Properties of the Metal-poor Thick Disk of the Milky Way and Insights into its Origins}",
      journal = {\apj},
         year = 2011,
        month = aug,
       volume = {737},
       number = {1},
          eid = {9},
        pages = {9},
          doi = {10.1088/0004-637X/737/1/9},
archivePrefix = {arXiv},
       eprint = {1105.3691},
 primaryClass = {astro-ph.GA},
       adsurl = {https://ui.adsabs.harvard.edu/abs/2011ApJ...737....9R}
}

@ARTICLE{Panphasia,
       author = {{Jenkins}, Adrian},
        title = "{A new way of setting the phases for cosmological multiscale Gaussian initial conditions}",
      journal = {\mnras},
         year = 2013,
        month = sep,
       volume = {434},
       number = {3},
        pages = {2094-2120},
          doi = {10.1093/mnras/stt1154},
archivePrefix = {arXiv},
       eprint = {1306.5968},
 primaryClass = {astro-ph.CO},
       adsurl = {https://ui.adsabs.harvard.edu/abs/2013MNRAS.434.2094J}
}

@ARTICLE{subfind,
       author = {{Springel}, Volker and {White}, Simon D.~M. and {Tormen}, Giuseppe and {Kauffmann}, Guinevere},
        title = "{Populating a cluster of galaxies - I. Results at [formmu2]z=0}",
      journal = {\mnras},
         year = 2001,
        month = dec,
       volume = {328},
       number = {3},
        pages = {726-750},
          doi = {10.1046/j.1365-8711.2001.04912.x},
archivePrefix = {arXiv},
       eprint = {astro-ph/0012055},
 primaryClass = {astro-ph},
       adsurl = {https://ui.adsabs.harvard.edu/abs/2001MNRAS.328..726S}
}

@ARTICLE{Planck2014,
       author = {{Planck Collaboration} and {Ade}, P.~A.~R. and {Aghanim}, N. and {Armitage-Caplan}, C. and {Arnaud}, M. and {Ashdown}, M. and {Atrio-Barandela}, F. and {Aumont}, J. and {Baccigalupi}, C. and {Banday}, A.~J. and {Barreiro}, R.~B. and {Bartlett}, J.~G. and {Battaner}, E. and {Benabed}, K. and {Beno{\^\i}t}, A. and {Benoit-L{\'e}vy}, A. and {Bernard}, J. -P. and {Bersanelli}, M. and {Bielewicz}, P. and {Bobin}, J. and {Bock}, J.~J. and {Bonaldi}, A. and {Bond}, J.~R. and {Borrill}, J. and {Bouchet}, F.~R. and {Bridges}, M. and {Bucher}, M. and {Burigana}, C. and {Butler}, R.~C. and {Calabrese}, E. and {Cappellini}, B. and {Cardoso}, J. -F. and {Catalano}, A. and {Challinor}, A. and {Chamballu}, A. and {Chary}, R. -R. and {Chen}, X. and {Chiang}, H.~C. and {Chiang}, L. -Y. and {Christensen}, P.~R. and {Church}, S. and {Clements}, D.~L. and {Colombi}, S. and {Colombo}, L.~P.~L. and {Couchot}, F. and {Coulais}, A. and {Crill}, B.~P. and {Curto}, A. and {Cuttaia}, F. and {Danese}, L. and {Davies}, R.~D. and {Davis}, R.~J. and {de Bernardis}, P. and {de Rosa}, A. and {de Zotti}, G. and {Delabrouille}, J. and {Delouis}, J. -M. and {D{\'e}sert}, F. -X. and {Dickinson}, C. and {Diego}, J.~M. and {Dolag}, K. and {Dole}, H. and {Donzelli}, S. and {Dor{\'e}}, O. and {Douspis}, M. and {Dunkley}, J. and {Dupac}, X. and {Efstathiou}, G. and {Elsner}, F. and {En{\ss}lin}, T.~A. and {Eriksen}, H.~K. and {Finelli}, F. and {Forni}, O. and {Frailis}, M. and {Fraisse}, A.~A. and {Franceschi}, E. and {Gaier}, T.~C. and {Galeotta}, S. and {Galli}, S. and {Ganga}, K. and {Giard}, M. and {Giardino}, G. and {Giraud-H{\'e}raud}, Y. and {Gjerl{\o}w}, E. and {Gonz{\'a}lez-Nuevo}, J. and {G{\'o}rski}, K.~M. and {Gratton}, S. and {Gregorio}, A. and {Gruppuso}, A. and {Gudmundsson}, J.~E. and {Haissinski}, J. and {Hamann}, J. and {Hansen}, F.~K. and {Hanson}, D. and {Harrison}, D. and {Henrot-Versill{\'e}}, S. and {Hern{\'a}ndez-Monteagudo}, C. and {Herranz}, D. and {Hildebrandt}, S.~R. and {Hivon}, E. and {Hobson}, M. and {Holmes}, W.~A. and {Hornstrup}, A. and {Hou}, Z. and {Hovest}, W. and {Huffenberger}, K.~M. and {Jaffe}, A.~H. and {Jaffe}, T.~R. and {Jewell}, J. and {Jones}, W.~C. and {Juvela}, M. and {Keih{\"a}nen}, E. and {Keskitalo}, R. and {Kisner}, T.~S. and {Kneissl}, R. and {Knoche}, J. and {Knox}, L. and {Kunz}, M. and {Kurki-Suonio}, H. and {Lagache}, G. and {L{\"a}hteenm{\"a}ki}, A. and {Lamarre}, J. -M. and {Lasenby}, A. and {Lattanzi}, M. and {Laureijs}, R.~J. and {Lawrence}, C.~R. and {Leach}, S. and {Leahy}, J.~P. and {Leonardi}, R. and {Le{\'o}n-Tavares}, J. and {Lesgourgues}, J. and {Lewis}, A. and {Liguori}, M. and {Lilje}, P.~B. and {Linden-V{\o}rnle}, M. and {L{\'o}pez-Caniego}, M. and {Lubin}, P.~M. and {Mac{\'\i}as-P{\'e}rez}, J.~F. and {Maffei}, B. and {Maino}, D. and {Mandolesi}, N. and {Maris}, M. and {Marshall}, D.~J. and {Martin}, P.~G. and {Mart{\'\i}nez-Gonz{\'a}lez}, E. and {Masi}, S. and {Massardi}, M. and {Matarrese}, S. and {Matthai}, F. and {Mazzotta}, P. and {Meinhold}, P.~R. and {Melchiorri}, A. and {Melin}, J. -B. and {Mendes}, L. and {Menegoni}, E. and {Mennella}, A. and {Migliaccio}, M. and {Millea}, M. and {Mitra}, S. and {Miville-Desch{\^e}nes}, M. -A. and {Moneti}, A. and {Montier}, L. and {Morgante}, G. and {Mortlock}, D. and {Moss}, A. and {Munshi}, D. and {Murphy}, J.~A. and {Naselsky}, P. and {Nati}, F. and {Natoli}, P. and {Netterfield}, C.~B. and {N{\o}rgaard-Nielsen}, H.~U. and {Noviello}, F. and {Novikov}, D. and {Novikov}, I. and {O'Dwyer}, I.~J. and {Osborne}, S. and {Oxborrow}, C.~A. and {Paci}, F. and {Pagano}, L. and {Pajot}, F. and {Paladini}, R. and {Paoletti}, D. and {Partridge}, B. and {Pasian}, F. and {Patanchon}, G. and {Pearson}, D. and {Pearson}, T.~J. and {Peiris}, H.~V. and {Perdereau}, O. and {Perotto}, L. and {Perrotta}, F. and {Pettorino}, V. and {Piacentini}, F. and {Piat}, M. and {Pierpaoli}, E. and {Pietrobon}, D. and {Plaszczynski}, S. and {Platania}, P. and {Pointecouteau}, E. and {Polenta}, G. and {Ponthieu}, N. and {Popa}, L. and {Poutanen}, T. and {Pratt}, G.~W. and {Pr{\'e}zeau}, G. and {Prunet}, S. and {Puget}, J. -L. and {Rachen}, J.~P. and {Reach}, W.~T. and {Rebolo}, R. and {Reinecke}, M. and {Remazeilles}, M. and {Renault}, C. and {Ricciardi}, S. and {Riller}, T. and {Ristorcelli}, I. and {Rocha}, G. and {Rosset}, C. and {Roudier}, G. and {Rowan-Robinson}, M. and {Rubi{\~n}o-Mart{\'\i}n}, J.~A. and {Rusholme}, B. and {Sandri}, M. and {Santos}, D. and {Savelainen}, M. and {Savini}, G. and {Scott}, D. and {Seiffert}, M.~D. and {Shellard}, E.~P.~S. and {Spencer}, L.~D. and {Starck}, J. -L. and {Stolyarov}, V. and {Stompor}, R. and {Sudiwala}, R. and {Sunyaev}, R. and {Sureau}, F. and {Sutton}, D. and {Suur-Uski}, A. -S. and {Sygnet}, J. -F. and {Tauber}, J.~A. and {Tavagnacco}, D. and {Terenzi}, L. and {Toffolatti}, L. and {Tomasi}, M. and {Tristram}, M. and {Tucci}, M. and {Tuovinen}, J. and {T{\"u}rler}, M. and {Umana}, G. and {Valenziano}, L. and {Valiviita}, J. and {Van Tent}, B. and {Vielva}, P. and {Villa}, F. and {Vittorio}, N. and {Wade}, L.~A. and {Wandelt}, B.~D. and {Wehus}, I.~K. and {White}, M. and {White}, S.~D.~M. and {Wilkinson}, A. and {Yvon}, D. and {Zacchei}, A. and {Zonca}, A.},
        title = "{Planck 2013 results. XVI. Cosmological parameters}",
      journal = {\aap},
         year = 2014,
        month = nov,
       volume = {571},
          eid = {A16},
        pages = {A16},
          doi = {10.1051/0004-6361/201321591},
archivePrefix = {arXiv},
       eprint = {1303.5076},
 primaryClass = {astro-ph.CO},
       adsurl = {https://ui.adsabs.harvard.edu/abs/2014A&A...571A..16P}
}

@ARTICLE{marinacci2014,
       author = {{Marinacci}, Federico and {Pakmor}, R{\"u}diger and {Springel}, Volker},
        title = "{The formation of disc galaxies in high-resolution moving-mesh cosmological simulations}",
      journal = {\mnras},
         year = 2014,
        month = jan,
       volume = {437},
       number = {2},
        pages = {1750-1775},
          doi = {10.1093/mnras/stt2003},
archivePrefix = {arXiv},
       eprint = {1305.5360},
 primaryClass = {astro-ph.CO},
       adsurl = {https://ui.adsabs.harvard.edu/abs/2014MNRAS.437.1750M}
}

@ARTICLE{vogelsberger2013,
       author = {{Vogelsberger}, Mark and {Genel}, Shy and {Sijacki}, Debora and {Torrey}, Paul and {Springel}, Volker and {Hernquist}, Lars},
        title = "{A model for cosmological simulations of galaxy formation physics}",
      journal = {\mnras},
         year = 2013,
        month = dec,
       volume = {436},
       number = {4},
        pages = {3031-3067},
          doi = {10.1093/mnras/stt1789},
archivePrefix = {arXiv},
       eprint = {1305.2913},
 primaryClass = {astro-ph.CO},
       adsurl = {https://ui.adsabs.harvard.edu/abs/2013MNRAS.436.3031V}
}

@ARTICLE{genel2014,
       author = {{Genel}, Shy and {Vogelsberger}, Mark and {Springel}, Volker and {Sijacki}, Debora and {Nelson}, Dylan and {Snyder}, Greg and {Rodriguez-Gomez}, Vicente and {Torrey}, Paul and {Hernquist}, Lars},
        title = "{Introducing the Illustris project: the evolution of galaxy populations across cosmic time}",
      journal = {\mnras},
         year = 2014,
        month = nov,
       volume = {445},
       number = {1},
        pages = {175-200},
          doi = {10.1093/mnras/stu1654},
archivePrefix = {arXiv},
       eprint = {1405.3749},
 primaryClass = {astro-ph.CO},
       adsurl = {https://ui.adsabs.harvard.edu/abs/2014MNRAS.445..175G}
}

@ARTICLE{springel2005,
       author = {{Springel}, Volker and {White}, Simon D.~M. and {Jenkins}, Adrian and {Frenk}, Carlos S. and {Yoshida}, Naoki and {Gao}, Liang and {Navarro}, Julio and {Thacker}, Robert and {Croton}, Darren and {Helly}, John and {Peacock}, John A. and {Cole}, Shaun and {Thomas}, Peter and {Couchman}, Hugh and {Evrard}, August and {Colberg}, J{\"o}rg and {Pearce}, Frazer},
        title = "{Simulations of the formation, evolution and clustering of galaxies and quasars}",
      journal = {\nat},
         year = 2005,
        month = jun,
       volume = {435},
       number = {7042},
        pages = {629-636},
          doi = {10.1038/nature03597},
archivePrefix = {arXiv},
       eprint = {astro-ph/0504097},
 primaryClass = {astro-ph},
       adsurl = {https://ui.adsabs.harvard.edu/abs/2005Natur.435..629S}
}

@ARTICLE{monachesi2019,
       author = {{Monachesi}, Antonela and {G{\'o}mez}, Facundo A. and {Grand}, Robert J.~J. and {Simpson}, Christine M. and {Kauffmann}, Guinevere and {Bustamante}, Sebasti{\'a}n and {Marinacci}, Federico and {Pakmor}, R{\"u}diger and {Springel}, Volker and {Frenk}, Carlos S. and {White}, Simon D.~M. and {Tissera}, Patricia B.},
        title = "{The Auriga stellar haloes: connecting stellar population properties with accretion and merging history}",
      journal = {\mnras},
         year = 2019,
        month = may,
       volume = {485},
       number = {2},
        pages = {2589-2616},
          doi = {10.1093/mnras/stz538},
archivePrefix = {arXiv},
       eprint = {1804.07798},
 primaryClass = {astro-ph.GA},
       adsurl = {https://ui.adsabs.harvard.edu/abs/2019MNRAS.485.2589M}
}

@ARTICLE{grand2018,
       author = {{Grand}, Robert J.~J. and {Bustamante}, Sebasti{\'a}n and {G{\'o}mez}, Facundo A. and {Kawata}, Daisuke and {Marinacci}, Federico and {Pakmor}, R{\"u}diger and {Rix}, Hans-Walter and {Simpson}, Christine M. and {Sparre}, Martin and {Springel}, Volker},
        title = "{Origin of chemically distinct discs in the Auriga cosmological simulations}",
      journal = {\mnras},
         year = 2018,
        month = mar,
       volume = {474},
       number = {3},
        pages = {3629-3639},
          doi = {10.1093/mnras/stx3025},
archivePrefix = {arXiv},
       eprint = {1708.07834},
 primaryClass = {astro-ph.GA},
       adsurl = {https://ui.adsabs.harvard.edu/abs/2018MNRAS.474.3629G}
}

@ARTICLE{fragkoudi2020,
       author = {{Fragkoudi}, F. and {Grand}, R.~J.~J. and {Pakmor}, R. and {Bl{\'a}zquez-Calero}, G. and {Gargiulo}, I. and {Gomez}, F. and {Marinacci}, F. and {Monachesi}, A. and {Ness}, M.~K. and {Perez}, I. and {Tissera}, P. and {White}, S.~D.~M.},
        title = "{Chemodynamics of barred galaxies in cosmological simulations: On the Milky Way's quiescent merger history and in-situ bulge}",
      journal = {\mnras},
         year = 2020,
        month = jun,
       volume = {494},
       number = {4},
        pages = {5936-5960},
          doi = {10.1093/mnras/staa1104},
archivePrefix = {arXiv},
       eprint = {1911.06826},
 primaryClass = {astro-ph.GA},
       adsurl = {https://ui.adsabs.harvard.edu/abs/2020MNRAS.494.5936F}
}

@ARTICLE{hayden2015,
       author = {{Hayden}, Michael R. and {Bovy}, Jo and {Holtzman}, Jon A. and {Nidever}, David L. and {Bird}, Jonathan C. and {Weinberg}, David H. and {Andrews}, Brett H. and {Majewski}, Steven R. and {Allende Prieto}, Carlos and {Anders}, Friedrich and {Beers}, Timothy C. and {Bizyaev}, Dmitry and {Chiappini}, Cristina and {Cunha}, Katia and {Frinchaboy}, Peter and {Garc{\'\i}a-Her{\'n}andez}, D.~A. and {Garc{\'\i}a P{\'e}rez}, Ana E. and {Girardi}, L{\'e}o and {Harding}, Paul and {Hearty}, Fred R. and {Johnson}, Jennifer A. and {M{\'e}sz{\'a}ros}, Szabolcs and {Minchev}, Ivan and {O'Connell}, Robert and {Pan}, Kaike and {Robin}, Annie C. and {Schiavon}, Ricardo P. and {Schneider}, Donald P. and {Schultheis}, Mathias and {Shetrone}, Matthew and {Skrutskie}, Michael and {Steinmetz}, Matthias and {Smith}, Verne and {Wilson}, John C. and {Zamora}, Olga and {Zasowski}, Gail},
        title = "{Chemical Cartography with APOGEE: Metallicity Distribution Functions and the Chemical Structure of the Milky Way Disk}",
      journal = {\apj},
         year = 2015,
        month = aug,
       volume = {808},
       number = {2},
          eid = {132},
        pages = {132},
          doi = {10.1088/0004-637X/808/2/132},
archivePrefix = {arXiv},
       eprint = {1503.02110},
 primaryClass = {astro-ph.GA},
       adsurl = {https://ui.adsabs.harvard.edu/abs/2015ApJ...808..132H}
}

@ARTICLE{beraldo2021,
       author = {{Beraldo e Silva}, Leandro and {Debattista}, Victor P. and {Nidever}, David and {Amarante}, Jo{\~a}o A.~S. and {Garver}, Bethany},
        title = "{Co-formation of the thin and thick discs revealed by APOGEE-DR16 and Gaia-DR2}",
      journal = {\mnras},
         year = 2021,
        month = mar,
       volume = {502},
       number = {1},
        pages = {260-272},
          doi = {10.1093/mnras/staa3966},
archivePrefix = {arXiv},
       eprint = {2009.03346},
 primaryClass = {astro-ph.GA},
       adsurl = {https://ui.adsabs.harvard.edu/abs/2021MNRAS.502..260B}
}

@ARTICLE{gent2024,
       author = {{Gent}, Matthew Raymond and {Eitner}, Philipp and {Serenelli}, Aldo and {Friske}, Jennifer K.~S. and {Koposov}, Sergey E. and {Laporte}, Chervin F.~P. and {Buck}, Tobias and {Bergemann}, Maria},
        title = "{The Prince and the Pauper: Evidence for the early high-redshift formation of the Galactic {\ensuremath{\alpha}}-poor disc population}",
      journal = {\aap},
         year = 2024,
        month = mar,
       volume = {683},
          eid = {A74},
        pages = {A74},
          doi = {10.1051/0004-6361/202244157},
archivePrefix = {arXiv},
       eprint = {2206.10949},
 primaryClass = {astro-ph.GA},
       adsurl = {https://ui.adsabs.harvard.edu/abs/2024A&A...683A..74G}
}

@ARTICLE{grand2024,
       author = {{Grand}, Robert J.~J. and {Fragkoudi}, Francesca and {G{\'o}mez}, Facundo A. and {Jenkins}, Adrian and {Marinacci}, Federico and {Pakmor}, R{\"u}diger and {Springel}, Volker},
        title = "{Overview and public data release of the augmented Auriga Project: cosmological simulations of dwarf and Milky Way-mass galaxies}",
      journal = {\mnras},
         year = 2024,
        month = aug,
       volume = {532},
       number = {2},
        pages = {1814-1831},
          doi = {10.1093/mnras/stae1598},
archivePrefix = {arXiv},
       eprint = {2401.08750},
 primaryClass = {astro-ph.GA},
       adsurl = {https://ui.adsabs.harvard.edu/abs/2024MNRAS.532.1814G}
}

@ARTICLE{khoperskov2021,
       author = {{Khoperskov}, Sergey and {Haywood}, Misha and {Snaith}, Owain and {Di Matteo}, Paola and {Lehnert}, Matthew and {Vasiliev}, Evgenii and {Naroenkov}, Sergey and {Berczik}, Peter},
        title = "{Bimodality of [{\ensuremath{\alpha}} Fe]-[Fe/H] distributions is a natural outcome of dissipative collapse and disc growth in Milky Way-type galaxies}",
      journal = {\mnras},
         year = 2021,
        month = mar,
       volume = {501},
       number = {4},
        pages = {5176-5196},
          doi = {10.1093/mnras/staa3996},
archivePrefix = {arXiv},
       eprint = {2006.10195},
 primaryClass = {astro-ph.GA},
       adsurl = {https://ui.adsabs.harvard.edu/abs/2021MNRAS.501.5176K}
}

@ARTICLE{clarke2019,
       author = {{Clarke}, Adam J. and {Debattista}, Victor P. and {Nidever}, David L. and {Loebman}, Sarah R. and {Simons}, Raymond C. and {Kassin}, Susan and {Du}, Min and {Ness}, Melissa and {Fisher}, Deanne B. and {Quinn}, Thomas R. and {Wadsley}, James and {Freeman}, Ken C. and {Popescu}, Cristina C.},
        title = "{The imprint of clump formation at high redshift - I. A disc {\ensuremath{\alpha}}-abundance dichotomy}",
      journal = {\mnras},
         year = 2019,
        month = apr,
       volume = {484},
       number = {3},
        pages = {3476-3490},
          doi = {10.1093/mnras/stz104},
archivePrefix = {arXiv},
       eprint = {1901.00931},
 primaryClass = {astro-ph.GA},
       adsurl = {https://ui.adsabs.harvard.edu/abs/2019MNRAS.484.3476C}
}

@ARTICLE{anguiano2020,
       author = {{Anguiano}, Borja and {Majewski}, Steven R. and {Hayes}, Christian R. and {Allende Prieto}, Carlos and {Cheng}, Xinlun and {Bidin}, Christian Moni and {Beaton}, Rachael L. and {Beers}, Timothy C. and {Minniti}, Dante},
        title = "{The Stellar Velocity Distribution Function in the Milky Way Galaxy}",
      journal = {\aj},
         year = 2020,
        month = jul,
       volume = {160},
       number = {1},
          eid = {43},
        pages = {43},
          doi = {10.3847/1538-3881/ab9813},
archivePrefix = {arXiv},
       eprint = {2005.14534},
 primaryClass = {astro-ph.GA},
       adsurl = {https://ui.adsabs.harvard.edu/abs/2020AJ....160...43A}
}

@ARTICLE{khoperskov2024,
       author = {{Khoperskov}, Sergey and {Steinmetz}, Matthias and {Haywood}, Misha and {van de Ven}, Glenn and {Krajnovi{\'c}}, Davor and {Ratcliffe}, Bridget and {Minchev}, Ivan and {Di Matteo}, Paola and {Kacharov}, Nikolay and {Marques}, L{\'e}a and {Valentini}, Marica and {de Jong}, Roelof S.},
        title = "{Rediscovering the Milky Way with an orbit superposition approach and APOGEE data: II. Chrono-chemo-kinematics of the disc}",
      journal = {\aap},
         year = 2025,
        month = aug,
       volume = {700},
          eid = {A89},
        pages = {A89},
          doi = {10.1051/0004-6361/202453305},
archivePrefix = {arXiv},
       eprint = {2411.16866},
 primaryClass = {astro-ph.GA},
       adsurl = {https://ui.adsabs.harvard.edu/abs/2025A&A...700A..89K}
}

@ARTICLE{faria2025,
       author = {{Faria}, Lawrence and {Patton}, David R. and {Courteau}, St{\'e}phane and {Ellison}, Sara and {Brown}, Westley},
        title = "{Interacting galaxies in the IllustrisTNG simulations - VIII. Pericentric star formation rate enhancements are driven both by increased fuelling and efficiency}",
      journal = {\mnras},
         year = 2025,
        month = feb,
       volume = {537},
       number = {2},
        pages = {915-930},
          doi = {10.1093/mnras/staf124},
archivePrefix = {arXiv},
       eprint = {2501.14031},
 primaryClass = {astro-ph.GA},
       adsurl = {https://ui.adsabs.harvard.edu/abs/2025MNRAS.537..915F}
}

@ARTICLE{queiroz2020,
       author = {{Queiroz}, A.~B.~A. and {Anders}, F. and {Chiappini}, C. and {Khalatyan}, A. and {Santiago}, B.~X. and {Steinmetz}, M. and {Valentini}, M. and {Miglio}, A. and {Bossini}, D. and {Barbuy}, B. and {Minchev}, I. and {Minniti}, D. and {Garc{\'\i}a Hern{\'a}ndez}, D.~A. and {Schultheis}, M. and {Beaton}, R.~L. and {Beers}, T.~C. and {Bizyaev}, D. and {Brownstein}, J.~R. and {Cunha}, K. and {Fern{\'a}ndez-Trincado}, J.~G. and {Frinchaboy}, P.~M. and {Lane}, R.~R. and {Majewski}, S.~R. and {Nataf}, D. and {Nitschelm}, C. and {Pan}, K. and {Roman-Lopes}, A. and {Sobeck}, J.~S. and {Stringfellow}, G. and {Zamora}, O.},
        title = "{From the bulge to the outer disc: StarHorse stellar parameters, distances, and extinctions for stars in APOGEE DR16 and other spectroscopic surveys}",
      journal = {\aap},
         year = 2020,
        month = jun,
       volume = {638},
          eid = {A76},
        pages = {A76},
          doi = {10.1051/0004-6361/201937364},
archivePrefix = {arXiv},
       eprint = {1912.09778},
 primaryClass = {astro-ph.GA},
       adsurl = {https://ui.adsabs.harvard.edu/abs/2020A&A...638A..76Q}
}

@ARTICLE{bournaud2009,
       author = {{Bournaud}, Fr{\'e}d{\'e}ric and {Elmegreen}, Bruce G. and {Martig}, Marie},
        title = "{The Thick Disks of Spiral Galaxies as Relics from Gas-rich, Turbulent, Clumpy Disks at High Redshift}",
      journal = {\apjl},
         year = 2009,
        month = dec,
       volume = {707},
       number = {1},
        pages = {L1-L5},
          doi = {10.1088/0004-637X/707/1/L1},
archivePrefix = {arXiv},
       eprint = {0910.3677},
 primaryClass = {astro-ph.CO},
       adsurl = {https://ui.adsabs.harvard.edu/abs/2009ApJ...707L...1B}
}

@ARTICLE{bensby2005,
       author = {{Bensby}, T. and {Feltzing}, S. and {Lundstr{\"o}m}, I. and {Ilyin}, I.},
        title = "{{\ensuremath{\alpha}}-, r-, and s-process element trends in the Galactic thin and thick disks}",
      journal = {\aap},
         year = 2005,
        month = apr,
       volume = {433},
       number = {1},
        pages = {185-203},
          doi = {10.1051/0004-6361:20040332},
archivePrefix = {arXiv},
       eprint = {astro-ph/0412132},
 primaryClass = {astro-ph},
       adsurl = {https://ui.adsabs.harvard.edu/abs/2005A&A...433..185B}
}

@ARTICLE{haywood2013,
       author = {{Haywood}, Misha and {Di Matteo}, Paola and {Lehnert}, Matthew D. and {Katz}, David and {G{\'o}mez}, Ana},
        title = "{The age structure of stellar populations in the solar vicinity. Clues of a two-phase formation history of the Milky Way disk}",
      journal = {\aap},
         year = 2013,
        month = dec,
       volume = {560},
          eid = {A109},
        pages = {A109},
          doi = {10.1051/0004-6361/201321397},
archivePrefix = {arXiv},
       eprint = {1305.4663},
 primaryClass = {astro-ph.GA},
       adsurl = {https://ui.adsabs.harvard.edu/abs/2013A&A...560A.109H}
}

@ARTICLE{hayden2017,
       author = {{Hayden}, M.~R. and {Recio-Blanco}, A. and {de Laverny}, P. and {Mikolaitis}, S. and {Worley}, C.~C.},
        title = "{The AMBRE project: The thick thin disk and thin thick disk of the Milky Way}",
      journal = {\aap},
         year = 2017,
        month = dec,
       volume = {608},
          eid = {L1},
        pages = {L1},
          doi = {10.1051/0004-6361/201731494},
archivePrefix = {arXiv},
       eprint = {1712.02358},
 primaryClass = {astro-ph.GA},
       adsurl = {https://ui.adsabs.harvard.edu/abs/2017A&A...608L...1H}
}

@ARTICLE{xiang2022,
       author = {{Xiang}, Maosheng and {Rix}, Hans-Walter},
        title = "{A time-resolved picture of our Milky Way's early formation history}",
      journal = {\nat},
         year = 2022,
        month = mar,
       volume = {603},
       number = {7902},
        pages = {599-603},
          doi = {10.1038/s41586-022-04496-5},
archivePrefix = {arXiv},
       eprint = {2203.12110},
 primaryClass = {astro-ph.GA},
       adsurl = {https://ui.adsabs.harvard.edu/abs/2022Natur.603..599X}
}

@ARTICLE{belokurov2018,
       author = {{Belokurov}, V. and {Erkal}, D. and {Evans}, N.~W. and {Koposov}, S.~E. and {Deason}, A.~J.},
        title = "{Co-formation of the disc and the stellar halo}",
      journal = {\mnras},
         year = 2018,
        month = jul,
       volume = {478},
       number = {1},
        pages = {611-619},
          doi = {10.1093/mnras/sty982},
archivePrefix = {arXiv},
       eprint = {1802.03414},
 primaryClass = {astro-ph.GA},
       adsurl = {https://ui.adsabs.harvard.edu/abs/2018MNRAS.478..611B}
}

@ARTICLE{helmi2018,
       author = {{Helmi}, Amina and {Babusiaux}, Carine and {Koppelman}, Helmer H. and {Massari}, Davide and {Veljanoski}, Jovan and {Brown}, Anthony G.~A.},
        title = "{The merger that led to the formation of the Milky Way's inner stellar halo and thick disk}",
      journal = {\nat},
         year = 2018,
        month = oct,
       volume = {563},
       number = {7729},
        pages = {85-88},
          doi = {10.1038/s41586-018-0625-x},
archivePrefix = {arXiv},
       eprint = {1806.06038},
 primaryClass = {astro-ph.GA},
       adsurl = {https://ui.adsabs.harvard.edu/abs/2018Natur.563...85H}
}

@ARTICLE{chiappini2001,
       author = {{Chiappini}, Cristina and {Matteucci}, Francesca and {Romano}, Donatella},
        title = "{Abundance Gradients and the Formation of the Milky Way}",
      journal = {\apj},
         year = 2001,
        month = jun,
       volume = {554},
       number = {2},
        pages = {1044-1058},
          doi = {10.1086/321427},
archivePrefix = {arXiv},
       eprint = {astro-ph/0102134},
 primaryClass = {astro-ph},
       adsurl = {https://ui.adsabs.harvard.edu/abs/2001ApJ...554.1044C}
}

@ARTICLE{spitoni2019,
       author = {{Spitoni}, E. and {Silva Aguirre}, V. and {Matteucci}, F. and {Calura}, F. and {Grisoni}, V.},
        title = "{Galactic Archaeology with asteroseismic ages: Evidence for delayed gas infall in the formation of the Milky Way disc}",
      journal = {\aap},
         year = 2019,
        month = mar,
       volume = {623},
          eid = {A60},
        pages = {A60},
          doi = {10.1051/0004-6361/201834188},
archivePrefix = {arXiv},
       eprint = {1809.00914},
 primaryClass = {astro-ph.GA},
       adsurl = {https://ui.adsabs.harvard.edu/abs/2019A&A...623A..60S}
}

@ARTICLE{spitoni2023,
       author = {{Spitoni}, E. and {Recio-Blanco}, A. and {de Laverny}, P. and {Palicio}, P.~A. and {Kordopatis}, G. and {Schultheis}, M. and {Contursi}, G. and {Poggio}, E. and {Romano}, D. and {Matteucci}, F.},
        title = "{Beyond the two-infall model. I. Indications for a recent gas infall with Gaia DR3 chemical abundances}",
      journal = {\aap},
         year = 2023,
        month = feb,
       volume = {670},
          eid = {A109},
        pages = {A109},
          doi = {10.1051/0004-6361/202244349},
archivePrefix = {arXiv},
       eprint = {2206.12436},
 primaryClass = {astro-ph.GA},
       adsurl = {https://ui.adsabs.harvard.edu/abs/2023A&A...670A.109S}
}

@ARTICLE{chandra2024,
       author = {{Chandra}, Vedant and {Semenov}, Vadim A. and {Rix}, Hans-Walter and {Conroy}, Charlie and {Bonaca}, Ana and {Naidu}, Rohan P. and {Andrae}, Ren{\'e} and {Li}, Jiadong and {Hernquist}, Lars},
        title = "{The Three-phase Evolution of the Milky Way}",
      journal = {\apj},
         year = 2024,
        month = sep,
       volume = {972},
       number = {1},
          eid = {112},
        pages = {112},
          doi = {10.3847/1538-4357/ad5b60},
archivePrefix = {arXiv},
       eprint = {2310.13050},
 primaryClass = {astro-ph.GA},
       adsurl = {https://ui.adsabs.harvard.edu/abs/2024ApJ...972..112C}
}

@ARTICLE{belokurov2022,
       author = {{Belokurov}, Vasily and {Kravtsov}, Andrey},
        title = "{From dawn till disc: Milky Way's turbulent youth revealed by the APOGEE+Gaia data}",
      journal = {\mnras},
         year = 2022,
        month = jul,
       volume = {514},
       number = {1},
        pages = {689-714},
          doi = {10.1093/mnras/stac1267},
archivePrefix = {arXiv},
       eprint = {2203.04980},
 primaryClass = {astro-ph.GA},
       adsurl = {https://ui.adsabs.harvard.edu/abs/2022MNRAS.514..689B}
}

@ARTICLE{mackereth2019,
       author = {{Mackereth}, J. Ted and {Schiavon}, Ricardo P. and {Pfeffer}, Joel and {Hayes}, Christian R. and {Bovy}, Jo and {Anguiano}, Borja and {Allende Prieto}, Carlos and {Hasselquist}, Sten and {Holtzman}, Jon and {Johnson}, Jennifer A. and {Majewski}, Steven R. and {O'Connell}, Robert and {Shetrone}, Matthew and {Tissera}, Patricia B. and {Fern{\'a}ndez-Trincado}, J.~G.},
        title = "{The origin of accreted stellar halo populations in the Milky Way using APOGEE, Gaia, and the EAGLE simulations}",
      journal = {\mnras},
         year = 2019,
        month = jan,
       volume = {482},
       number = {3},
        pages = {3426-3442},
          doi = {10.1093/mnras/sty2955},
archivePrefix = {arXiv},
       eprint = {1808.00968},
 primaryClass = {astro-ph.GA},
       adsurl = {https://ui.adsabs.harvard.edu/abs/2019MNRAS.482.3426M}
}

@ARTICLE{Kruijssen2019,
       author = {{Kruijssen}, J.~M. Diederik and {Pfeffer}, Joel L. and {Reina-Campos}, Marta and {Crain}, Robert A. and {Bastian}, Nate},
        title = "{The formation and assembly history of the Milky Way revealed by its globular cluster population}",
      journal = {\mnras},
         year = 2019,
        month = jul,
       volume = {486},
       number = {3},
        pages = {3180-3202},
          doi = {10.1093/mnras/sty1609},
archivePrefix = {arXiv},
       eprint = {1806.05680},
 primaryClass = {astro-ph.GA},
       adsurl = {https://ui.adsabs.harvard.edu/abs/2019MNRAS.486.3180K}
}

@ARTICLE{massari2019,
       author = {{Massari}, D. and {Koppelman}, H.~H. and {Helmi}, A.},
        title = "{Origin of the system of globular clusters in the Milky Way}",
      journal = {\aap},
         year = 2019,
        month = oct,
       volume = {630},
          eid = {L4},
        pages = {L4},
          doi = {10.1051/0004-6361/201936135},
archivePrefix = {arXiv},
       eprint = {1906.08271},
 primaryClass = {astro-ph.GA},
       adsurl = {https://ui.adsabs.harvard.edu/abs/2019A&A...630L...4M}
}

@ARTICLE{li2025,
       author = {{Li}, Fei and {Rahman}, Mubdi and {Murray}, Norman and {Kere{\v{s}}}, Du{\v{s}}an and {Wetzel}, Andrew and {Faucher-Gigu{\`e}re}, Claude-Andr{\'e} and {Hopkins}, Philip F. and {Moreno}, Jorge},
        title = "{The Effect of Galaxy Interactions on Starbursts in Milky Way-mass Galaxies in FIRE Simulations}",
      journal = {\apj},
         year = 2025,
        month = jan,
       volume = {979},
       number = {1},
          eid = {7},
        pages = {7},
          doi = {10.3847/1538-4357/ad94ef},
archivePrefix = {arXiv},
       eprint = {2411.10373},
 primaryClass = {astro-ph.GA},
       adsurl = {https://ui.adsabs.harvard.edu/abs/2025ApJ...979....7L}
}

@ARTICLE{palla2020,
       author = {{Palla}, M. and {Matteucci}, F. and {Spitoni}, E. and {Vincenzo}, F. and {Grisoni}, V.},
        title = "{Chemical evolution of the Milky Way: constraints on the formation of the thick and thin discs}",
      journal = {\mnras},
         year = 2020,
        month = oct,
       volume = {498},
       number = {2},
        pages = {1710-1725},
          doi = {10.1093/mnras/staa2437},
archivePrefix = {arXiv},
       eprint = {2008.07484},
 primaryClass = {astro-ph.GA},
       adsurl = {https://ui.adsabs.harvard.edu/abs/2020MNRAS.498.1710P}
}

@ARTICLE{ciuca2024,
       author = {{Ciuc{\u{a}}}, Ioana and {Kawata}, Daisuke and {Ting}, Yuan-Sen and {Grand}, Robert J.~J. and {Miglio}, Andrea and {Hayden}, Michael and {Baba}, Junichi and {Fragkoudi}, Francesca and {Monty}, Stephanie and {Buder}, Sven and {Freeman}, Ken},
        title = "{Chasing the impact of the Gaia-Sausage-Enceladus merger on the formation of the Milky Way thick disc}",
      journal = {\mnras},
         year = 2024,
        month = feb,
       volume = {528},
       number = {1},
        pages = {L122-L126},
          doi = {10.1093/mnrasl/slad033},
archivePrefix = {arXiv},
       eprint = {2211.01006},
 primaryClass = {astro-ph.GA},
       adsurl = {https://ui.adsabs.harvard.edu/abs/2024MNRAS.528L.122C}
}

@ARTICLE{belokurov2020,
       author = {{Belokurov}, Vasily and {Sanders}, Jason L. and {Fattahi}, Azadeh and {Smith}, Martin C. and {Deason}, Alis J. and {Evans}, N. Wyn and {Grand}, Robert J.~J.},
        title = "{The biggest splash}",
      journal = {\mnras},
         year = 2020,
        month = may,
       volume = {494},
       number = {3},
        pages = {3880-3898},
          doi = {10.1093/mnras/staa876},
archivePrefix = {arXiv},
       eprint = {1909.04679},
 primaryClass = {astro-ph.GA},
       adsurl = {https://ui.adsabs.harvard.edu/abs/2020MNRAS.494.3880B}
}

@ARTICLE{fuhrmann1998,
       author = {{Fuhrmann}, Klaus},
        title = "{Nearby stars of the Galactic disk and halo}",
      journal = {\aap},
         year = 1998,
        month = oct,
       volume = {338},
        pages = {161-183},
       adsurl = {https://ui.adsabs.harvard.edu/abs/1998A&A...338..161F}
}

@ARTICLE{frankel2018,
       author = {{Frankel}, Neige and {Rix}, Hans-Walter and {Ting}, Yuan-Sen and {Ness}, Melissa and {Hogg}, David W.},
        title = "{Measuring Radial Orbit Migration in the Galactic Disk}",
      journal = {\apj},
         year = 2018,
        month = oct,
       volume = {865},
       number = {2},
          eid = {96},
        pages = {96},
          doi = {10.3847/1538-4357/aadba5},
archivePrefix = {arXiv},
       eprint = {1805.09198},
 primaryClass = {astro-ph.GA},
       adsurl = {https://ui.adsabs.harvard.edu/abs/2018ApJ...865...96F}
}

@ARTICLE{frankel2020,
       author = {{Frankel}, Neige and {Sanders}, Jason and {Ting}, Yuan-Sen and {Rix}, Hans-Walter},
        title = "{Keeping It Cool: Much Orbit Migration, yet Little Heating, in the Galactic Disk}",
      journal = {\apj},
         year = 2020,
        month = jun,
       volume = {896},
       number = {1},
          eid = {15},
        pages = {15},
          doi = {10.3847/1538-4357/ab910c},
archivePrefix = {arXiv},
       eprint = {2002.04622},
 primaryClass = {astro-ph.GA},
       adsurl = {https://ui.adsabs.harvard.edu/abs/2020ApJ...896...15F}
}

@INPROCEEDINGS{patil2024,
       author = {{Patil}, Aarya and {Bovy}, Jo},
        title = "{Decoding the age-chemical structure of the Milky Way disk: An application of Copulas and Elicitable Maps}",
    booktitle = {American Astronomical Society Meeting Abstracts},
         year = 2024,
       series = {American Astronomical Society Meeting Abstracts},
       volume = {243},
        month = feb,
          eid = {129.03},
        pages = {129.03},
       adsurl = {https://ui.adsabs.harvard.edu/abs/2024AAS...24312903P}
}

@ARTICLE{genel2013,
       author = {{Genel}, Shy and {Vogelsberger}, Mark and {Nelson}, Dylan and {Sijacki}, Debora and {Springel}, Volker and {Hernquist}, Lars},
        title = "{Following the flow: tracer particles in astrophysical fluid simulations}",
      journal = {\mnras},
         year = 2013,
        month = oct,
       volume = {435},
       number = {2},
        pages = {1426-1442},
          doi = {10.1093/mnras/stt1383},
archivePrefix = {arXiv},
       eprint = {1305.2195},
 primaryClass = {astro-ph.IM},
       adsurl = {https://ui.adsabs.harvard.edu/abs/2013MNRAS.435.1426G}
}

@ARTICLE{nidever2014,
       author = {{Nidever}, David L. and {Bovy}, Jo and {Bird}, Jonathan C. and {Andrews}, Brett H. and {Hayden}, Michael and {Holtzman}, Jon and {Majewski}, Steven R. and {Smith}, Verne and {Robin}, Annie C. and {Garc{\'\i}a P{\'e}rez}, Ana E. and {Cunha}, Katia and {Allende Prieto}, Carlos and {Zasowski}, Gail and {Schiavon}, Ricardo P. and {Johnson}, Jennifer A. and {Weinberg}, David H. and {Feuillet}, Diane and {Schneider}, Donald P. and {Shetrone}, Matthew and {Sobeck}, Jennifer and {Garc{\'\i}a-Hern{\'a}ndez}, D.~A. and {Zamora}, O. and {Rix}, Hans-Walter and {Beers}, Timothy C. and {Wilson}, John C. and {O'Connell}, Robert W. and {Minchev}, Ivan and {Chiappini}, Cristina and {Anders}, Friedrich and {Bizyaev}, Dmitry and {Brewington}, Howard and {Ebelke}, Garrett and {Frinchaboy}, Peter M. and {Ge}, Jian and {Kinemuchi}, Karen and {Malanushenko}, Elena and {Malanushenko}, Viktor and {Marchante}, Moses and {M{\'e}sz{\'a}ros}, Szabolcs and {Oravetz}, Daniel and {Pan}, Kaike and {Simmons}, Audrey and {Skrutskie}, Michael F.},
        title = "{Tracing Chemical Evolution over the Extent of the Milky Way's Disk with APOGEE Red Clump Stars}",
      journal = {\apj},
         year = 2014,
        month = nov,
       volume = {796},
       number = {1},
          eid = {38},
        pages = {38},
          doi = {10.1088/0004-637X/796/1/38},
archivePrefix = {arXiv},
       eprint = {1409.3566},
 primaryClass = {astro-ph.GA},
       adsurl = {https://ui.adsabs.harvard.edu/abs/2014ApJ...796...38N}
}

@ARTICLE{minchev2014,
       author = {{Minchev}, I. and {Chiappini}, C. and {Martig}, M.},
        title = "{Chemodynamical evolution of the Milky Way disk. II. Variations with Galactic radius and height above the disk plane}",
      journal = {\aap},
         year = 2014,
        month = dec,
       volume = {572},
          eid = {A92},
        pages = {A92},
          doi = {10.1051/0004-6361/201423487},
archivePrefix = {arXiv},
       eprint = {1401.5796},
 primaryClass = {astro-ph.GA},
       adsurl = {https://ui.adsabs.harvard.edu/abs/2014A&A...572A..92M}
}

@ARTICLE{sellwood2002,
       author = {{Sellwood}, J.~A. and {Binney}, J.~J.},
        title = "{Radial mixing in galactic discs}",
      journal = {\mnras},
         year = 2002,
        month = nov,
       volume = {336},
       number = {3},
        pages = {785-796},
          doi = {10.1046/j.1365-8711.2002.05806.x},
archivePrefix = {arXiv},
       eprint = {astro-ph/0203510},
 primaryClass = {astro-ph},
       adsurl = {https://ui.adsabs.harvard.edu/abs/2002MNRAS.336..785S}
}

@ARTICLE{rovskar2008,
       author = {{Ro{\v{s}}kar}, Rok and {Debattista}, Victor P. and {Quinn}, Thomas R. and {Stinson}, Gregory S. and {Wadsley}, James},
        title = "{Riding the Spiral Waves: Implications of Stellar Migration for the Properties of Galactic Disks}",
      journal = {\apjl},
         year = 2008,
        month = sep,
       volume = {684},
       number = {2},
        pages = {L79},
          doi = {10.1086/592231},
archivePrefix = {arXiv},
       eprint = {0808.0206},
 primaryClass = {astro-ph},
       adsurl = {https://ui.adsabs.harvard.edu/abs/2008ApJ...684L..79R}
}

@ARTICLE{minchev2010,
       author = {{Minchev}, I. and {Famaey}, B.},
        title = "{A New Mechanism for Radial Migration in Galactic Disks: Spiral-Bar Resonance Overlap}",
      journal = {\apj},
         year = 2010,
        month = oct,
       volume = {722},
       number = {1},
        pages = {112-121},
          doi = {10.1088/0004-637X/722/1/112},
archivePrefix = {arXiv},
       eprint = {0911.1794},
 primaryClass = {astro-ph.GA},
       adsurl = {https://ui.adsabs.harvard.edu/abs/2010ApJ...722..112M}
}

@ARTICLE{minchev2011,
       author = {{Minchev}, I. and {Famaey}, B. and {Combes}, F. and {Di Matteo}, P. and {Mouhcine}, M. and {Wozniak}, H.},
        title = "{Radial migration in galactic disks caused by resonance overlap of multiple patterns: Self-consistent simulations}",
      journal = {\aap},
         year = 2011,
        month = mar,
       volume = {527},
          eid = {A147},
        pages = {A147},
          doi = {10.1051/0004-6361/201015139},
archivePrefix = {arXiv},
       eprint = {1006.0484},
 primaryClass = {astro-ph.GA},
       adsurl = {https://ui.adsabs.harvard.edu/abs/2011A&A...527A.147M}
}

@ARTICLE{carr2022,
       author = {{Carr}, Christopher and {Johnston}, Kathryn V. and {Laporte}, Chervin F.~P. and {Ness}, Melissa K.},
        title = "{Migration and heating in the galactic disc from encounters between Sagittarius and the Milky Way}",
      journal = {\mnras},
         year = 2022,
        month = nov,
       volume = {516},
       number = {4},
        pages = {5067-5083},
          doi = {10.1093/mnras/stac2403},
archivePrefix = {arXiv},
       eprint = {2201.04133},
 primaryClass = {astro-ph.GA},
       adsurl = {https://ui.adsabs.harvard.edu/abs/2022MNRAS.516.5067C}
}

@ARTICLE{minchev2014b,
       author = {{Minchev}, I. and {Chiappini}, C. and {Martig}, M. and {Steinmetz}, M. and {de Jong}, R.~S. and {Boeche}, C. and {Scannapieco}, C. and {Zwitter}, T. and {Wyse}, R.~F.~G. and {Binney}, J.~J. and {Bland-Hawthorn}, J. and {Bienaym{\'e}}, O. and {Famaey}, B. and {Freeman}, K.~C. and {Gibson}, B.~K. and {Grebel}, E.~K. and {Gilmore}, G. and {Helmi}, A. and {Kordopatis}, G. and {Lee}, Y.~S. and {Munari}, U. and {Navarro}, J.~F. and {Parker}, Q.~A. and {Quillen}, A.~C. and {Reid}, W.~A. and {Siebert}, A. and {Siviero}, A. and {Seabroke}, G. and {Watson}, F. and {Williams}, M.},
        title = "{A New Stellar Chemo-Kinematic Relation Reveals the Merger History of the Milky Way Disk}",
      journal = {\apjl},
         year = 2014,
        month = jan,
       volume = {781},
       number = {1},
          eid = {L20},
        pages = {L20},
          doi = {10.1088/2041-8205/781/1/L20},
archivePrefix = {arXiv},
       eprint = {1310.5145},
 primaryClass = {astro-ph.GA},
       adsurl = {https://ui.adsabs.harvard.edu/abs/2014ApJ...781L..20M}
}

@ARTICLE{schonrich2009a,
       author = {{Sch{\"o}nrich}, Ralph and {Binney}, James},
        title = "{Chemical evolution with radial mixing}",
      journal = {\mnras},
         year = 2009,
        month = jun,
       volume = {396},
       number = {1},
        pages = {203-222},
          doi = {10.1111/j.1365-2966.2009.14750.x},
archivePrefix = {arXiv},
       eprint = {0809.3006},
 primaryClass = {astro-ph},
       adsurl = {https://ui.adsabs.harvard.edu/abs/2009MNRAS.396..203S}
}

@ARTICLE{schonrich2009,
       author = {{Sch{\"o}nrich}, Ralph and {Binney}, James},
        title = "{Origin and structure of the Galactic disc(s)}",
      journal = {\mnras},
         year = 2009,
        month = nov,
       volume = {399},
       number = {3},
        pages = {1145-1156},
          doi = {10.1111/j.1365-2966.2009.15365.x},
archivePrefix = {arXiv},
       eprint = {0907.1899},
 primaryClass = {astro-ph.GA},
       adsurl = {https://ui.adsabs.harvard.edu/abs/2009MNRAS.399.1145S}
}

@ARTICLE{snaith2015,
       author = {{Snaith}, O. and {Haywood}, M. and {Di Matteo}, P. and {Lehnert}, M.~D. and {Combes}, F. and {Katz}, D. and {G{\'o}mez}, A.},
        title = "{Reconstructing the star formation history of the Milky Way disc(s) from chemical abundances}",
      journal = {\aap},
         year = 2015,
        month = jun,
       volume = {578},
          eid = {A87},
        pages = {A87},
          doi = {10.1051/0004-6361/201424281},
archivePrefix = {arXiv},
       eprint = {1410.3829},
 primaryClass = {astro-ph.GA},
       adsurl = {https://ui.adsabs.harvard.edu/abs/2015A&A...578A..87S}
}

@ARTICLE{soderblom2010,
       author = {{Soderblom}, David R.},
        title = "{The Ages of Stars}",
      journal = {\araa},
         year = 2010,
        month = sep,
       volume = {48},
        pages = {581-629},
          doi = {10.1146/annurev-astro-081309-130806},
archivePrefix = {arXiv},
       eprint = {1003.6074},
 primaryClass = {astro-ph.SR},
       adsurl = {https://ui.adsabs.harvard.edu/abs/2010ARA&A..48..581S}
}

@ARTICLE{fuhrmann2011,
       author = {{Fuhrmann}, Klaus},
        title = "{Nearby stars of the Galactic disc and halo - V}",
      journal = {\mnras},
         year = 2011,
        month = jul,
       volume = {414},
       number = {4},
        pages = {2893-2922},
          doi = {10.1111/j.1365-2966.2011.18476.x},
       adsurl = {https://ui.adsabs.harvard.edu/abs/2011MNRAS.414.2893F}
}

@ARTICLE{mackereth2017,
       author = {{Mackereth}, J. Ted and {Bovy}, Jo and {Schiavon}, Ricardo P. and {Zasowski}, Gail and {Cunha}, Katia and {Frinchaboy}, Peter M. and {Garc{\'\i}a Perez}, Ana E. and {Hayden}, Michael R. and {Holtzman}, Jon and {Majewski}, Steven R. and {M{\'e}sz{\'a}ros}, Szabolcs and {Nidever}, David L. and {Pinsonneault}, Marc and {Shetrone}, Matthew D.},
        title = "{The age-metallicity structure of the Milky Way disc using APOGEE}",
      journal = {\mnras},
         year = 2017,
        month = nov,
       volume = {471},
       number = {3},
        pages = {3057-3078},
          doi = {10.1093/mnras/stx1774},
archivePrefix = {arXiv},
       eprint = {1706.00018},
 primaryClass = {astro-ph.GA},
       adsurl = {https://ui.adsabs.harvard.edu/abs/2017MNRAS.471.3057M}
}

@ARTICLE{moster2013,
       author = {{Moster}, Benjamin P. and {Naab}, Thorsten and {White}, Simon D.~M.},
        title = "{Galactic star formation and accretion histories from matching galaxies to dark matter haloes}",
      journal = {\mnras},
         year = 2013,
        month = feb,
       volume = {428},
       number = {4},
        pages = {3121-3138},
          doi = {10.1093/mnras/sts261},
archivePrefix = {arXiv},
       eprint = {1205.5807},
 primaryClass = {astro-ph.CO},
       adsurl = {https://ui.adsabs.harvard.edu/abs/2013MNRAS.428.3121M}
}

@ARTICLE{kruijssen2020,
       author = {{Kruijssen}, J.~M. Diederik and {Pfeffer}, Joel L. and {Chevance}, M{\'e}lanie and {Bonaca}, Ana and {Trujillo-Gomez}, Sebastian and {Bastian}, Nate and {Reina-Campos}, Marta and {Crain}, Robert A. and {Hughes}, Meghan E.},
        title = "{Kraken reveals itself - the merger history of the Milky Way reconstructed with the E-MOSAICS simulations}",
      journal = {\mnras},
         year = 2020,
        month = oct,
       volume = {498},
       number = {2},
        pages = {2472-2491},
          doi = {10.1093/mnras/staa2452},
archivePrefix = {arXiv},
       eprint = {2003.01119},
 primaryClass = {astro-ph.GA},
       adsurl = {https://ui.adsabs.harvard.edu/abs/2020MNRAS.498.2472K}
}

@ARTICLE{vincezo2019,
       author = {{Vincenzo}, Fiorenzo and {Spitoni}, Emanuele and {Calura}, Francesco and {Matteucci}, Francesca and {Silva Aguirre}, Victor and {Miglio}, Andrea and {Cescutti}, Gabriele},
        title = "{The Fall of a Giant. Chemical evolution of Enceladus, alias the Gaia Sausage}",
      journal = {\mnras},
         year = 2019,
        month = jul,
       volume = {487},
       number = {1},
        pages = {L47-L52},
          doi = {10.1093/mnrasl/slz070},
archivePrefix = {arXiv},
       eprint = {1903.03465},
 primaryClass = {astro-ph.GA},
       adsurl = {https://ui.adsabs.harvard.edu/abs/2019MNRAS.487L..47V}
}

@ARTICLE{lane2023,
       author = {{Lane}, James M.~M. and {Bovy}, Jo and {Mackereth}, J. Ted},
        title = "{The stellar mass of the Gaia-Sausage/Enceladus accretion remnant}",
      journal = {\mnras},
         year = 2023,
        month = nov,
       volume = {526},
       number = {1},
        pages = {1209-1234},
          doi = {10.1093/mnras/stad2834},
archivePrefix = {arXiv},
       eprint = {2306.03084},
 primaryClass = {astro-ph.GA},
       adsurl = {https://ui.adsabs.harvard.edu/abs/2023MNRAS.526.1209L}
}

@ARTICLE{spitoni2024,
       author = {{Spitoni}, E. and {Matteucci}, F. and {Gratton}, R. and {Ratcliffe}, B. and {Minchev}, I. and {Cescutti}, G.},
        title = "{(Re)mind the gap: A hiatus in star formation history unveiled by APOGEE DR17}",
      journal = {\aap},
         year = 2024,
        month = oct,
       volume = {690},
          eid = {A208},
        pages = {A208},
          doi = {10.1051/0004-6361/202450754},
archivePrefix = {arXiv},
       eprint = {2405.11025},
 primaryClass = {astro-ph.GA},
       adsurl = {https://ui.adsabs.harvard.edu/abs/2024A&A...690A.208S}
}

@ARTICLE{johnson2021,
       author = {{Johnson}, James W. and {Weinberg}, David H. and {Vincenzo}, Fiorenzo and {Bird}, Jonathan C. and {Loebman}, Sarah R. and {Brooks}, Alyson M. and {Quinn}, Thomas R. and {Christensen}, Charlotte R. and {Griffith}, Emily J.},
        title = "{Stellar migration and chemical enrichment in the milky way disc: a hybrid model}",
      journal = {\mnras},
         year = 2021,
        month = dec,
       volume = {508},
       number = {3},
        pages = {4484-4511},
          doi = {10.1093/mnras/stab2718},
archivePrefix = {arXiv},
       eprint = {2103.09838},
 primaryClass = {astro-ph.GA},
       adsurl = {https://ui.adsabs.harvard.edu/abs/2021MNRAS.508.4484J}
}

@ARTICLE{sharma2021,
       author = {{Sharma}, Sanjib and {Hayden}, Michael R. and {Bland-Hawthorn}, Joss},
        title = "{Chemical enrichment and radial migration in the Galactic disc - the origin of the [{\ensuremath{\alpha}}Fe] double sequence}",
      journal = {\mnras},
         year = 2021,
        month = nov,
       volume = {507},
       number = {4},
        pages = {5882-5901},
          doi = {10.1093/mnras/stab2015},
archivePrefix = {arXiv},
       eprint = {2005.03646},
 primaryClass = {astro-ph.GA},
       adsurl = {https://ui.adsabs.harvard.edu/abs/2021MNRAS.507.5882S}
}

@ARTICLE{chen2023,
       author = {{Chen}, Boquan and {Hayden}, Michael R. and {Sharma}, Sanjib and {Bland-Hawthorn}, Joss and {Kobayashi}, Chiaki and {Karakas}, Amanda I.},
        title = "{Chemical evolution with radial mixing redux: a detailed model for formation and evolution of the Milky Way}",
      journal = {\mnras},
         year = 2023,
        month = aug,
       volume = {523},
       number = {3},
        pages = {3791-3811},
          doi = {10.1093/mnras/stad1568},
archivePrefix = {arXiv},
       eprint = {2204.11413},
 primaryClass = {astro-ph.GA},
       adsurl = {https://ui.adsabs.harvard.edu/abs/2023MNRAS.523.3791C}
}

@ARTICLE{noguchi2018,
       author = {{Noguchi}, Masafumi},
        title = "{The formation of solar-neighbourhood stars in two generations separated by 5 billion years}",
      journal = {\nat},
         year = 2018,
        month = jul,
       volume = {559},
       number = {7715},
        pages = {585-588},
          doi = {10.1038/s41586-018-0329-2},
archivePrefix = {arXiv},
       eprint = {1809.02299},
 primaryClass = {astro-ph.GA},
       adsurl = {https://ui.adsabs.harvard.edu/abs/2018Natur.559..585N}
}

@ARTICLE{house2011,
       author = {{House}, E.~L. and {Brook}, C.~B. and {Gibson}, B.~K. and {S{\'a}nchez-Bl{\'a}zquez}, P. and {Courty}, S. and {Few}, C.~G. and {Governato}, F. and {Kawata}, D. and {Ro{\v{s}}kar}, R. and {Steinmetz}, M. and {Stinson}, G.~S. and {Teyssier}, R.},
        title = "{Disc heating: comparing the Milky Way with cosmological simulations}",
      journal = {\mnras},
         year = 2011,
        month = aug,
       volume = {415},
       number = {3},
        pages = {2652-2664},
          doi = {10.1111/j.1365-2966.2011.18891.x},
archivePrefix = {arXiv},
       eprint = {1104.2037},
 primaryClass = {astro-ph.CO},
       adsurl = {https://ui.adsabs.harvard.edu/abs/2011MNRAS.415.2652H}
}

@ARTICLE{sanderson2020,
       author = {{Sanderson}, Robyn E. and {Wetzel}, Andrew and {Loebman}, Sarah and {Sharma}, Sanjib and {Hopkins}, Philip F. and {Garrison-Kimmel}, Shea and {Faucher-Gigu{\`e}re}, Claude-Andr{\'e} and {Kere{\v{s}}}, Du{\v{s}}an and {Quataert}, Eliot},
        title = "{Synthetic Gaia Surveys from the FIRE Cosmological Simulations of Milky Way-mass Galaxies}",
      journal = {\apjs},
         year = 2020,
        month = jan,
       volume = {246},
       number = {1},
          eid = {6},
        pages = {6},
          doi = {10.3847/1538-4365/ab5b9d},
archivePrefix = {arXiv},
       eprint = {1806.10564},
 primaryClass = {astro-ph.GA},
       adsurl = {https://ui.adsabs.harvard.edu/abs/2020ApJS..246....6S}
}

@ARTICLE{haywood2018,
       author = {{Haywood}, Misha and {Di Matteo}, Paola and {Lehnert}, Matthew and {Snaith}, Owain and {Fragkoudi}, Francesca and {Khoperskov}, Sergey},
        title = "{Phylogeny of the Milky Way's inner disk and bulge populations: Implications for gas accretion, (the lack of) inside-out thick disk formation, and quenching}",
      journal = {\aap},
         year = 2018,
        month = oct,
       volume = {618},
          eid = {A78},
        pages = {A78},
          doi = {10.1051/0004-6361/201731363},
archivePrefix = {arXiv},
       eprint = {1802.09887},
 primaryClass = {astro-ph.GA},
       adsurl = {https://ui.adsabs.harvard.edu/abs/2018A&A...618A..78H}
}

@ARTICLE{ruizlara2020,
       author = {{Ruiz-Lara}, Tom{\'a}s and {Gallart}, Carme and {Bernard}, Edouard J. and {Cassisi}, Santi},
        title = "{The recurrent impact of the Sagittarius dwarf on the star formation history of the Milky Way}",
      journal = {Nature Astronomy},
         year = 2020,
        month = may,
       volume = {4},
        pages = {965-973},
          doi = {10.1038/s41550-020-1097-0},
archivePrefix = {arXiv},
       eprint = {2003.12577},
 primaryClass = {astro-ph.GA},
       adsurl = {https://ui.adsabs.harvard.edu/abs/2020NatAs...4..965R}
}

@ARTICLE{lian2020,
       author = {{Lian}, Jianhui and {Zasowski}, Gail and {Hasselquist}, Sten and {Nataf}, David M. and {Thomas}, Daniel and {Moni Bidin}, Christian and {Fern{\'a}ndez-Trincado}, Jos{\'e} G. and {Garcia-Hernandez}, D.~A. and {Lane}, Richard R. and {Majewski}, Steven R. and {Roman-Lopes}, Alexandre and {Schultheis}, Mathias},
        title = "{The Milky Way's bulge star formation history as constrained from its bimodal chemical abundance distribution}",
      journal = {\mnras},
         year = 2020,
        month = sep,
       volume = {497},
       number = {3},
        pages = {3557-3570},
          doi = {10.1093/mnras/staa2205},
archivePrefix = {arXiv},
       eprint = {2007.12179},
 primaryClass = {astro-ph.GA},
       adsurl = {https://ui.adsabs.harvard.edu/abs/2020MNRAS.497.3557L}
}

@ARTICLE{beane2024a,
       author = {{Beane}, Angus},
        title = "{Rising from the Ashes: A Metallicity-dependent Star Formation Gap Splits the Milky Way's {\ensuremath{\alpha}} Sequences}",
      journal = {\apj},
         year = 2025,
        month = apr,
       volume = {982},
       number = {2},
          eid = {106},
        pages = {106},
          doi = {10.3847/1538-4357/adb83e},
archivePrefix = {arXiv},
       eprint = {2407.07985},
 primaryClass = {astro-ph.GA},
       adsurl = {https://ui.adsabs.harvard.edu/abs/2025ApJ...982..106B}
}

@ARTICLE{beane2024b,
       author = {{Beane}, Angus and {Johnson}, James W. and {Semenov}, Vadim A. and {Hernquist}, Lars and {Chandra}, Vedant and {Conroy}, Charlie},
        title = "{Rising from the Ashes. II. The Bar-driven Abundance Bimodality of the Milky Way}",
      journal = {\apj},
         year = 2025,
        month = jun,
       volume = {985},
       number = {2},
          eid = {221},
        pages = {221},
          doi = {10.3847/1538-4357/adceab},
archivePrefix = {arXiv},
       eprint = {2410.21580},
 primaryClass = {astro-ph.GA},
       adsurl = {https://ui.adsabs.harvard.edu/abs/2025ApJ...985..221B}
}

@ARTICLE{buck2020,
       author = {{Buck}, Tobias},
        title = "{On the origin of the chemical bimodality of disc stars: a tale of merger and migration}",
      journal = {\mnras},
         year = 2020,
        month = feb,
       volume = {491},
       number = {4},
        pages = {5435-5446},
          doi = {10.1093/mnras/stz3289},
archivePrefix = {arXiv},
       eprint = {1909.09162},
 primaryClass = {astro-ph.GA},
       adsurl = {https://ui.adsabs.harvard.edu/abs/2020MNRAS.491.5435B}
}

@ARTICLE{grisoni2017,
       author = {{Grisoni}, V. and {Spitoni}, E. and {Matteucci}, F. and {Recio-Blanco}, A. and {de Laverny}, P. and {Hayden}, M. and {Mikolaitis}, {\^{S}}. and {Worley}, C.~C.},
        title = "{The AMBRE project: chemical evolution models for the Milky Way thick and thin discs}",
      journal = {\mnras},
         year = 2017,
        month = dec,
       volume = {472},
       number = {3},
        pages = {3637-3647},
          doi = {10.1093/mnras/stx2201},
archivePrefix = {arXiv},
       eprint = {1706.02614},
 primaryClass = {astro-ph.GA},
       adsurl = {https://ui.adsabs.harvard.edu/abs/2017MNRAS.472.3637G}
}

@ARTICLE{tinsley1979,
       author = {{Tinsley}, B.~M.},
        title = "{Stellar lifetimes and abundance ratios in chemical evolution.}",
      journal = {\apj},
         year = 1979,
        month = may,
       volume = {229},
        pages = {1046-1056},
          doi = {10.1086/157039},
       adsurl = {https://ui.adsabs.harvard.edu/abs/1979ApJ...229.1046T}
}

@ARTICLE{tinsley1980,
       author = {{Tinsley}, B.~M.},
        title = "{Evolution of the Stars and Gas in Galaxies}",
      journal = {\fcp},
         year = 1980,
        month = jan,
       volume = {5},
        pages = {287-388},
          doi = {10.48550/arXiv.2203.02041},
archivePrefix = {arXiv},
       eprint = {2203.02041},
 primaryClass = {astro-ph.GA},
       adsurl = {https://ui.adsabs.harvard.edu/abs/1980FCPh....5..287T}
}

@ARTICLE{mcwilliam1997,
       author = {{McWilliam}, Andrew},
        title = "{Abundance Ratios and Galactic Chemical Evolution}",
      journal = {\araa},
         year = 1997,
        month = jan,
       volume = {35},
        pages = {503-556},
          doi = {10.1146/annurev.astro.35.1.503},
       adsurl = {https://ui.adsabs.harvard.edu/abs/1997ARA&A..35..503M}
}

@ARTICLE{jonsson2018,
       author = {{J{\"o}nsson}, Henrik and {Allende Prieto}, Carlos and {Holtzman}, Jon A. and {Feuillet}, Diane K. and {Hawkins}, Keith and {Cunha}, Katia and {M{\'e}sz{\'a}ros}, Szabolcs and {Hasselquist}, Sten and {Fern{\'a}ndez-Trincado}, J.~G. and {Garc{\'\i}a-Hern{\'a}ndez}, D.~A. and {Bizyaev}, Dmitry and {Carrera}, Ricardo and {Majewski}, Steven R. and {Pinsonneault}, Marc H. and {Shetrone}, Matthew and {Smith}, Verne and {Sobeck}, Jennifer and {Souto}, Diogo and {Stringfellow}, Guy S. and {Teske}, Johanna and {Zamora}, Olga},
        title = "{APOGEE Data Releases 13 and 14: Stellar Parameter and Abundance Comparisons with Independent Analyses}",
      journal = {\aj},
         year = 2018,
        month = sep,
       volume = {156},
       number = {3},
          eid = {126},
        pages = {126},
          doi = {10.3847/1538-3881/aad4f5},
archivePrefix = {arXiv},
       eprint = {1807.09784},
 primaryClass = {astro-ph.SR},
       adsurl = {https://ui.adsabs.harvard.edu/abs/2018AJ....156..126J}
}

@ARTICLE{amarante2020b,
       author = {{Amarante}, Jo{\~a}o A.~S. and {Beraldo e Silva}, Leandro and {Debattista}, Victor P. and {Smith}, Martin C.},
        title = "{The Splash without a Merger}",
      journal = {\apjl},
         year = 2020,
        month = mar,
       volume = {891},
       number = {2},
          eid = {L30},
        pages = {L30},
          doi = {10.3847/2041-8213/ab78a4},
archivePrefix = {arXiv},
       eprint = {1912.12690},
 primaryClass = {astro-ph.GA},
       adsurl = {https://ui.adsabs.harvard.edu/abs/2020ApJ...891L..30A}
}

@ARTICLE{tan2024,
       author = {{Tan}, Vivian Yun Yan and {Muzzin}, Adam and {Marchesini}, Danilo and {Sok}, Visal and {Sarrouh}, Ghassan T.~E. and {Marsan}, Z. Cemile},
        title = "{A Measurement of the Assembly of Milky Way Analogs at Redshifts 0.5 < z < 2 with Resolved Stellar Mass and Star Formation Rate Profiles}",
      journal = {\apj},
         year = 2024,
        month = apr,
       volume = {964},
       number = {2},
          eid = {177},
        pages = {177},
          doi = {10.3847/1538-4357/ad2c90},
archivePrefix = {arXiv},
       eprint = {2402.12433},
 primaryClass = {astro-ph.GA},
       adsurl = {https://ui.adsabs.harvard.edu/abs/2024ApJ...964..177T}
}

@ARTICLE{tan2024b,
       author = {{Tan}, Vivian Yun Yan and {Muzzin}, Adam and {Sarrouh}, Ghassan T.~E. and {Antwi-Danso}, Jacqueline and {Sok}, Visal and {Jagga}, Naadiyah and {Rihtar{\v{s}}i{\v{c}}}, Gregor and {Brown}, Westley and {Abraham}, Roberto and {Asada}, Yoshihisa and {Desprez}, Guillaume and {Iyer}, Kartheik and {Martis}, Nicholas S. and {M{\'e}rida}, Rosa M. and {Mowla}, Lamiya A. and {Noirot}, Ga{\"e}l and {Omori}, Kiyoaki Christopher and {Sawicki}, Marcin and {Tripodi}, Roberta and {Willott}, Chris J.},
        title = "{Resolved Mass Assembly and Star Formation in Milky Way Progenitors since z = 5 from JWST/CANUCS: From Clumps and Mergers to Well-ordered Disks}",
      journal = {\apj},
         year = 2025,
        month = nov,
       volume = {994},
       number = {1},
          eid = {94},
        pages = {94},
          doi = {10.3847/1538-4357/ae0ffe},
archivePrefix = {arXiv},
       eprint = {2412.07829},
 primaryClass = {astro-ph.GA},
       adsurl = {https://ui.adsabs.harvard.edu/abs/2025ApJ...994...94T}
}

@ARTICLE{rusta2024,
       author = {{Rusta}, Elka and {Salvadori}, Stefania and {Gelli}, Viola and {Koutsouridou}, Ioanna and {Marconi}, Alessandro},
        title = "{Linking High-z and Low-z: Are We Observing the Progenitors of the Milky Way with JWST?}",
      journal = {\apjl},
         year = 2024,
        month = oct,
       volume = {974},
       number = {2},
          eid = {L35},
        pages = {L35},
          doi = {10.3847/2041-8213/ad833d},
archivePrefix = {arXiv},
       eprint = {2407.06255},
 primaryClass = {astro-ph.GA},
       adsurl = {https://ui.adsabs.harvard.edu/abs/2024ApJ...974L..35R}
}

@ARTICLE{mowla2024,
       author = {{Mowla}, Lamiya and {Iyer}, Kartheik and {Asada}, Yoshihisa and {Desprez}, Guillaume and {Tan}, Vivian Yun Yan and {Martis}, Nicholas and {Sarrouh}, Ghassan and {Strait}, Victoria and {Abraham}, Roberto and {Brada{\v{c}}}, Maru{\v{s}}a and {Brammer}, Gabriel and {Muzzin}, Adam and {Pacifici}, Camilla and {Ravindranath}, Swara and {Sawicki}, Marcin and {Willott}, Chris and {Estrada-Carpenter}, Vince and {Jahan}, Nusrath and {Noirot}, Ga{\"e}l and {Matharu}, Jasleen and {Rihtar{\v{s}}i{\v{c}}}, Gregor and {Zabl}, Johannes},
        title = "{Formation of a low-mass galaxy from star clusters in a 600-million-year-old Universe}",
      journal = {\nat},
         year = 2024,
        month = dec,
       volume = {636},
       number = {8042},
        pages = {332-336},
          doi = {10.1038/s41586-024-08293-0},
archivePrefix = {arXiv},
       eprint = {2402.08696},
 primaryClass = {astro-ph.GA},
       adsurl = {https://ui.adsabs.harvard.edu/abs/2024Natur.636..332M}
}

@ARTICLE{bovy2012,
       author = {{Bovy}, Jo and {Rix}, Hans-Walter and {Hogg}, David W.},
        title = "{The Milky Way Has No Distinct Thick Disk}",
      journal = {\apj},
         year = 2012,
        month = jun,
       volume = {751},
       number = {2},
          eid = {131},
        pages = {131},
          doi = {10.1088/0004-637X/751/2/131},
archivePrefix = {arXiv},
       eprint = {1111.6585},
 primaryClass = {astro-ph.GA},
       adsurl = {https://ui.adsabs.harvard.edu/abs/2012ApJ...751..131B}
}

@ARTICLE{naidu2021,
       author = {{Naidu}, Rohan P. and {Conroy}, Charlie and {Bonaca}, Ana and {Zaritsky}, Dennis and {Weinberger}, Rainer and {Ting}, Yuan-Sen and {Caldwell}, Nelson and {Tacchella}, Sandro and {Han}, Jiwon Jesse and {Speagle}, Joshua S. and {Cargile}, Phillip A.},
        title = "{Reconstructing the Last Major Merger of the Milky Way with the H3 Survey}",
      journal = {\apj},
         year = 2021,
        month = dec,
       volume = {923},
       number = {1},
          eid = {92},
        pages = {92},
          doi = {10.3847/1538-4357/ac2d2d},
archivePrefix = {arXiv},
       eprint = {2103.03251},
 primaryClass = {astro-ph.GA},
       adsurl = {https://ui.adsabs.harvard.edu/abs/2021ApJ...923...92N}
}

@ARTICLE{tacconi2013,
       author = {{Tacconi}, L.~J. and {Neri}, R. and {Genzel}, R. and {Combes}, F. and {Bolatto}, A. and {Cooper}, M.~C. and {Wuyts}, S. and {Bournaud}, F. and {Burkert}, A. and {Comerford}, J. and {Cox}, P. and {Davis}, M. and {F{\"o}rster Schreiber}, N.~M. and {Garc{\'\i}a-Burillo}, S. and {Gracia-Carpio}, J. and {Lutz}, D. and {Naab}, T. and {Newman}, S. and {Omont}, A. and {Saintonge}, A. and {Shapiro Griffin}, K. and {Shapley}, A. and {Sternberg}, A. and {Weiner}, B.},
        title = "{Phibss: Molecular Gas Content and Scaling Relations in z \raisebox{-0.5ex}\textasciitilde 1-3 Massive, Main-sequence Star-forming Galaxies}",
      journal = {\apj},
         year = 2013,
        month = may,
       volume = {768},
       number = {1},
          eid = {74},
        pages = {74},
          doi = {10.1088/0004-637X/768/1/74},
archivePrefix = {arXiv},
       eprint = {1211.5743},
 primaryClass = {astro-ph.CO},
       adsurl = {https://ui.adsabs.harvard.edu/abs/2013ApJ...768...74T}
}

@ARTICLE{borbolato2025,
       author = {{Borbolato}, Lais and {Rossi}, Silvia and {Perottoni}, H{\'e}lio D. and {Limberg}, Guilherme and {Amarante}, Jo{\~a}o A.~S. and {Queiroz}, Anna B.~A. and {Chiappini}, Cristina and {Anders}, Friedrich and {Santucci}, Rafael M. and {Barbosa}, Fabr{\'\i}cia O. and {Nogueira-Santos}, Jo{\~a}o V.},
        title = "{Early Coformation of the Milky Way's Thin and Thick Disks at Redshift z > 2}",
      journal = {\apj},
         year = 2025,
        month = nov,
       volume = {994},
       number = {1},
          eid = {126},
        pages = {126},
          doi = {10.3847/1538-4357/ae0c96},
archivePrefix = {arXiv},
       eprint = {2504.00135},
 primaryClass = {astro-ph.GA},
       adsurl = {https://ui.adsabs.harvard.edu/abs/2025ApJ...994..126B}
}

@ARTICLE{PLATO,
       author = {{Rauer}, Heike and {Aerts}, Conny and {Cabrera}, Juan and {Deleuil}, Magali and {Erikson}, Anders and {Gizon}, Laurent and {Goupil}, Mariejo and {Heras}, Ana and {Walloschek}, Thomas and {Lorenzo-Alvarez}, Jose and {Marliani}, Filippo and {Martin-Garcia}, C{\'e}sar and {Mas-Hesse}, J. Miguel and {O'Rourke}, Laurence and {Osborn}, Hugh and {Pagano}, Isabella and {Piotto}, Giampaolo and {Pollacco}, Don and {Ragazzoni}, Roberto and {Ramsay}, Gavin and {Udry}, St{\'e}phane and {Appourchaux}, Thierry and {Benz}, Willy and {Brandeker}, Alexis and {G{\"u}del}, Manuel and {Janot-Pacheco}, Eduardo and {Kabath}, Petr and {Kjeldsen}, Hans and {Min}, Michiel and {Santos}, Nuno and {Smith}, Alan and {Suarez}, Juan-Carlos and {Werner}, Stephanie C. and {Aboudan}, Alessio and {Abreu}, Manuel and {Acu{\~n}a}, Lorena and {Adams}, Moritz and {Adibekyan}, Vardan and {Affer}, Laura and {Agneray}, Fran{\c{c}}ois and {Agnor}, Craig and {Aguirre B{\o}rsen-Koch}, Victor and {Ahmed}, Saad and {Aigrain}, Suzanne and {Al-Bahlawan}, Ashraf and {Alcacera Gil}, Ma de los Angeles and {Alei}, Eleonora and {Alencar}, Silvia and {Alexander}, Richard and {Alfonso-Garz{\'o}n}, Julia and {Alibert}, Yann and {Allende Prieto}, Carlos and {Almeida}, Leonardo and {Alonso Sobrino}, Roi and {Altavilla}, Giuseppe and {Althaus}, Christian and {Alvarez Trujillo}, Luis Alonso and {Amarsi}, Anish and {Ammler-von Eiff}, Matthias and {Am{\^o}res}, Eduardo and {Andrade}, Laerte and {Antoniadis-Karnavas}, Alexandros and {Ant{\'o}nio}, Carlos and {Aparicio del Moral}, Beatriz and {Appolloni}, Matteo and {Arena}, Claudio and {Armstrong}, David and {Aroca Aliaga}, Jose and {Asplund}, Martin and {Audenaert}, Jeroen and {Auricchio}, Natalia and {Avelino}, Pedro and {Baeke}, Ann and {Bailli{\'e}}, Kevin and {Balado}, Ana and {Ballber Balaguer{\'o}}, Pau and {Balestra}, Andrea and {Ball}, Warrick and {Ballans}, Herve and {Ballot}, Jerome and {Barban}, Caroline and {Barbary}, Ga{\"e}le and {Barbieri}, Mauro and {Barcel{\'o} Forteza}, Sebasti{\`a} and {Barker}, Adrian and {Barklem}, Paul and {Barnes}, Sydney and {Barrado Navascues}, David and {Barragan}, Oscar and {Baruteau}, Cl{\'e}ment and {Basu}, Sarbani and {Baudin}, Frederic and {Baumeister}, Philipp and {Bayliss}, Daniel and {Bazot}, Michael and {Beck}, Paul G. and {Belkacem}, Kevin and {Bellinger}, Earl and {Benatti}, Serena and {Benomar}, Othman and {B{\'e}rard}, Diane and {Bergemann}, Maria and {Bergomi}, Maria and {Bernardo}, Pierre and {Biazzo}, Katia and {Bignamini}, Andrea and {Bigot}, Lionel and {Billot}, Nicolas and {Binet}, Martin and {Biondi}, David and {Biondi}, Federico and {Birch}, Aaron C. and {Bitsch}, Bertram and {Bluhm Ceballos}, Paz Victoria and {B{\'o}di}, Attila and {Bogn{\'a}r}, Zs{\'o}fia and {Boisse}, Isabelle and {Bolmont}, Emeline and {Bonanno}, Alfio and {Bonavita}, Mariangela and {Bonfanti}, Andrea and {Bonfils}, Xavier and {Bonito}, Rosaria and {Bonomo}, Aldo Stefano and {B{\"o}rner}, Anko and {Boro Saikia}, Sudeshna and {Borreguero Mart{\'\i}n}, Elisa and {Borsa}, Francesco and {Borsato}, Luca and {Bossini}, Diego and {Bouchy}, Francois and {Bou{\'e}}, Gwena{\"e}l and {Boufleur}, Rodrigo and {Boumier}, Patrick and {Bourrier}, Vincent and {Bowman}, Dominic M. and {Bozzo}, Enrico and {Bradley}, Louisa and {Bray}, John and {Bressan}, Alessandro and {Breton}, Sylvain and {Brienza}, Daniele and {Brito}, Ana and {Brogi}, Matteo and {Brown}, Beverly and {Brown}, David J.~A. and {Brun}, Allan Sacha and {Bruno}, Giovanni and {Bruns}, Michael and {Buchhave}, Lars A. and {Bugnet}, Lisa and {Buldgen}, Ga{\"e}l and {Burgess}, Patrick and {Busatta}, Andrea and {Busso}, Giorgia and {Buzasi}, Derek and {Caballero}, Jos{\'e} A. and {Cabral}, Alexandre and {Cabrero Gomez}, Juan-Francisco and {Calderone}, Flavia and {Cameron}, Robert and {Cameron}, Andrew and {Campante}, Tiago and {Campos Gestal}, N{\'e}stor and {Canto Martins}, Bruno Leonardo and {Cara}, Christophe and {Carone}, Ludmila and {Carrasco}, Josep Manel and {Casagrande}, Luca and {Casewell}, Sarah L. and {Cassisi}, Santi and {Castellani}, Marco and {Castro}, Matthieu and {Catala}, Claude and {Catal{\'a}n Fern{\'a}ndez}, Irene and {Catelan}, M{\'a}rcio and {Cegla}, Heather and {Cerruti}, Chiara and {Cessa}, Virginie and {Chadid}, Merieme and {Chaplin}, William and {Charpinet}, Stephane and {Chiappini}, Cristina and {Chiarucci}, Simone and {Chiavassa}, Andrea and {Chinellato}, Simonetta and {Chirulli}, Giovanni and {Christensen-Dalsgaard}, J{\o}rgen and {Church}, Ross and {Claret}, Antonio and {Clarke}, Cathie and {Claudi}, Riccardo and {Clermont}, Lionel and {Coelho}, Hugo and {Coelho}, Joao and {Cogato}, Fabrizio and {Colom{\'e}}, Josep and {Condamin}, Mathieu and {Conde Garc{\'\i}a}, Fernando and {Conseil}, Simon},
        title = "{The PLATO mission}",
      journal = {Experimental Astronomy},
         year = 2025,
        month = jun,
       volume = {59},
       number = {3},
          eid = {26},
        pages = {26},
          doi = {10.1007/s10686-025-09985-9},
archivePrefix = {arXiv},
       eprint = {2406.05447},
 primaryClass = {astro-ph.IM},
       adsurl = {https://ui.adsabs.harvard.edu/abs/2025ExA....59...26R}
}

@ARTICLE{jones2023,
       author = {{Jones}, Tucker and {Sanders}, Ryan and {Chen}, Yuguang and {Wang}, Xin and {Morishita}, Takahiro and {Roberts-Borsani}, Guido and {Treu}, Tommaso and {Dressler}, Alan and {Merlin}, Emiliano and {Paris}, Diego and {Santini}, Paola and {Bergamini}, Pietro and {Henry}, A. and {Huntzinger}, Erin and {Nanayakkara}, Themiya and {Boyett}, Kristan and {Bradac}, Marusa and {Brammer}, Gabriel and {Calabr{\'o}}, Antonello and {Glazebrook}, Karl and {Grasha}, Kathryn and {Mascia}, Sara and {Pentericci}, Laura and {Trenti}, Michele and {Vulcani}, Benedetta},
        title = "{Early Results from GLASS-JWST. XXI. Rapid Asembly of a Galaxy at z = 6.23 Revealed by Its C/O Abundance}",
      journal = {\apjl},
         year = 2023,
        month = jul,
       volume = {951},
       number = {1},
          eid = {L17},
        pages = {L17},
          doi = {10.3847/2041-8213/acd938},
archivePrefix = {arXiv},
       eprint = {2301.07126},
 primaryClass = {astro-ph.GA},
       adsurl = {https://ui.adsabs.harvard.edu/abs/2023ApJ...951L..17J}
}

@ARTICLE{gaia,
       author = {{Gaia Collaboration} and {Prusti}, T. and {de Bruijne}, J.~H.~J. and {Brown}, A.~G.~A. and {Vallenari}, A. and {Babusiaux}, C. and {Bailer-Jones}, C.~A.~L. and {Bastian}, U. and {Biermann}, M. and {Evans}, D.~W. and {Eyer}, L. and {Jansen}, F. and {Jordi}, C. and {Klioner}, S.~A. and {Lammers}, U. and {Lindegren}, L. and {Luri}, X. and {Mignard}, F. and {Milligan}, D.~J. and {Panem}, C. and {Poinsignon}, V. and {Pourbaix}, D. and {Randich}, S. and {Sarri}, G. and {Sartoretti}, P. and {Siddiqui}, H.~I. and {Soubiran}, C. and {Valette}, V. and {van Leeuwen}, F. and {Walton}, N.~A. and {Aerts}, C. and {Arenou}, F. and {Cropper}, M. and {Drimmel}, R. and {H{\o}g}, E. and {Katz}, D. and {Lattanzi}, M.~G. and {O'Mullane}, W. and {Grebel}, E.~K. and {Holland}, A.~D. and {Huc}, C. and {Passot}, X. and {Bramante}, L. and {Cacciari}, C. and {Casta{\~n}eda}, J. and {Chaoul}, L. and {Cheek}, N. and {De Angeli}, F. and {Fabricius}, C. and {Guerra}, R. and {Hern{\'a}ndez}, J. and {Jean-Antoine-Piccolo}, A. and {Masana}, E. and {Messineo}, R. and {Mowlavi}, N. and {Nienartowicz}, K. and {Ord{\'o}{\~n}ez-Blanco}, D. and {Panuzzo}, P. and {Portell}, J. and {Richards}, P.~J. and {Riello}, M. and {Seabroke}, G.~M. and {Tanga}, P. and {Th{\'e}venin}, F. and {Torra}, J. and {Els}, S.~G. and {Gracia-Abril}, G. and {Comoretto}, G. and {Garcia-Reinaldos}, M. and {Lock}, T. and {Mercier}, E. and {Altmann}, M. and {Andrae}, R. and {Astraatmadja}, T.~L. and {Bellas-Velidis}, I. and {Benson}, K. and {Berthier}, J. and {Blomme}, R. and {Busso}, G. and {Carry}, B. and {Cellino}, A. and {Clementini}, G. and {Cowell}, S. and {Creevey}, O. and {Cuypers}, J. and {Davidson}, M. and {De Ridder}, J. and {de Torres}, A. and {Delchambre}, L. and {Dell'Oro}, A. and {Ducourant}, C. and {Fr{\'e}mat}, Y. and {Garc{\'\i}a-Torres}, M. and {Gosset}, E. and {Halbwachs}, J. -L. and {Hambly}, N.~C. and {Harrison}, D.~L. and {Hauser}, M. and {Hestroffer}, D. and {Hodgkin}, S.~T. and {Huckle}, H.~E. and {Hutton}, A. and {Jasniewicz}, G. and {Jordan}, S. and {Kontizas}, M. and {Korn}, A.~J. and {Lanzafame}, A.~C. and {Manteiga}, M. and {Moitinho}, A. and {Muinonen}, K. and {Osinde}, J. and {Pancino}, E. and {Pauwels}, T. and {Petit}, J. -M. and {Recio-Blanco}, A. and {Robin}, A.~C. and {Sarro}, L.~M. and {Siopis}, C. and {Smith}, M. and {Smith}, K.~W. and {Sozzetti}, A. and {Thuillot}, W. and {van Reeven}, W. and {Viala}, Y. and {Abbas}, U. and {Abreu Aramburu}, A. and {Accart}, S. and {Aguado}, J.~J. and {Allan}, P.~M. and {Allasia}, W. and {Altavilla}, G. and {{\'A}lvarez}, M.~A. and {Alves}, J. and {Anderson}, R.~I. and {Andrei}, A.~H. and {Anglada Varela}, E. and {Antiche}, E. and {Antoja}, T. and {Ant{\'o}n}, S. and {Arcay}, B. and {Atzei}, A. and {Ayache}, L. and {Bach}, N. and {Baker}, S.~G. and {Balaguer-N{\'u}{\~n}ez}, L. and {Barache}, C. and {Barata}, C. and {Barbier}, A. and {Barblan}, F. and {Baroni}, M. and {Barrado y Navascu{\'e}s}, D. and {Barros}, M. and {Barstow}, M.~A. and {Becciani}, U. and {Bellazzini}, M. and {Bellei}, G. and {Bello Garc{\'\i}a}, A. and {Belokurov}, V. and {Bendjoya}, P. and {Berihuete}, A. and {Bianchi}, L. and {Bienaym{\'e}}, O. and {Billebaud}, F. and {Blagorodnova}, N. and {Blanco-Cuaresma}, S. and {Boch}, T. and {Bombrun}, A. and {Borrachero}, R. and {Bouquillon}, S. and {Bourda}, G. and {Bouy}, H. and {Bragaglia}, A. and {Breddels}, M.~A. and {Brouillet}, N. and {Br{\"u}semeister}, T. and {Bucciarelli}, B. and {Budnik}, F. and {Burgess}, P. and {Burgon}, R. and {Burlacu}, A. and {Busonero}, D. and {Buzzi}, R. and {Caffau}, E. and {Cambras}, J. and {Campbell}, H. and {Cancelliere}, R. and {Cantat-Gaudin}, T. and {Carlucci}, T. and {Carrasco}, J.~M. and {Castellani}, M. and {Charlot}, P. and {Charnas}, J. and {Charvet}, P. and {Chassat}, F. and {Chiavassa}, A. and {Clotet}, M. and {Cocozza}, G. and {Collins}, R.~S. and {Collins}, P. and {Costigan}, G.},
        title = "{The Gaia mission}",
      journal = {\aap},
         year = 2016,
        month = nov,
       volume = {595},
          eid = {A1},
        pages = {A1},
          doi = {10.1051/0004-6361/201629272},
archivePrefix = {arXiv},
       eprint = {1609.04153},
 primaryClass = {astro-ph.IM},
       adsurl = {https://ui.adsabs.harvard.edu/abs/2016A&A...595A...1G}
}

@ARTICLE{apogee,
       author = {{Majewski}, Steven R. and {Schiavon}, Ricardo P. and {Frinchaboy}, Peter M. and {Allende Prieto}, Carlos and {Barkhouser}, Robert and {Bizyaev}, Dmitry and {Blank}, Basil and {Brunner}, Sophia and {Burton}, Adam and {Carrera}, Ricardo and {Chojnowski}, S. Drew and {Cunha}, K{\'a}tia and {Epstein}, Courtney and {Fitzgerald}, Greg and {Garc{\'\i}a P{\'e}rez}, Ana E. and {Hearty}, Fred R. and {Henderson}, Chuck and {Holtzman}, Jon A. and {Johnson}, Jennifer A. and {Lam}, Charles R. and {Lawler}, James E. and {Maseman}, Paul and {M{\'e}sz{\'a}ros}, Szabolcs and {Nelson}, Matthew and {Nguyen}, Duy Coung and {Nidever}, David L. and {Pinsonneault}, Marc and {Shetrone}, Matthew and {Smee}, Stephen and {Smith}, Verne V. and {Stolberg}, Todd and {Skrutskie}, Michael F. and {Walker}, Eric and {Wilson}, John C. and {Zasowski}, Gail and {Anders}, Friedrich and {Basu}, Sarbani and {Beland}, Stephane and {Blanton}, Michael R. and {Bovy}, Jo and {Brownstein}, Joel R. and {Carlberg}, Joleen and {Chaplin}, William and {Chiappini}, Cristina and {Eisenstein}, Daniel J. and {Elsworth}, Yvonne and {Feuillet}, Diane and {Fleming}, Scott W. and {Galbraith-Frew}, Jessica and {Garc{\'\i}a}, Rafael A. and {Garc{\'\i}a-Hern{\'a}ndez}, D. An{\'\i}bal and {Gillespie}, Bruce A. and {Girardi}, L{\'e}o and {Gunn}, James E. and {Hasselquist}, Sten and {Hayden}, Michael R. and {Hekker}, Saskia and {Ivans}, Inese and {Kinemuchi}, Karen and {Klaene}, Mark and {Mahadevan}, Suvrath and {Mathur}, Savita and {Mosser}, Beno{\^\i}t and {Muna}, Demitri and {Munn}, Jeffrey A. and {Nichol}, Robert C. and {O'Connell}, Robert W. and {Parejko}, John K. and {Robin}, A.~C. and {Rocha-Pinto}, Helio and {Schultheis}, Matthias and {Serenelli}, Aldo M. and {Shane}, Neville and {Silva Aguirre}, Victor and {Sobeck}, Jennifer S. and {Thompson}, Benjamin and {Troup}, Nicholas W. and {Weinberg}, David H. and {Zamora}, Olga},
        title = "{The Apache Point Observatory Galactic Evolution Experiment (APOGEE)}",
      journal = {\aj},
         year = 2017,
        month = sep,
       volume = {154},
       number = {3},
          eid = {94},
        pages = {94},
          doi = {10.3847/1538-3881/aa784d},
archivePrefix = {arXiv},
       eprint = {1509.05420},
 primaryClass = {astro-ph.IM},
       adsurl = {https://ui.adsabs.harvard.edu/abs/2017AJ....154...94M}
}

@ARTICLE{lamost,
       author = {{Deng}, Li-Cai and {Newberg}, Heidi Jo and {Liu}, Chao and {Carlin}, Jeffrey L. and {Beers}, Timothy C. and {Chen}, Li and {Chen}, Yu-Qin and {Christlieb}, Norbert and {Grillmair}, Carl J. and {Guhathakurta}, Puragra and {Han}, Zhan-Wen and {Hou}, Jin-Liang and {Lee}, Hsu-Tai and {L{\'e}pine}, S{\'e}bastien and {Li}, Jing and {Liu}, Xiao-Wei and {Pan}, Kai-Ke and {Sellwood}, J.~A. and {Wang}, Bo and {Wang}, Hong-Chi and {Yang}, Fan and {Yanny}, Brian and {Zhang}, Hao-Tong and {Zhang}, Yue-Yang and {Zheng}, Zheng and {Zhu}, Zi},
        title = "{LAMOST Experiment for Galactic Understanding and Exploration (LEGUE) {\textemdash} The survey's science plan}",
      journal = {Research in Astronomy and Astrophysics},
         year = 2012,
        month = jul,
       volume = {12},
       number = {7},
        pages = {735-754},
          doi = {10.1088/1674-4527/12/7/003},
archivePrefix = {arXiv},
       eprint = {1206.3578},
 primaryClass = {astro-ph.GA},
       adsurl = {https://ui.adsabs.harvard.edu/abs/2012RAA....12..735D}
}

@ARTICLE{rave,
       author = {{Steinmetz}, M. and {Zwitter}, T. and {Siebert}, A. and {Watson}, F.~G. and {Freeman}, K.~C. and {Munari}, U. and {Campbell}, R. and {Williams}, M. and {Seabroke}, G.~M. and {Wyse}, R.~F.~G. and {Parker}, Q.~A. and {Bienaym{\'e}}, O. and {Roeser}, S. and {Gibson}, B.~K. and {Gilmore}, G. and {Grebel}, E.~K. and {Helmi}, A. and {Navarro}, J.~F. and {Burton}, D. and {Cass}, C.~J.~P. and {Dawe}, J.~A. and {Fiegert}, K. and {Hartley}, M. and {Russell}, K.~S. and {Saunders}, W. and {Enke}, H. and {Bailin}, J. and {Binney}, J. and {Bland-Hawthorn}, J. and {Boeche}, C. and {Dehnen}, W. and {Eisenstein}, D.~J. and {Evans}, N.~W. and {Fiorucci}, M. and {Fulbright}, J.~P. and {Gerhard}, O. and {Jauregi}, U. and {Kelz}, A. and {Mijovi{\'c}}, L. and {Minchev}, I. and {Parmentier}, G. and {Pe{\~n}arrubia}, J. and {Quillen}, A.~C. and {Read}, M.~A. and {Ruchti}, G. and {Scholz}, R. -D. and {Siviero}, A. and {Smith}, M.~C. and {Sordo}, R. and {Veltz}, L. and {Vidrih}, S. and {von Berlepsch}, R. and {Boyle}, B.~J. and {Schilbach}, E.},
        title = "{The Radial Velocity Experiment (RAVE): First Data Release}",
      journal = {\aj},
         year = 2006,
        month = oct,
       volume = {132},
       number = {4},
        pages = {1645-1668},
          doi = {10.1086/506564},
archivePrefix = {arXiv},
       eprint = {astro-ph/0606211},
 primaryClass = {astro-ph},
       adsurl = {https://ui.adsabs.harvard.edu/abs/2006AJ....132.1645S}
}

@ARTICLE{tsukui2024,
       author = {{Tsukui}, Takafumi and {Wisnioski}, Emily and {Bland-Hawthorn}, Joss and {Freeman}, Ken},
        title = "{The emergence of galactic thin and thick discs across cosmic history}",
      journal = {\mnras},
         year = 2025,
        month = jul,
       volume = {540},
       number = {4},
        pages = {3493-3522},
          doi = {10.1093/mnras/staf604},
archivePrefix = {arXiv},
       eprint = {2409.15909},
 primaryClass = {astro-ph.GA},
       adsurl = {https://ui.adsabs.harvard.edu/abs/2025MNRAS.540.3493T}
}

@ARTICLE{yoachim2006,
       author = {{Yoachim}, Peter and {Dalcanton}, Julianne J.},
        title = "{Structural Parameters of Thin and Thick Disks in Edge-on Disk Galaxies}",
      journal = {\aj},
         year = 2006,
        month = jan,
       volume = {131},
       number = {1},
        pages = {226-249},
          doi = {10.1086/497970},
archivePrefix = {arXiv},
       eprint = {astro-ph/0508460},
 primaryClass = {astro-ph},
       adsurl = {https://ui.adsabs.harvard.edu/abs/2006AJ....131..226Y}
}

@ARTICLE{mould2005,
       author = {{Mould}, Jeremy},
        title = "{Red Thick Disks of Nearby Galaxies}",
      journal = {\aj},
         year = 2005,
        month = feb,
       volume = {129},
       number = {2},
        pages = {698-711},
          doi = {10.1086/427248},
archivePrefix = {arXiv},
       eprint = {astro-ph/0411231},
 primaryClass = {astro-ph},
       adsurl = {https://ui.adsabs.harvard.edu/abs/2005AJ....129..698M}
}

@INPROCEEDINGS{nidever2024,
       author = {{Nidever}, David L. and {Gilbert}, Karoline and {Tollerud}, Erik and {Siders}, Charles and {Escala}, Ivanna and {Prieto}, Carlos Allende and {Smith}, Verne and {Cunha}, Katia and {Debattista}, Victor P. and {Ting}, Yuan-Sen and {Kirby}, Evan N.},
        title = "{The Prevalence of the {\ensuremath{\alpha}}-bimodality: First JWST {\ensuremath{\alpha}}-abundance Results in M31}",
    booktitle = {Early Disk-Galaxy Formation from JWST to the Milky Way},
         year = 2024,
       editor = {{Tabatabaei}, Fatemeh and {Barbuy}, Beatriz and {Ting}, Yuan-Sen},
       series = {IAU Symposium},
       volume = {377},
        month = jan,
        pages = {115-122},
          doi = {10.1017/S1743921323002016},
archivePrefix = {arXiv},
       eprint = {2306.04688},
 primaryClass = {astro-ph.GA},
       adsurl = {https://ui.adsabs.harvard.edu/abs/2024IAUS..377..115N}
}

@ARTICLE{loebman2011,
       author = {{Loebman}, Sarah R. and {Ro{\v{s}}kar}, Rok and {Debattista}, Victor P. and {Ivezi{\'c}}, {\v{Z}}eljko and {Quinn}, Thomas R. and {Wadsley}, James},
        title = "{The Genesis of the Milky Way's Thick Disk Via Stellar Migration}",
      journal = {\apj},
         year = 2011,
        month = aug,
       volume = {737},
       number = {1},
          eid = {8},
        pages = {8},
          doi = {10.1088/0004-637X/737/1/8},
archivePrefix = {arXiv},
       eprint = {1009.5997},
 primaryClass = {astro-ph.GA},
       adsurl = {https://ui.adsabs.harvard.edu/abs/2011ApJ...737....8L}
}

@ARTICLE{loebman2016,
       author = {{Loebman}, Sarah R. and {Debattista}, Victor P. and {Nidever}, David L. and {Hayden}, Michael R. and {Holtzman}, Jon A. and {Clarke}, Adam J. and {Ro{\v{s}}kar}, Rok and {Valluri}, Monica},
        title = "{Imprints of Radial Migration on the Milky Way{\textquoteright}s Metallicity Distribution Functions}",
      journal = {\apjl},
         year = 2016,
        month = feb,
       volume = {818},
       number = {1},
          eid = {L6},
        pages = {L6},
          doi = {10.3847/2041-8205/818/1/L6},
archivePrefix = {arXiv},
       eprint = {1511.06369},
 primaryClass = {astro-ph.GA},
       adsurl = {https://ui.adsabs.harvard.edu/abs/2016ApJ...818L...6L}
}

@ARTICLE{grand2015,
       author = {{Grand}, Robert J.~J. and {Kawata}, Daisuke and {Cropper}, Mark},
        title = "{Impact of radial migration on stellar and gas radial metallicity distribution}",
      journal = {\mnras},
         year = 2015,
        month = mar,
       volume = {447},
       number = {4},
        pages = {4018-4027},
          doi = {10.1093/mnras/stv016},
archivePrefix = {arXiv},
       eprint = {1410.3836},
 primaryClass = {astro-ph.GA},
       adsurl = {https://ui.adsabs.harvard.edu/abs/2015MNRAS.447.4018G}
}

@ARTICLE{gaiadr2,
       author = {{Lindegren}, L. and {Hern{\'a}ndez}, J. and {Bombrun}, A. and {Klioner}, S. and {Bastian}, U. and {Ramos-Lerate}, M. and {de Torres}, A. and {Steidelm{\"u}ller}, H. and {Stephenson}, C. and {Hobbs}, D. and {Lammers}, U. and {Biermann}, M. and {Geyer}, R. and {Hilger}, T. and {Michalik}, D. and {Stampa}, U. and {McMillan}, P.~J. and {Casta{\~n}eda}, J. and {Clotet}, M. and {Comoretto}, G. and {Davidson}, M. and {Fabricius}, C. and {Gracia}, G. and {Hambly}, N.~C. and {Hutton}, A. and {Mora}, A. and {Portell}, J. and {van Leeuwen}, F. and {Abbas}, U. and {Abreu}, A. and {Altmann}, M. and {Andrei}, A. and {Anglada}, E. and {Balaguer-N{\'u}{\~n}ez}, L. and {Barache}, C. and {Becciani}, U. and {Bertone}, S. and {Bianchi}, L. and {Bouquillon}, S. and {Bourda}, G. and {Br{\"u}semeister}, T. and {Bucciarelli}, B. and {Busonero}, D. and {Buzzi}, R. and {Cancelliere}, R. and {Carlucci}, T. and {Charlot}, P. and {Cheek}, N. and {Crosta}, M. and {Crowley}, C. and {de Bruijne}, J. and {de Felice}, F. and {Drimmel}, R. and {Esquej}, P. and {Fienga}, A. and {Fraile}, E. and {Gai}, M. and {Garralda}, N. and {Gonz{\'a}lez-Vidal}, J.~J. and {Guerra}, R. and {Hauser}, M. and {Hofmann}, W. and {Holl}, B. and {Jordan}, S. and {Lattanzi}, M.~G. and {Lenhardt}, H. and {Liao}, S. and {Licata}, E. and {Lister}, T. and {L{\"o}ffler}, W. and {Marchant}, J. and {Martin-Fleitas}, J. -M. and {Messineo}, R. and {Mignard}, F. and {Morbidelli}, R. and {Poggio}, E. and {Riva}, A. and {Rowell}, N. and {Salguero}, E. and {Sarasso}, M. and {Sciacca}, E. and {Siddiqui}, H. and {Smart}, R.~L. and {Spagna}, A. and {Steele}, I. and {Taris}, F. and {Torra}, J. and {van Elteren}, A. and {van Reeven}, W. and {Vecchiato}, A.},
        title = "{Gaia Data Release 2. The astrometric solution}",
      journal = {\aap},
         year = 2018,
        month = aug,
       volume = {616},
          eid = {A2},
        pages = {A2},
          doi = {10.1051/0004-6361/201832727},
archivePrefix = {arXiv},
       eprint = {1804.09366},
 primaryClass = {astro-ph.IM},
       adsurl = {https://ui.adsabs.harvard.edu/abs/2018A&A...616A...2L}
}

@ARTICLE{gaiadr3,
       author = {{Gaia Collaboration} and {Brown}, A.~G.~A. and {Vallenari}, A. and {Prusti}, T. and {de Bruijne}, J.~H.~J. and {Babusiaux}, C. and {Biermann}, M. and {Creevey}, O.~L. and {Evans}, D.~W. and {Eyer}, L. and {Hutton}, A. and {Jansen}, F. and {Jordi}, C. and {Klioner}, S.~A. and {Lammers}, U. and {Lindegren}, L. and {Luri}, X. and {Mignard}, F. and {Panem}, C. and {Pourbaix}, D. and {Randich}, S. and {Sartoretti}, P. and {Soubiran}, C. and {Walton}, N.~A. and {Arenou}, F. and {Bailer-Jones}, C.~A.~L. and {Bastian}, U. and {Cropper}, M. and {Drimmel}, R. and {Katz}, D. and {Lattanzi}, M.~G. and {van Leeuwen}, F. and {Bakker}, J. and {Cacciari}, C. and {Casta{\~n}eda}, J. and {De Angeli}, F. and {Ducourant}, C. and {Fabricius}, C. and {Fouesneau}, M. and {Fr{\'e}mat}, Y. and {Guerra}, R. and {Guerrier}, A. and {Guiraud}, J. and {Jean-Antoine Piccolo}, A. and {Masana}, E. and {Messineo}, R. and {Mowlavi}, N. and {Nicolas}, C. and {Nienartowicz}, K. and {Pailler}, F. and {Panuzzo}, P. and {Riclet}, F. and {Roux}, W. and {Seabroke}, G.~M. and {Sordo}, R. and {Tanga}, P. and {Th{\'e}venin}, F. and {Gracia-Abril}, G. and {Portell}, J. and {Teyssier}, D. and {Altmann}, M. and {Andrae}, R. and {Bellas-Velidis}, I. and {Benson}, K. and {Berthier}, J. and {Blomme}, R. and {Brugaletta}, E. and {Burgess}, P.~W. and {Busso}, G. and {Carry}, B. and {Cellino}, A. and {Cheek}, N. and {Clementini}, G. and {Damerdji}, Y. and {Davidson}, M. and {Delchambre}, L. and {Dell'Oro}, A. and {Fern{\'a}ndez-Hern{\'a}ndez}, J. and {Galluccio}, L. and {Garc{\'\i}a-Lario}, P. and {Garcia-Reinaldos}, M. and {Gonz{\'a}lez-N{\'u}{\~n}ez}, J. and {Gosset}, E. and {Haigron}, R. and {Halbwachs}, J. -L. and {Hambly}, N.~C. and {Harrison}, D.~L. and {Hatzidimitriou}, D. and {Heiter}, U. and {Hern{\'a}ndez}, J. and {Hestroffer}, D. and {Hodgkin}, S.~T. and {Holl}, B. and {Jan{\ss}en}, K. and {Jevardat de Fombelle}, G. and {Jordan}, S. and {Krone-Martins}, A. and {Lanzafame}, A.~C. and {L{\"o}ffler}, W. and {Lorca}, A. and {Manteiga}, M. and {Marchal}, O. and {Marrese}, P.~M. and {Moitinho}, A. and {Mora}, A. and {Muinonen}, K. and {Osborne}, P. and {Pancino}, E. and {Pauwels}, T. and {Petit}, J. -M. and {Recio-Blanco}, A. and {Richards}, P.~J. and {Riello}, M. and {Rimoldini}, L. and {Robin}, A.~C. and {Roegiers}, T. and {Rybizki}, J. and {Sarro}, L.~M. and {Siopis}, C. and {Smith}, M. and {Sozzetti}, A. and {Ulla}, A. and {Utrilla}, E. and {van Leeuwen}, M. and {van Reeven}, W. and {Abbas}, U. and {Abreu Aramburu}, A. and {Accart}, S. and {Aerts}, C. and {Aguado}, J.~J. and {Ajaj}, M. and {Altavilla}, G. and {{\'A}lvarez}, M.~A. and {{\'A}lvarez Cid-Fuentes}, J. and {Alves}, J. and {Anderson}, R.~I. and {Anglada Varela}, E. and {Antoja}, T. and {Audard}, M. and {Baines}, D. and {Baker}, S.~G. and {Balaguer-N{\'u}{\~n}ez}, L. and {Balbinot}, E. and {Balog}, Z. and {Barache}, C. and {Barbato}, D. and {Barros}, M. and {Barstow}, M.~A. and {Bartolom{\'e}}, S. and {Bassilana}, J. -L. and {Bauchet}, N. and {Baudesson-Stella}, A. and {Becciani}, U. and {Bellazzini}, M. and {Bernet}, M. and {Bertone}, S. and {Bianchi}, L. and {Blanco-Cuaresma}, S. and {Boch}, T. and {Bombrun}, A. and {Bossini}, D. and {Bouquillon}, S. and {Bragaglia}, A. and {Bramante}, L. and {Breedt}, E. and {Bressan}, A. and {Brouillet}, N. and {Bucciarelli}, B. and {Burlacu}, A. and {Busonero}, D. and {Butkevich}, A.~G. and {Buzzi}, R. and {Caffau}, E. and {Cancelliere}, R. and {C{\'a}novas}, H. and {Cantat-Gaudin}, T. and {Carballo}, R. and {Carlucci}, T. and {Carnerero}, M.~I. and {Carrasco}, J.~M. and {Casamiquela}, L. and {Castellani}, M. and {Castro-Ginard}, A. and {Castro Sampol}, P. and {Chaoul}, L. and {Charlot}, P. and {Chemin}, L. and {Chiavassa}, A. and {Cioni}, M. -R.~L. and {Comoretto}, G. and {Cooper}, W.~J. and {Cornez}, T. and {Cowell}, S. and {Crifo}, F. and {Crosta}, M. and {Crowley}, C. and {Dafonte}, C. and {Dapergolas}, A. and {David}, M. and {David}, P.},
        title = "{Gaia Early Data Release 3. Summary of the contents and survey properties}",
      journal = {\aap},
         year = 2021,
        month = may,
       volume = {649},
          eid = {A1},
        pages = {A1},
          doi = {10.1051/0004-6361/202039657},
archivePrefix = {arXiv},
       eprint = {2012.01533},
 primaryClass = {astro-ph.GA},
       adsurl = {https://ui.adsabs.harvard.edu/abs/2021A&A...649A...1G}
}

@ARTICLE{segue,
       author = {{Yanny}, Brian and {Rockosi}, Constance and {Newberg}, Heidi Jo and {Knapp}, Gillian R. and {Adelman-McCarthy}, Jennifer K. and {Alcorn}, Bonnie and {Allam}, Sahar and {Allende Prieto}, Carlos and {An}, Deokkeun and {Anderson}, Kurt S.~J. and {Anderson}, Scott and {Bailer-Jones}, Coryn A.~L. and {Bastian}, Steve and {Beers}, Timothy C. and {Bell}, Eric and {Belokurov}, Vasily and {Bizyaev}, Dmitry and {Blythe}, Norm and {Bochanski}, John J. and {Boroski}, William N. and {Brinchmann}, Jarle and {Brinkmann}, J. and {Brewington}, Howard and {Carey}, Larry and {Cudworth}, Kyle M. and {Evans}, Michael and {Evans}, N.~W. and {Gates}, Evalyn and {G{\"a}nsicke}, B.~T. and {Gillespie}, Bruce and {Gilmore}, Gerald and {Nebot Gomez-Moran}, Ada and {Grebel}, Eva K. and {Greenwell}, Jim and {Gunn}, James E. and {Jordan}, Cathy and {Jordan}, Wendell and {Harding}, Paul and {Harris}, Hugh and {Hendry}, John S. and {Holder}, Diana and {Ivans}, Inese I. and {Ivezi{\v{c}}}, {\v{Z}}eljko and {Jester}, Sebastian and {Johnson}, Jennifer A. and {Kent}, Stephen M. and {Kleinman}, Scot and {Kniazev}, Alexei and {Krzesinski}, Jurek and {Kron}, Richard and {Kuropatkin}, Nikolay and {Lebedeva}, Svetlana and {Lee}, Young Sun and {French Leger}, R. and {L{\'e}pine}, S{\'e}bastien and {Levine}, Steve and {Lin}, Huan and {Long}, Daniel C. and {Loomis}, Craig and {Lupton}, Robert and {Malanushenko}, Olena and {Malanushenko}, Viktor and {Margon}, Bruce and {Martinez-Delgado}, David and {McGehee}, Peregrine and {Monet}, Dave and {Morrison}, Heather L. and {Munn}, Jeffrey A. and {Neilsen}, Jr., Eric H. and {Nitta}, Atsuko and {Norris}, John E. and {Oravetz}, Dan and {Owen}, Russell and {Padmanabhan}, Nikhil and {Pan}, Kaike and {Peterson}, R.~S. and {Pier}, Jeffrey R. and {Platson}, Jared and {Re Fiorentin}, Paola and {Richards}, Gordon T. and {Rix}, Hans-Walter and {Schlegel}, David J. and {Schneider}, Donald P. and {Schreiber}, Matthias R. and {Schwope}, Axel and {Sibley}, Valena and {Simmons}, Audrey and {Snedden}, Stephanie A. and {Allyn Smith}, J. and {Stark}, Larry and {Stauffer}, Fritz and {Steinmetz}, M. and {Stoughton}, C. and {SubbaRao}, Mark and {Szalay}, Alex and {Szkody}, Paula and {Thakar}, Aniruddha R. and {Sivarani}, Thirupathi and {Tucker}, Douglas and {Uomoto}, Alan and {Vanden Berk}, Dan and {Vidrih}, Simon and {Wadadekar}, Yogesh and {Watters}, Shannon and {Wilhelm}, Ron and {Wyse}, Rosemary F.~G. and {Yarger}, Jean and {Zucker}, Dan},
        title = "{SEGUE: A Spectroscopic Survey of 240,000 Stars with g = 14-20}",
      journal = {\aj},
         year = 2009,
        month = may,
       volume = {137},
       number = {5},
        pages = {4377-4399},
          doi = {10.1088/0004-6256/137/5/4377},
archivePrefix = {arXiv},
       eprint = {0902.1781},
 primaryClass = {astro-ph.GA},
       adsurl = {https://ui.adsabs.harvard.edu/abs/2009AJ....137.4377Y}
}

@ARTICLE{gaia-eso,
       author = {{Gilmore}, G. and {Randich}, S. and {Asplund}, M. and {Binney}, J. and {Bonifacio}, P. and {Drew}, J. and {Feltzing}, S. and {Ferguson}, A. and {Jeffries}, R. and {Micela}, G. and {Negueruela}, I. and {Prusti}, T. and {Rix}, H. -W. and {Vallenari}, A. and {Alfaro}, E. and {Allende-Prieto}, C. and {Babusiaux}, C. and {Bensby}, T. and {Blomme}, R. and {Bragaglia}, A. and {Flaccomio}, E. and {Fran{\c{c}}ois}, P. and {Irwin}, M. and {Koposov}, S. and {Korn}, A. and {Lanzafame}, A. and {Pancino}, E. and {Paunzen}, E. and {Recio-Blanco}, A. and {Sacco}, G. and {Smiljanic}, R. and {Van Eck}, S. and {Walton}, N. and {Aden}, D. and {Aerts}, C. and {Affer}, L. and {Alcala}, J. -M. and {Altavilla}, G. and {Alves}, J. and {Antoja}, T. and {Arenou}, F. and {Argiroffi}, C. and {Asensio Ramos}, A. and {Bailer-Jones}, C. and {Balaguer-Nunez}, L. and {Bayo}, A. and {Barbuy}, B. and {Barisevicius}, G. and {Barrado y Navascues}, D. and {Battistini}, C. and {Bellas Velidis}, I. and {Bellazzini}, M. and {Belokurov}, V. and {Bergemann}, M. and {Bertelli}, G. and {Biazzo}, K. and {Bienayme}, O. and {Bland-Hawthorn}, J. and {Boeche}, C. and {Bonito}, S. and {Boudreault}, S. and {Bouvier}, J. and {Brandao}, I. and {Brown}, A. and {de Bruijne}, J. and {Burleigh}, M. and {Caballero}, J. and {Caffau}, E. and {Calura}, F. and {Capuzzo-Dolcetta}, R. and {Caramazza}, M. and {Carraro}, G. and {Casagrande}, L. and {Casewell}, S. and {Chapman}, S. and {Chiappini}, C. and {Chorniy}, Y. and {Christlieb}, N. and {Cignoni}, M. and {Cocozza}, G. and {Colless}, M. and {Collet}, R. and {Collins}, M. and {Correnti}, M. and {Covino}, E. and {Crnojevic}, D. and {Cropper}, M. and {Cunha}, M. and {Damiani}, F. and {David}, M. and {Delgado}, A. and {Duffau}, S. and {Edvardsson}, B. and {Eldridge}, J. and {Enke}, H. and {Eriksson}, K. and {Evans}, N.~W. and {Eyer}, L. and {Famaey}, B. and {Fellhauer}, M. and {Ferreras}, I. and {Figueras}, F. and {Fiorentino}, G. and {Flynn}, C. and {Folha}, D. and {Franciosini}, E. and {Frasca}, A. and {Freeman}, K. and {Fremat}, Y. and {Friel}, E. and {Gaensicke}, B. and {Gameiro}, J. and {Garzon}, F. and {Geier}, S. and {Geisler}, D. and {Gerhard}, O. and {Gibson}, B. and {Gomboc}, A. and {Gomez}, A. and {Gonzalez-Fernandez}, C. and {Gonzalez Hernandez}, J. and {Gosset}, E. and {Grebel}, E. and {Greimel}, R. and {Groenewegen}, M. and {Grundahl}, F. and {Guarcello}, M. and {Gustafsson}, B. and {Hadrava}, P. and {Hatzidimitriou}, D. and {Hambly}, N. and {Hammersley}, P. and {Hansen}, C. and {Haywood}, M. and {Heber}, U. and {Heiter}, U. and {Held}, E. and {Helmi}, A. and {Hensler}, G. and {Herrero}, A. and {Hill}, V. and {Hodgkin}, S. and {Huelamo}, N. and {Huxor}, A. and {Ibata}, R. and {Jackson}, R. and {de Jong}, R. and {Jonker}, P. and {Jordan}, S. and {Jordi}, C. and {Jorissen}, A. and {Katz}, D. and {Kawata}, D. and {Keller}, S. and {Kharchenko}, N. and {Klement}, R. and {Klutsch}, A. and {Knude}, J. and {Koch}, A. and {Kochukhov}, O. and {Kontizas}, M. and {Koubsky}, P. and {Lallement}, R. and {de Laverny}, P. and {van Leeuwen}, F. and {Lemasle}, B. and {Lewis}, G. and {Lind}, K. and {Lindstrom}, H.~P.~E. and {Lobel}, A. and {Lopez Santiago}, J. and {Lucas}, P. and {Ludwig}, H. and {Lueftinger}, T. and {Magrini}, L. and {Maiz Apellaniz}, J. and {Maldonado}, J. and {Marconi}, G. and {Marino}, A. and {Martayan}, C. and {Martinez-Valpuesta}, I. and {Matijevic}, G. and {McMahon}, R. and {Messina}, S. and {Meyer}, M. and {Miglio}, A. and {Mikolaitis}, S. and {Minchev}, I. and {Minniti}, D. and {Moitinho}, A. and {Momany}, Y. and {Monaco}, L. and {Montalto}, M. and {Monteiro}, M.~J. and {Monier}, R. and {Montes}, D. and {Mora}, A. and {Moraux}, E. and {Morel}, T. and {Mowlavi}, N.},
        title = "{The Gaia-ESO Public Spectroscopic Survey}",
      journal = {The Messenger},
         year = 2012,
        month = mar,
       volume = {147},
        pages = {25-31},
       adsurl = {https://ui.adsabs.harvard.edu/abs/2012Msngr.147...25G}
}

@ARTICLE{galah,
       author = {{De Silva}, G.~M. and {Freeman}, K.~C. and {Bland-Hawthorn}, J. and {Martell}, S. and {de Boer}, E. Wylie and {Asplund}, M. and {Keller}, S. and {Sharma}, S. and {Zucker}, D.~B. and {Zwitter}, T. and {Anguiano}, B. and {Bacigalupo}, C. and {Bayliss}, D. and {Beavis}, M.~A. and {Bergemann}, M. and {Campbell}, S. and {Cannon}, R. and {Carollo}, D. and {Casagrande}, L. and {Casey}, A.~R. and {Da Costa}, G. and {D'Orazi}, V. and {Dotter}, A. and {Duong}, L. and {Heger}, A. and {Ireland}, M.~J. and {Kafle}, P.~R. and {Kos}, J. and {Lattanzio}, J. and {Lewis}, G.~F. and {Lin}, J. and {Lind}, K. and {Munari}, U. and {Nataf}, D.~M. and {O'Toole}, S. and {Parker}, Q. and {Reid}, W. and {Schlesinger}, K.~J. and {Sheinis}, A. and {Simpson}, J.~D. and {Stello}, D. and {Ting}, Y. -S. and {Traven}, G. and {Watson}, F. and {Wittenmyer}, R. and {Yong}, D. and {{\v{Z}}erjal}, M.},
        title = "{The GALAH survey: scientific motivation}",
      journal = {\mnras},
         year = 2015,
        month = may,
       volume = {449},
       number = {3},
        pages = {2604-2617},
          doi = {10.1093/mnras/stv327},
archivePrefix = {arXiv},
       eprint = {1502.04767},
 primaryClass = {astro-ph.GA},
       adsurl = {https://ui.adsabs.harvard.edu/abs/2015MNRAS.449.2604D}
}

@ARTICLE{segue-2,
       author = {{Rockosi}, Constance M. and {Lee}, Young Sun and {Morrison}, Heather L. and {Yanny}, Brian and {Johnson}, Jennifer A. and {Lucatello}, Sara and {Sobeck}, Jennifer and {Beers}, Timothy C. and {Allende Prieto}, Carlos and {An}, Deokkeun and {Bizyaev}, Dmitry and {Blanton}, Michael R. and {Casagrande}, Luca and {Eisenstein}, Daniel J. and {Gould}, Andrew and {Gunn}, James E. and {Harding}, Paul and {Ivans}, Inese I. and {Jacobson}, H.~R. and {Janesh}, William and {Knapp}, Gillian R. and {Kollmeier}, Juna A. and {L{\'e}pine}, S{\'e}bastien and {L{\'o}pez-Corredoira}, Mart{\'\i}n and {Ma}, Zhibo and {Newberg}, Heidi J. and {Pan}, Kaike and {Prchlik}, Jakub and {Sayers}, Conor and {Schlesinger}, Katharine J. and {Simmerer}, Jennifer and {Weinberg}, David H.},
        title = "{SEGUE-2: Old Milky Way Stars Near and Far}",
      journal = {\apjs},
         year = 2022,
        month = apr,
       volume = {259},
       number = {2},
          eid = {60},
        pages = {60},
          doi = {10.3847/1538-4365/ac5323},
       adsurl = {https://ui.adsabs.harvard.edu/abs/2022ApJS..259...60R}
}

@ARTICLE{anders2014,
       author = {{Anders}, F. and {Chiappini}, C. and {Santiago}, B.~X. and {Rocha-Pinto}, H.~J. and {Girardi}, L. and {da Costa}, L.~N. and {Maia}, M.~A.~G. and {Steinmetz}, M. and {Minchev}, I. and {Schultheis}, M. and {Boeche}, C. and {Miglio}, A. and {Montalb{\'a}n}, J. and {Schneider}, D.~P. and {Beers}, T.~C. and {Cunha}, K. and {Allende Prieto}, C. and {Balbinot}, E. and {Bizyaev}, D. and {Brauer}, D.~E. and {Brinkmann}, J. and {Frinchaboy}, P.~M. and {Garc{\'\i}a P{\'e}rez}, A.~E. and {Hayden}, M.~R. and {Hearty}, F.~R. and {Holtzman}, J. and {Johnson}, J.~A. and {Kinemuchi}, K. and {Majewski}, S.~R. and {Malanushenko}, E. and {Malanushenko}, V. and {Nidever}, D.~L. and {O'Connell}, R.~W. and {Pan}, K. and {Robin}, A.~C. and {Schiavon}, R.~P. and {Shetrone}, M. and {Skrutskie}, M.~F. and {Smith}, V.~V. and {Stassun}, K. and {Zasowski}, G.},
        title = "{Chemodynamics of the Milky Way. I. The first year of APOGEE data}",
      journal = {\aap},
         year = 2014,
        month = apr,
       volume = {564},
          eid = {A115},
        pages = {A115},
          doi = {10.1051/0004-6361/201323038},
archivePrefix = {arXiv},
       eprint = {1311.4549},
 primaryClass = {astro-ph.GA},
       adsurl = {https://ui.adsabs.harvard.edu/abs/2014A&A...564A.115A}
}

@ARTICLE{bovy2016,
       author = {{Bovy}, Jo and {Rix}, Hans-Walter and {Schlafly}, Edward F. and {Nidever}, David L. and {Holtzman}, Jon A. and {Shetrone}, Matthew and {Beers}, Timothy C.},
        title = "{The Stellar Population Structure of the Galactic Disk}",
      journal = {\apj},
         year = 2016,
        month = may,
       volume = {823},
       number = {1},
          eid = {30},
        pages = {30},
          doi = {10.3847/0004-637X/823/1/30},
archivePrefix = {arXiv},
       eprint = {1509.05796},
 primaryClass = {astro-ph.GA},
       adsurl = {https://ui.adsabs.harvard.edu/abs/2016ApJ...823...30B}
}

@ARTICLE{robin1996,
       author = {{Robin}, A.~C. and {Haywood}, M. and {Creze}, M. and {Ojha}, D.~K. and {Bienayme}, O.},
        title = "{The thick disc of the Galaxy: sequel of a merging event.}",
      journal = {\aap},
         year = 1996,
        month = jan,
       volume = {305},
        pages = {125},
          doi = {10.48550/arXiv.astro-ph/9504090},
archivePrefix = {arXiv},
       eprint = {astro-ph/9504090},
 primaryClass = {astro-ph},
       adsurl = {https://ui.adsabs.harvard.edu/abs/1996A&A...305..125R}
}

@ARTICLE{juric2008,
       author = {{Juri{\'c}}, Mario and {Ivezi{\'c}}, {\v{Z}}eljko and {Brooks}, Alyson and {Lupton}, Robert H. and {Schlegel}, David and {Finkbeiner}, Douglas and {Padmanabhan}, Nikhil and {Bond}, Nicholas and {Sesar}, Branimir and {Rockosi}, Constance M. and {Knapp}, Gillian R. and {Gunn}, James E. and {Sumi}, Takahiro and {Schneider}, Donald P. and {Barentine}, J.~C. and {Brewington}, Howard J. and {Brinkmann}, J. and {Fukugita}, Masataka and {Harvanek}, Michael and {Kleinman}, S.~J. and {Krzesinski}, Jurek and {Long}, Dan and {Neilsen}, Jr., Eric H. and {Nitta}, Atsuko and {Snedden}, Stephanie A. and {York}, Donald G.},
        title = "{The Milky Way Tomography with SDSS. I. Stellar Number Density Distribution}",
      journal = {\apj},
         year = 2008,
        month = feb,
       volume = {673},
       number = {2},
        pages = {864-914},
          doi = {10.1086/523619},
archivePrefix = {arXiv},
       eprint = {astro-ph/0510520},
 primaryClass = {astro-ph},
       adsurl = {https://ui.adsabs.harvard.edu/abs/2008ApJ...673..864J}
}

@ARTICLE{minchev2013,
       author = {{Minchev}, I. and {Chiappini}, C. and {Martig}, M.},
        title = "{Chemodynamical evolution of the Milky Way disk. I. The solar vicinity}",
      journal = {\aap},
         year = 2013,
        month = oct,
       volume = {558},
          eid = {A9},
        pages = {A9},
          doi = {10.1051/0004-6361/201220189},
archivePrefix = {arXiv},
       eprint = {1208.1506},
 primaryClass = {astro-ph.GA},
       adsurl = {https://ui.adsabs.harvard.edu/abs/2013A&A...558A...9M}
}

@ARTICLE{linden2017,
       author = {{Linden}, Sean T. and {Pryal}, Matthew and {Hayes}, Christian R. and {Troup}, Nicholas W. and {Majewski}, Steven R. and {Andrews}, Brett H. and {Beers}, Timothy C. and {Carrera}, Ricardo and {Cunha}, Katia and {Fern{\'a}ndez-Trincado}, J.~G. and {Frinchaboy}, Peter and {Geisler}, Doug and {Lane}, Richard R. and {Nitschelm}, Christian and {Pan}, Kaike and {Allende Prieto}, Carlos and {Roman-Lopes}, Alexandre and {Smith}, Verne V. and {Sobeck}, Jennifer and {Tang}, Baitian and {Villanova}, Sandro and {Zasowski}, Gail},
        title = "{Timing the Evolution of the Galactic Disk with NGC 6791: An Open Cluster with Peculiar High-{\ensuremath{\alpha}} Chemistry as Seen by APOGEE}",
      journal = {\apj},
         year = 2017,
        month = jun,
       volume = {842},
       number = {1},
          eid = {49},
        pages = {49},
          doi = {10.3847/1538-4357/aa6f17},
archivePrefix = {arXiv},
       eprint = {1704.07305},
 primaryClass = {astro-ph.GA},
       adsurl = {https://ui.adsabs.harvard.edu/abs/2017ApJ...842...49L}
}

@ARTICLE{sahlholdt2022,
       author = {{Sahlholdt}, Christian L. and {Feltzing}, Sofia and {Feuillet}, Diane K.},
        title = "{Characterizing epochs of star formation across the Milky Way disc using age-metallicity distributions of GALAH stars}",
      journal = {\mnras},
         year = 2022,
        month = mar,
       volume = {510},
       number = {4},
        pages = {4669-4688},
          doi = {10.1093/mnras/stab3681},
archivePrefix = {arXiv},
       eprint = {2112.08218},
 primaryClass = {astro-ph.GA},
       adsurl = {https://ui.adsabs.harvard.edu/abs/2022MNRAS.510.4669S}
}

@ARTICLE{deng2024,
       author = {{Deng}, Mingji and {Du}, Cuihua and {Yang}, Yanbin and {Liao}, Jiwei and {Ye}, Dashuang},
        title = "{A Potential Dynamical Origin of the Galactic Disk Warp: The Gaia{\textendash}Sausage{\textendash}Enceladus Major Merger}",
      journal = {\apj},
         year = 2024,
        month = nov,
       volume = {975},
       number = {1},
          eid = {28},
        pages = {28},
          doi = {10.3847/1538-4357/ad7799},
archivePrefix = {arXiv},
       eprint = {2409.03264},
 primaryClass = {astro-ph.GA},
       adsurl = {https://ui.adsabs.harvard.edu/abs/2024ApJ...975...28D}
}

@ARTICLE{dodge2023,
       author = {{Dodge}, Benjamin C. and {Slone}, Oren and {Lisanti}, Mariangela and {Cohen}, Timothy},
        title = "{Dynamics of stellar disc tilting from satellite mergers}",
      journal = {\mnras},
         year = 2023,
        month = jan,
       volume = {518},
       number = {2},
        pages = {2870-2884},
          doi = {10.1093/mnras/stac3249},
archivePrefix = {arXiv},
       eprint = {2207.02861},
 primaryClass = {astro-ph.GA},
       adsurl = {https://ui.adsabs.harvard.edu/abs/2023MNRAS.518.2870D}
}

@ARTICLE{dillamore2022,
       author = {{Dillamore}, Adam M. and {Belokurov}, Vasily and {Font}, Andreea S. and {McCarthy}, Ian G.},
        title = "{Merger-induced galaxy transformations in the ARTEMIS simulations}",
      journal = {\mnras},
         year = 2022,
        month = jun,
       volume = {513},
       number = {2},
        pages = {1867-1886},
          doi = {10.1093/mnras/stac1038},
archivePrefix = {arXiv},
       eprint = {2109.13244},
 primaryClass = {astro-ph.GA},
       adsurl = {https://ui.adsabs.harvard.edu/abs/2022MNRAS.513.1867D}
}

@ARTICLE{buck2023,
       author = {{Buck}, Tobias and {Obreja}, Aura and {Ratcliffe}, Bridget and {Lu}, Yuxi(Lucy) and {Minchev}, Ivan and {Macci{\`o}}, Andrea V.},
        title = "{The impact of early massive mergers on the chemical evolution of Milky Way-like galaxies: insights from NIHAO-UHD simulations}",
      journal = {\mnras},
         year = 2023,
        month = jul,
       volume = {523},
       number = {1},
        pages = {1565-1576},
          doi = {10.1093/mnras/stad1503},
archivePrefix = {arXiv},
       eprint = {2305.13759},
 primaryClass = {astro-ph.GA},
       adsurl = {https://ui.adsabs.harvard.edu/abs/2023MNRAS.523.1565B}
}

@ARTICLE{livermore2012,
       author = {{Livermore}, R.~C. and {Jones}, T. and {Richard}, J. and {Bower}, R.~G. and {Ellis}, R.~S. and {Swinbank}, A.~M. and {Rigby}, J.~R. and {Smail}, Ian and {Arribas}, S. and {Rodriguez Zaurin}, J. and {Colina}, L. and {Ebeling}, H. and {Crain}, R.~A.},
        title = "{Hubble Space Telescope H{\ensuremath{\alpha}} imaging of star-forming galaxies at z ≃ 1-1.5: evolution in the size and luminosity of giant H II regions}",
      journal = {\mnras},
         year = 2012,
        month = nov,
       volume = {427},
       number = {1},
        pages = {688-702},
          doi = {10.1111/j.1365-2966.2012.21900.x},
archivePrefix = {arXiv},
       eprint = {1209.5741},
 primaryClass = {astro-ph.CO},
       adsurl = {https://ui.adsabs.harvard.edu/abs/2012MNRAS.427..688L}
}

@ARTICLE{elmegreen2007,
       author = {{Elmegreen}, Debra Meloy and {Elmegreen}, Bruce G. and {Ravindranath}, Swara and {Coe}, Daniel A.},
        title = "{Resolved Galaxies in the Hubble Ultra Deep Field: Star Formation in Disks at High Redshift}",
      journal = {\apj},
         year = 2007,
        month = apr,
       volume = {658},
       number = {2},
        pages = {763-777},
          doi = {10.1086/511667},
archivePrefix = {arXiv},
       eprint = {astro-ph/0701121},
 primaryClass = {astro-ph},
       adsurl = {https://ui.adsabs.harvard.edu/abs/2007ApJ...658..763E}
}

@ARTICLE{dekel2009,
       author = {{Dekel}, Avishai and {Sari}, Re'em and {Ceverino}, Daniel},
        title = "{Formation of Massive Galaxies at High Redshift: Cold Streams, Clumpy Disks, and Compact Spheroids}",
      journal = {\apj},
         year = 2009,
        month = sep,
       volume = {703},
       number = {1},
        pages = {785-801},
          doi = {10.1088/0004-637X/703/1/785},
archivePrefix = {arXiv},
       eprint = {0901.2458},
 primaryClass = {astro-ph.GA},
       adsurl = {https://ui.adsabs.harvard.edu/abs/2009ApJ...703..785D}
}

@ARTICLE{richard2010,
       author = {{Richard}, Simon and {Brook}, Chris B. and {Martel}, Hugo and {Kawata}, Daisuke and {Gibson}, Brad K. and {Sanchez-Blazquez}, Patricia},
        title = "{Structure, kinematics and chemical enrichment patterns after major gas-rich disc-disc mergers}",
      journal = {\mnras},
         year = 2010,
        month = mar,
       volume = {402},
       number = {3},
        pages = {1489-1503},
          doi = {10.1111/j.1365-2966.2009.16008.x},
archivePrefix = {arXiv},
       eprint = {0911.1801},
 primaryClass = {astro-ph.CO},
       adsurl = {https://ui.adsabs.harvard.edu/abs/2010MNRAS.402.1489R}
}

@ARTICLE{parul2025,
       author = {{Parul}, Hanna and {Bailin}, Jeremy and {Loebman}, Sarah R. and {Wetzel}, Andrew and {Barry}, Megan and {Bhattarai}, Binod},
        title = "{Effect of gas accretion on {\ensuremath{\alpha}}-element bimodality in Milky Way-mass galaxies in the FIRE-2 simulations}",
      journal = {\mnras},
         year = 2025,
        month = feb,
       volume = {537},
       number = {2},
        pages = {1571-1585},
          doi = {10.1093/mnras/staf137},
archivePrefix = {arXiv},
       eprint = {2501.12342},
 primaryClass = {astro-ph.GA},
       adsurl = {https://ui.adsabs.harvard.edu/abs/2025MNRAS.537.1571P}
}

@ARTICLE{miglio2021,
       author = {{Miglio}, A. and {Chiappini}, C. and {Mackereth}, J.~T. and {Davies}, G.~R. and {Brogaard}, K. and {Casagrande}, L. and {Chaplin}, W.~J. and {Girardi}, L. and {Kawata}, D. and {Khan}, S. and {Izzard}, R. and {Montalb{\'a}n}, J. and {Mosser}, B. and {Vincenzo}, F. and {Bossini}, D. and {Noels}, A. and {Rodrigues}, T. and {Valentini}, M. and {Mandel}, I.},
        title = "{Age dissection of the Milky Way discs: Red giants in the Kepler field}",
      journal = {\aap},
         year = 2021,
        month = jan,
       volume = {645},
          eid = {A85},
        pages = {A85},
          doi = {10.1051/0004-6361/202038307},
archivePrefix = {arXiv},
       eprint = {2004.14806},
 primaryClass = {astro-ph.GA},
       adsurl = {https://ui.adsabs.harvard.edu/abs/2021A&A...645A..85M}
}

@ARTICLE{bonaca2020,
       author = {{Bonaca}, Ana and {Conroy}, Charlie and {Cargile}, Phillip A. and {Naidu}, Rohan P. and {Johnson}, Benjamin D. and {Zaritsky}, Dennis and {Ting}, Yuan-Sen and {Caldwell}, Nelson and {Han}, Jiwon Jesse and {van Dokkum}, Pieter},
        title = "{Timing the Early Assembly of the Milky Way with the H3 Survey}",
      journal = {\apjl},
         year = 2020,
        month = jul,
       volume = {897},
       number = {1},
          eid = {L18},
        pages = {L18},
          doi = {10.3847/2041-8213/ab9caa},
archivePrefix = {arXiv},
       eprint = {2004.11384},
 primaryClass = {astro-ph.GA},
       adsurl = {https://ui.adsabs.harvard.edu/abs/2020ApJ...897L..18B}
}

@ARTICLE{spitoni2021,
       author = {{Spitoni}, E. and {Verma}, K. and {Silva Aguirre}, V. and {Vincenzo}, F. and {Matteucci}, F. and {Vai{\v{c}}ekauskait{\.{e}}}, B. and {Palla}, M. and {Grisoni}, V. and {Calura}, F.},
        title = "{APOGEE DR16: A multi-zone chemical evolution model for the Galactic disc based on MCMC methods}",
      journal = {\aap},
         year = 2021,
        month = mar,
       volume = {647},
          eid = {A73},
        pages = {A73},
          doi = {10.1051/0004-6361/202039864},
archivePrefix = {arXiv},
       eprint = {2101.08803},
 primaryClass = {astro-ph.GA},
       adsurl = {https://ui.adsabs.harvard.edu/abs/2021A&A...647A..73S}
}

@ARTICLE{angles2017,
       author = {{Angl{\'e}s-Alc{\'a}zar}, Daniel and {Faucher-Gigu{\`e}re}, Claude-Andr{\'e} and {Kere{\v{s}}}, Du{\v{s}}an and {Hopkins}, Philip F. and {Quataert}, Eliot and {Murray}, Norman},
        title = "{The cosmic baryon cycle and galaxy mass assembly in the FIRE simulations}",
      journal = {\mnras},
         year = 2017,
        month = oct,
       volume = {470},
       number = {4},
        pages = {4698-4719},
          doi = {10.1093/mnras/stx1517},
archivePrefix = {arXiv},
       eprint = {1610.08523},
 primaryClass = {astro-ph.GA},
       adsurl = {https://ui.adsabs.harvard.edu/abs/2017MNRAS.470.4698A}
}

@ARTICLE{queiroz2023,
       author = {{Queiroz}, A.~B.~A. and {Anders}, F. and {Chiappini}, C. and {Khalatyan}, A. and {Santiago}, B.~X. and {Nepal}, S. and {Steinmetz}, M. and {Gallart}, C. and {Valentini}, M. and {Dal Ponte}, M. and {Barbuy}, B. and {P{\'e}rez-Villegas}, A. and {Masseron}, T. and {Fern{\'a}ndez-Trincado}, J.~G. and {Khoperskov}, S. and {Minchev}, I. and {Fern{\'a}ndez-Alvar}, E. and {Lane}, R.~R. and {Nitschelm}, C.},
        title = "{StarHorse results for spectroscopic surveys and Gaia DR3: Chrono-chemical populations in the solar vicinity, the genuine thick disk, and young alpha-rich stars}",
      journal = {\aap},
         year = 2023,
        month = may,
       volume = {673},
          eid = {A155},
        pages = {A155},
          doi = {10.1051/0004-6361/202245399},
archivePrefix = {arXiv},
       eprint = {2303.09926},
 primaryClass = {astro-ph.GA},
       adsurl = {https://ui.adsabs.harvard.edu/abs/2023A&A...673A.155Q}
}

@ARTICLE{snaith2016,
       author = {{Snaith}, O.~N. and {Bailin}, J. and {Gibson}, B.~K. and {Bell}, E.~F. and {Stinson}, G. and {Valluri}, M. and {Wadsley}, J. and {Couchman}, H.},
        title = "{The history of stellar metallicity in a simulated disc galaxy}",
      journal = {\mnras},
         year = 2016,
        month = mar,
       volume = {456},
       number = {3},
        pages = {3119-3141},
          doi = {10.1093/mnras/stv2788},
archivePrefix = {arXiv},
       eprint = {1512.02680},
 primaryClass = {astro-ph.GA},
       adsurl = {https://ui.adsabs.harvard.edu/abs/2016MNRAS.456.3119S}
}

@ARTICLE{mackereth2018,
       author = {{Mackereth}, J. Ted and {Crain}, Robert A. and {Schiavon}, Ricardo P. and {Schaye}, Joop and {Theuns}, Tom and {Schaller}, Matthieu},
        title = "{The origin of diverse {\ensuremath{\alpha}}-element abundances in galaxy discs}",
      journal = {\mnras},
         year = 2018,
        month = jul,
       volume = {477},
       number = {4},
        pages = {5072-5089},
          doi = {10.1093/mnras/sty972},
archivePrefix = {arXiv},
       eprint = {1801.03593},
 primaryClass = {astro-ph.GA},
       adsurl = {https://ui.adsabs.harvard.edu/abs/2018MNRAS.477.5072M}
}

@ARTICLE{elmegreen2005,
       author = {{Elmegreen}, Bruce G. and {Elmegreen}, Debra Meloy},
        title = "{Stellar Populations in 10 Clump-Cluster Galaxies of the Hubble Ultra Deep Field}",
      journal = {\apj},
         year = 2005,
        month = jul,
       volume = {627},
       number = {2},
        pages = {632-646},
          doi = {10.1086/430514},
archivePrefix = {arXiv},
       eprint = {astro-ph/0504032},
 primaryClass = {astro-ph},
       adsurl = {https://ui.adsabs.harvard.edu/abs/2005ApJ...627..632E}
}

@ARTICLE{wuyts2012,
       author = {{Wuyts}, Stijn and {F{\"o}rster Schreiber}, Natascha M. and {Genzel}, Reinhard and {Guo}, Yicheng and {Barro}, Guillermo and {Bell}, Eric F. and {Dekel}, Avishai and {Faber}, Sandra M. and {Ferguson}, Henry C. and {Giavalisco}, Mauro and {Grogin}, Norman A. and {Hathi}, Nimish P. and {Huang}, Kuang-Han and {Kocevski}, Dale D. and {Koekemoer}, Anton M. and {Koo}, David C. and {Lotz}, Jennifer and {Lutz}, Dieter and {McGrath}, Elizabeth and {Newman}, Jeffrey A. and {Rosario}, David and {Saintonge}, Amelie and {Tacconi}, Linda J. and {Weiner}, Benjamin J. and {van der Wel}, Arjen},
        title = "{Smooth(er) Stellar Mass Maps in CANDELS: Constraints on the Longevity of Clumps in High-redshift Star-forming Galaxies}",
      journal = {\apj},
         year = 2012,
        month = jul,
       volume = {753},
       number = {2},
          eid = {114},
        pages = {114},
          doi = {10.1088/0004-637X/753/2/114},
archivePrefix = {arXiv},
       eprint = {1203.2611},
 primaryClass = {astro-ph.CO},
       adsurl = {https://ui.adsabs.harvard.edu/abs/2012ApJ...753..114W}
}

@ARTICLE{dubay2024,
       author = {{Dubay}, Liam O. and {Johnson}, Jennifer A. and {Johnson}, James W.},
        title = "{Galactic Chemical Evolution Models Favor an Extended Type Ia Supernova Delay-time Distribution}",
      journal = {\apj},
         year = 2024,
        month = sep,
       volume = {973},
       number = {1},
          eid = {55},
        pages = {55},
          doi = {10.3847/1538-4357/ad61df},
archivePrefix = {arXiv},
       eprint = {2404.08059},
 primaryClass = {astro-ph.GA},
       adsurl = {https://ui.adsabs.harvard.edu/abs/2024ApJ...973...55D}
}

@ARTICLE{chronos,
       author = {{Michel}, Eric and {Haywood}, Misha and {Famaey}, Benoit and {Mosser}, Benoit and {Samadi}, Reza and {Monteiro}, Mario J.~P.~F.~G. and {Kjeldsen}, Hans and {Belkacem}, Kevin and {Miglio}, Andr{\'e}a and {Garcia}, Rafael and {Katz}, David and {Suarez}, Juan Carlos and {Deheuvels}, S{\'e}bastien and {Campante}, Tiago and {Cunha}, Margarida and {Aguirre}, Victor Silva and {Ballot}, Jer{\^o}me and {Moya}, Andy},
        title = "{Chronos - take the pulse of our galactic neighbourhood}",
      journal = {Experimental Astronomy},
         year = 2021,
        month = jun,
       volume = {51},
       number = {3},
        pages = {945-962},
          doi = {10.1007/s10686-021-09733-9},
archivePrefix = {arXiv},
       eprint = {1908.10977},
 primaryClass = {astro-ph.SR},
       adsurl = {https://ui.adsabs.harvard.edu/abs/2021ExA....51..945M}
}




\appendix

\section{Full star formation history plot} \label{AppendixSFH}

\begin{figure*}
\centering
  \setlength\tabcolsep{2pt}%
    \includegraphics[keepaspectratio, trim={0.0cm 0.0cm 0.0cm 0.0cm}, width=1\linewidth]{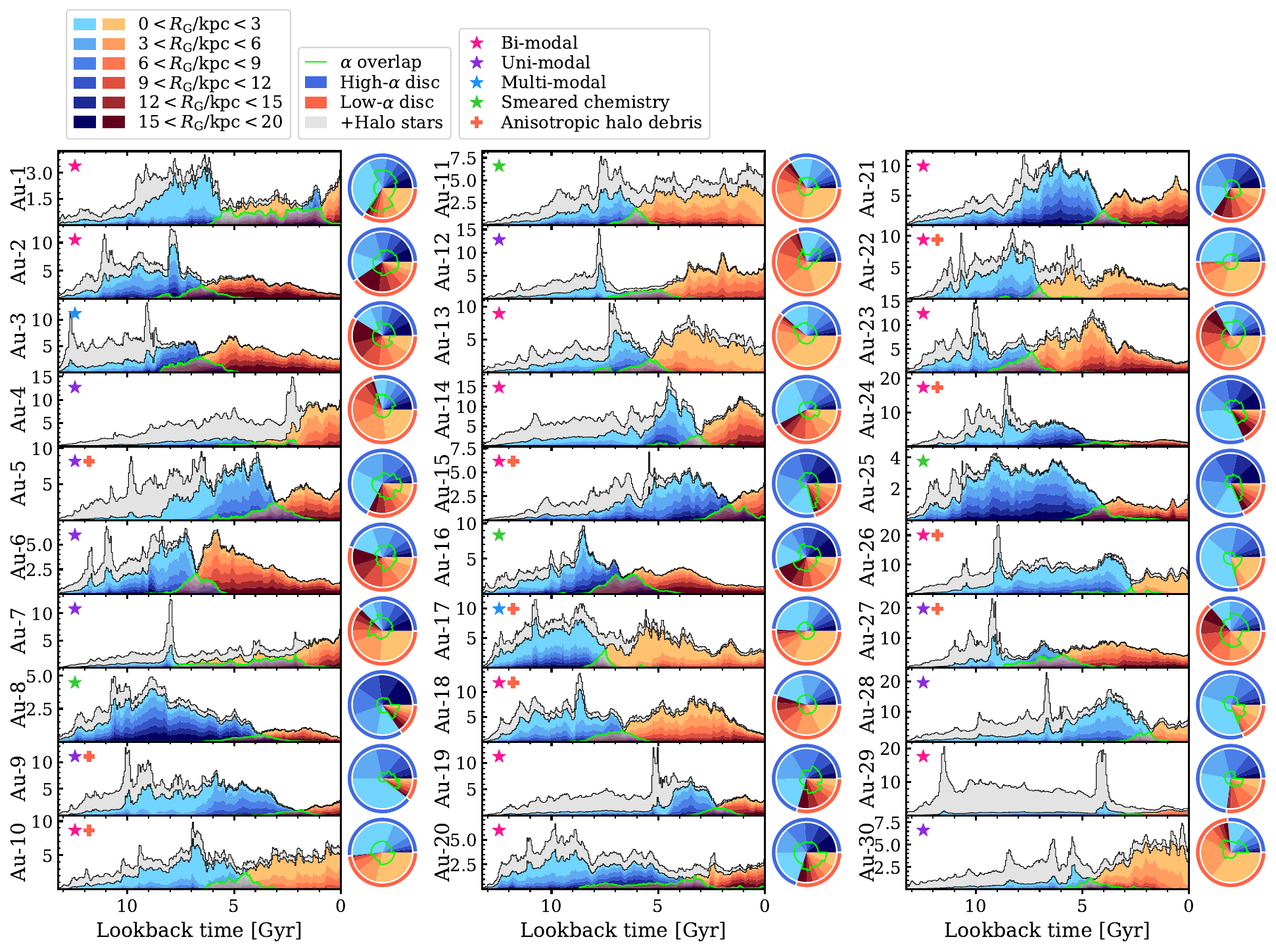}\\
\caption{The same as in Figure \ref{fig:SFH}, but now for the entire {\sc auriga} suite.}
\label{fig:SFH_full}
\end{figure*}

We include star formation histories for the full {\sc auriga} suite in Figure \ref{fig:SFH_full}.

\section{Estimating the gas dilution in the MW due to the GSE} \label{AppendixB}

\renewcommand{\arraystretch}{1.5}
\begin{table}
\setlength{\tabcolsep}{4pt} 
\resizebox{\columnwidth}{!}{
\begin{tabular}{l|ccc|c} 
\toprule
Model & $M_{\star}$ [M$_{\odot}$] & $M_{\rm gas}$ [M$_{\odot}$] & $M_{\rm DM}$ [M$_{\odot}$] & Gas [Fe/H] \\
\midrule
MW & $4.92^{+7.97}_{-3.37} \times 10^9$ & $4.92^{+8.28}_{-3.36} \times 10^9$ & $4.92^{+4.97}_{-2.96} \times 10^{11}$ & $0.00\pm0.05$ \\
\textcolor{GSE1}{\textbf{GSE 1}} & $3.69^{+31.5}_{-0.33} \times 10^7$ & $1.47^{+13.3}_{-1.30} \times 10^8$ & $3.37^{+8.57}_{-2.89} \times 10^{10}$ & $-1.22\pm0.23^{\rm \,\hyperlink{cite.limberg2022}{a}}$ \\
\textcolor{GSE2}{\textbf{GSE 2}} & $1.45^{+0.92}_{-0.51} \times 10^{8\rm \,\hyperlink{cite.lane2023}{b}}$ & $5.80^{+5.59}_{-2.38} \times 10^8$ & $7.07^{+6.00}_{-3.59} \times 10^{10}$ & $-1.04^{+0.20}_{-0.18}$ \\
\textcolor{GSE3}{\textbf{GSE 3}} & $6 \times 10^{8\rm \,\hyperlink{cite.helmi2018}{c} }$ & $2.40^{+1.56}_{-0.76} \times 10^9$ & $1.53^{+1.08}_{-0.65} \times 10^{11}$ & $-1.6^{\rm \,\hyperlink{cite.helmi2018}{c}}$ \\
\bottomrule
\end{tabular}
}
\caption{The model properties for the proto-MW and GSE merger at a time of 10\,Gyr ago or $z\approx2$. Sources are as follows: a) \citet{limberg2022}, b) \citet{lane2023}, c) \citet{helmi2018}. The methods used to derive these model parameters are described in the main text of Section \ref{AppendixB}.}
\label{tab:MZR}
\end{table}

\begin{figure}
  \setlength\tabcolsep{2pt}%
    \includegraphics[keepaspectratio, trim={0.0cm 0.5cm 0.0cm 0.0cm}, width=\columnwidth]{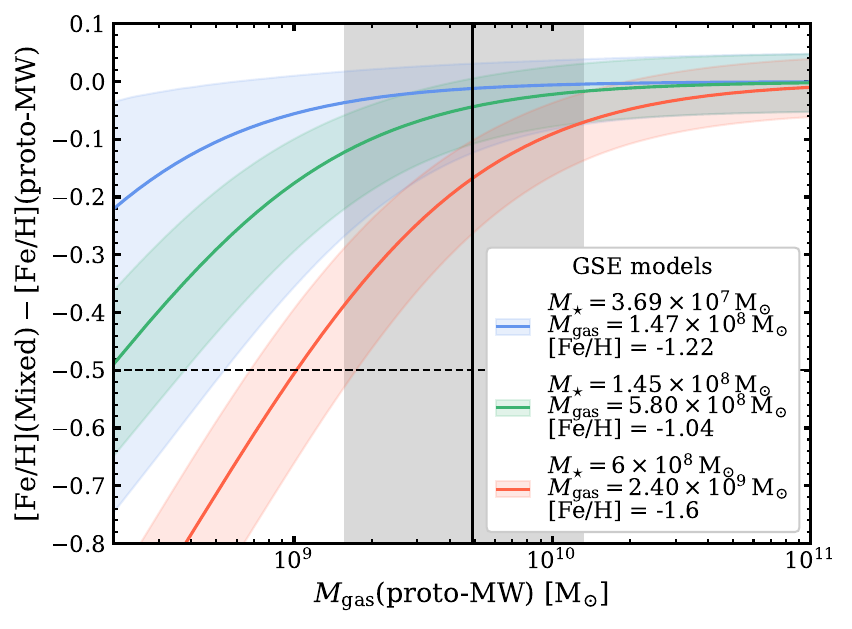}\\
\caption{The metallicity dilution resulting from three different GSE merger models (see legend and further description in the main text) mixing with a range of gas masses at a metallicity of $0.00\pm0.05$ (approximately the metal-rich end of the MW high-$\alpha$ sequence). The shaded bands represent the 1$\sigma$ uncertainty in each case. The vertical black line is an estimate of the gas mass in the proto-MW at the time of the GSE merger ($10\,$Gyr ago) where once again the shaded band represents the 1$\sigma$ uncertainty. A dashed horizontal line marks a dilution of 0.5\,dex, which we consider the minimum dilution required to meet the requirements of the two-infall model.}
\label{fig:MZR}
\end{figure}

We extend our analysis by estimating the metallicity dilution resulting from the GSE infall event. Our model parameters are described in Table \ref{tab:MZR}, including the stellar, gas and halo mass, as well as the gas metallicity for both the proto-MW and GSE. We now describe our methods for deriving these model parameters.

We assume that the GSE infall occurred 10\,Gyr ago, which corresponds to a redshift of $z\approx2$. The stellar mass of the proto-MW at this time can be estimated following comparisons with MW analogue galaxies in observational surveys \citep[i.e.][]{dokum2013, patel2013b, papovich2015, tan2024b}. We opt to employ the relation with 1$\sigma$ scatter shown in figure 2 of \citet{tan2024b}, based upon a selection of MW analogues out to $z=5$ from the CANUCS survey \citep{willott2022}. The gas mass is determined assuming a gas fraction of 0.4-0.6 (loosely based on measurements of star-forming massive galaxies at $z=2$ \citealp{tacconi2013}), where $f_{\rm gas} = M_{\rm gas}/(M_{\rm gas}+M_{\star})$. The corresponding halo mass is found following the redshift-dependent stellar mass-halo mass relation of \citet{moster2013} (hereafter Moster13 SMHM).

We estimate the gas metallicity by considering that the stars born at the metal-rich end of the MW high-$\alpha$ sequence will reflect the ambient gas metallicity before the GSE infall. From visual inspection of the figures in \citet{queiroz2020}, we take $\rm{[Fe/H]}=0.00\pm0.05$ as a rough guideline. Then, the amount of dilution required to reach the metal-poor end of the low-$\alpha$ sequence would be around -0.5\,dex. This is far less than what is presented in the chemical evolution models of \citep{spitoni2019, spitoni2021, spitoni2023}, but we consider it as a conservative minimum required dilution.

An alternative estimate of the metallicity can be obtained using the stellar mass–metallicity relation from \citet{kirby2013} (given by their equation 4, hereafter Kirby13 MZR), which gives $\rm{[Fe/H]}(\rm{proto\text{-}MW})=-0.58^{+0.27}_{-0.16}$. Although this relation is calibrated on galaxy observations at $z=0$, it nonetheless captures the spread and median of observed stellar metallicities within the high-$\alpha$ sequence. However, it is not suitable for our purposes, as we are interested in the chemical evolution stage near the \textit{end} of the high-$\alpha$ sequence where metallicities approach $\rm{[Fe/H]}\approx0$.

For the properties of the GSE progenitor galaxy, we construct three models based upon literature values. We assume the gas metallicity matches the stellar metallicity. Average stellar enrichment lags behind the gas enrichment, and so this choice should be considered as a low estimate of the gas metallicity. We calculate the gas mass assuming high gas fractions of 0.7-0.9, reflecting the high gas fractions expected for ancient dwarf galaxies \citep[i.e.][]{popping2012}. Once again, the halo masses are approximated from the Moster13 SMHM. Our models are described as follows:
\begin{itemize}
\setlength{\itemindent}{0.5em}
    \item GSE 1: We take the GSE stellar metallicity value of $\rm{[Fe/H]}=-1.22\pm0.23$ from \citet{limberg2022} and find the corresponding stellar mass from the Kirby13 MZR. We note that \citet{limberg2022} also quote their own stellar mass estimate of $\sim1.3\times10^9\,\rm{M}_{\odot}$, but we determine that this is likely too high because the entire MW stellar halo is only $\sim1.4\times10^9\,\rm{M}_{\odot}$ \citep{deason2019}.
    \item GSE 2: We take the GSE stellar mass value of $M_{\star} = 1.45^{+0.92}_{-0.51} \times 10^8\,\rm{M}_{\odot}$ from \citet{lane2023} and find the corresponding stellar metallicity from the inverted Kirby13 MZR.
    \item GSE 3: Here, we take the direct GSE stellar mass and metallicity values reported in \citet{helmi2018} as $M_{\star} = 6 \times 10^8\,\rm{M}_{\odot}$ and $\rm{[Fe/H]}=-1.6$, and note that these values are significantly offset from the Kirby13 MZR. This model can be considered as a particularly high-mass and metal-poor estimate for the GSE, thereby maximising any resulting dilution.
\end{itemize}

The uncertainties on the parameters in Table \ref{tab:MZR} are either taken from their respective sources, or are approximated as the $\pm\sigma$ percentiles from a Monte Carlo method using $10^6$ samples. Symmetric uncertainties are sampled from a normal distribution, whereas the upper and lower bounds for asymmetric uncertainties are calculated independently with a two-piece normal distribution provided by the {\sc twopiece} {\sc python} package. The distributions in $f_{\rm gas}$ are sampled from a uniform distribution. The Kirby13 MZR and Moster13 SMHM have a further 0.17 and 0.15\,dex scatter respectively, which we include in addition to the parametric uncertainties of each relation. The GSE parameters from \citet{helmi2018} do not include uncertainties, and so we caution the reader that the true uncertainties will be greater than those shown here.

We calculate the gas dilution by averaging the metallicities of the GSE model and the proto-MW, weighting by their respective gas masses. It is conceivable that the GSE gas did not mix homogeneously with the proto-MW gas, and so we choose to calculate the dilution over a range of proto-MW gas masses in Figure \ref{fig:MZR}. There, the vertical black line indicates our estimate for the true gas mass of the proto-MW from Table \ref{tab:MZR}, with associated uncertainty represented as a shaded band. This figure shows that the most likely outcome is for a dilution magnitude of $<0.2$, very similar to the maximum dilution magnitudes that we estimated for {\sc auriga} in panels (b) and (c) of Figure \ref{fig:mixing}, but far from the required 0.5\,dex needed in the MW. The dilution magnitude remains relatively low even in cases where the GSE gas has a greater mass than the proto-MW gas, and this is due to the logarithmic nature of the [Fe/H] ratio.

When the uncertainties are accounted for, the dilution magnitude can exceed 0.5\,dex for the GSE 3 model. However, this is assuming a very high GSE gas mass and a very low proto-MW gas mass simultaneously, which would be disfavoured by current estimates of the total merger mass ratio (of $\leq 1:4$). Also, a higher GSE stellar mass with a fixed halo mass would favour a lower gas fraction. These two effects are not considered by our uncertainty estimates. For the other two models, greater dilution magnitudes can only be achieved assuming that the GSE gas mixes with an infinitesimal volume of proto-MW gas, which is not realistic.

The GSE may have deposited its gas around the edge of the proto-MW gas disc, where the gas densities are lower and higher dilutions can be achieved more easily. An example of this scenario is described in {\sc auriga} by \citet{grand2018}, for a late-time minor merger in the higher-resolution version of Au-23. This merger has a wide pericentre passage of $>40\,$kpc, which enables its gas to be gradually stripped away and accreted onto the edge of the gas disc. This is a less likely possibility for the GSE, which must have had a highly radial infall trajectory with a low pericentre.

Our results are consistent with the study by \citet{palla2020}. They model the gas contribution from the GSE, calibrated to match the abundances reported in \citet{helmi2018}. The results indicate that the merger supplied approximately 15 per cent of the total gas during the second-infall epoch, remarkably close to the fractions shown in our Figure \ref{fig:tracers} in the case of a $1:8$ mass ratio merger. They also find a minimal metal dilution, no matter the assumed accretion time or the chemical composition of the merger gas.

\bsp	
\label{lastpage}
1\end{document}